\documentclass[11pt]{article}
\pdfoutput=1

\usepackage{amssymb}
\usepackage{amsmath}
\usepackage{bbold}
\usepackage{color}
\usepackage{colordvi}
\usepackage{dsfont}
\usepackage{fancybox}
\usepackage[footnotesize]{caption2}
\usepackage{graphicx}
\usepackage[center,footnotesize,hang]{subfigure}
\usepackage{bbm}
\usepackage{url}
\usepackage{multirow}
\usepackage{array}
\usepackage{arydshln}
\usepackage{pifont}
\usepackage{mathrsfs}
\usepackage{fancybox}
\usepackage{cite}
\usepackage[colorlinks]{hyperref}
\usepackage{enumitem}  
\usepackage{float} %
\usepackage{longtable}
\usepackage{enumitem}
\newcommand{\PreserveBackslash}[1]{\let\temp=\\#1\let\\=\temp}
\newcolumntype{C}[1]{>{\PreserveBackslash\centering}p{#1}}
\newcolumntype{R}[1]{>{\PreserveBackslash\raggedleft}p{#1}}
\newcolumntype{L}[1]{>{\PreserveBackslash\raggedright}p{#1}}
\allowdisplaybreaks  

\newcommand{\bq}{\begin{eqnarray}}
\newcommand{\nq}{\end{eqnarray}}

\def\bvec#1{\raise1.5ex\hbox{$\rightarrow$}\mkern-16.5mu #1}

\begin{document}
\title{
\begin{flushright}
\hfill\mbox{{\small\tt USTC-ICTS/PCFT-20-11}} \\[5mm]
\begin{minipage}{0.2\linewidth}
\normalsize
\end{minipage}
\end{flushright}
{\Large \bf
Modular Invariant Models of Leptons at Level 7
\\[2mm]}}
\date{}

\author{
Gui-Jun~Ding$^{1,2}$\footnote{E-mail: {\tt
dinggj@ustc.edu.cn}},  \
Stephen~F.~King$^{3}$\footnote{E-mail: {\tt king@soton.ac.uk}; ORCID: https://orcid.org/0000-0002-4351-7507}, \
Cai-Chang Li$^{4,5,6}$\footnote{E-mail: {\tt
lcc0915@mail.ustc.edu.cn}},  \
Ye-Ling Zhou$^{3}$\footnote{E-mail: {\tt
ye-ling.zhou@soton.ac.uk}; ORCID: https://orcid.org/0000-0002-3664-9472}  \
\\*[20pt]
\centerline{
\begin{minipage}{\linewidth}
\begin{center}
$^1${\it \small Peng Huanwu Center for Fundamental Theory, Hefei, Anhui 230026, China} \\[2mm]
$^2${\it \small
Interdisciplinary Center for Theoretical Study and  Department of Modern Physics,\\
University of Science and Technology of China, Hefei, Anhui 230026, China}\\[2mm]
$^3${\it \small
Physics and Astronomy, University of Southampton, Southampton, SO17 1BJ, U.K.}\\[2mm]
$^4${\it\small School of Physics, Northwest University, Xi'an 710127, China}\\[2mm]
$^5${\it\small Shaanxi Key Laboratory for Theoretical Physics Frontiers, Xi'an 710127, China}\\[2mm]
$^6${\it\small NSFC-SPTP Peng Huanwu Center for Fundamental Theory, Xi'an 710127, China}
\end{center}
\end{minipage}}
\\[10mm]}
\maketitle
\thispagestyle{empty}

\begin{abstract}
We consider for the first time level 7 modular invariant flavour models where the lepton mixing originates from the breaking of modular symmetry and couplings responsible for lepton masses are modular forms. The latter are decomposed into irreducible multiplets of the finite modular group $\Gamma_7$, which is isomorphic to $PSL(2,Z_{7})$, the projective special linear group of two dimensional matrices over the finite Galois field of seven elements, containing 168 elements, sometimes written as $PSL_2(7)$ or $\Sigma(168)$. At weight 2, there are 26 linearly independent modular forms, organised into a triplet, a septet and two octets of $\Gamma_7$. A full list of modular forms up to weight 8 are provided. Assuming the absence of flavons, the simplest modular-invariant models based on $\Gamma_7$ are constructed, in which neutrinos gain masses via either the Weinberg operator or the type-I seesaw mechanism, and their predictions compared to experiment.
\end{abstract}
\newpage

\section{\label{sec:introduction}Introduction}

The puzzle of quark and lepton masses and mixing, left unanswered by the Standard Model (SM), may be addressed by introducing some family symmetry, which is generally non-Abelian and may be associated with a finite discrete group.
Modular symmetry has been suggested as the origin of such a flavour symmetry, with
neutrino masses as complex analytic functions called modular forms~\cite{Feruglio:2017spp}.
Finite non-Abelian discrete family symmetry emerges from the modular symmetry at various positive integer {\em levels},
with each level associated with a particular flavour group. At each level, the physical fields carry various {\em modular weights}
which do not have to add up to zero in the coupling terms of the effective Lagrangian since
the effective Yukawa couplings may be modular forms, which are holomorphic functions of a complex modulus field
$\tau$~\cite{Feruglio:2017spp}. This may allow flavon fields to be removed,
with higher-dimensional operators in the superpotential being completely determined by modular invariance and supersymmetry. The neutrino masses and mixing parameters may be predicted in terms of a few input parameters, although the predictive power of this framework may be reduced by the K$\ddot{\mathrm{a}}$hler potential which is less  constrained by modular symmetry~\cite{Chen:2019ewa}.

The finite modular groups $\Gamma_2\cong S_3$~\cite{Kobayashi:2018vbk,Kobayashi:2018wkl,Kobayashi:2019rzp,Okada:2019xqk}, $\Gamma_3\cong A_4$~\cite{Feruglio:2017spp,Criado:2018thu,Kobayashi:2018vbk,Kobayashi:2018scp,Okada:2018yrn,Kobayashi:2018wkl,Novichkov:2018yse,Nomura:2019yft,Ding:2019zxk,Gui-JunDing:2019wap,Zhang:2019ngf}, $\Gamma_4\cong S_4$~\cite{Penedo:2018nmg,Novichkov:2018ovf,Kobayashi:2019mna,Gui-JunDing:2019wap,Criado:2019tzk,Wang:2019ovr} and $\Gamma_5\cong A_5$~\cite{Novichkov:2018nkm,Ding:2019xna,Criado:2019tzk} have been considered. For example, simple $A_4$ modular models can reproduce the measured neutrino masses and mixing angles~\cite{Feruglio:2017spp,Kobayashi:2018scp,Ding:2019zxk}.
The quark masses and mixing angles may also be included together with leptons in an $A_4$ modular invariant model~\cite{Okada:2019uoy}, and it has been shown how natural fermion mass hierarchies can arise as a result
of a weighton field~\cite{King:2020qaj}. The modular invariance approach has been extended to include odd weight modular forms which can be decomposed into irreducible representations of the the homogeneous finite modular group $\Gamma'_{N}$~\cite{Liu:2019khw}, and the modular symmetry $\Gamma'_{3}\cong T'$ has been discussed, including the new possibility of texture zeroes~\cite{Lu:2019vgm}. Also modular symmetry may be combined with generalized CP symmetry, where the modulus transforms as $\tau\rightarrow-\tau^{*}$ under the CP transformation~\cite{Novichkov:2019sqv,Baur:2019kwi,Acharya:1995ag,Dent:2001cc,Giedt:2002ns}. The formalism of the single modulus has been generalized to the case of a direct product of multiple moduli~\cite{deMedeirosVarzielas:2019cyj,King:2019vhv},
which is motivated by the additional extra dimensions in superstring theory, assuming toroidal compactification. Indeed, from a top-down perspective, modular symmetry naturally appears in string constructions~\cite{Kobayashi:2018rad,Kobayashi:2018bff,Baur:2019kwi,Baur:2019iai,Kobayashi:2020hoc}.
It has been realised that, if the VEV of the modulus $\tau$ takes some special value, a residual subgroup of the finite modular symmetry group $\Gamma_N$ would be preserved. The phenomenological implications of the residual modular symmetry have been discussed in the context of modular $A_4$~\cite{Novichkov:2018yse,Gui-JunDing:2019wap}, $S_4$~\cite{Gui-JunDing:2019wap,Novichkov:2018ovf} and $A_5$~\cite{Novichkov:2018nkm} symmetries.
If the modular symmetry is broken down to a residual $Z_3$ (or $Z_5$) subgroup in charged lepton sector and to a $Z_2$ subgroup in the neutrino sector, the trimaximal TM1 and TM2 mixing patterns can be obtained~\cite{Novichkov:2018yse,Novichkov:2018ovf}. Moreover, the dynamics of modular symmetry could potentially be tested in present and forthcoming neutrino oscillation experiments~\cite{Ding:2020yen}.

In this paper we consider the level 7 finite modular group $\Gamma_7\cong PSL(2,Z_{7})$, the projective special linear group of two dimensional matrices over the finite Galois field of seven elements. This is the smallest simple discrete group which contains complex triplets and sextet representations. It contains 168 elements and is sometimes written as $PSL_2(7)$ or $\Sigma(168)$~\cite{Fairbairn:1964sga,Ludl:2009ft}. The relationship of this group to some other family symmetries that have been used in the literature is discussed in~\cite{Luhn:2007sy,Luhn:2007yr,Luhn:2008sa,King:2009mk,King:2009tj}.
It has been proposed as a finite modular group in~\cite{deAdelhartToorop:2011re}, whose notations we follow. Thus, we consider for the first time level 7 modular invariant flavour models based on $\Gamma_7$ where the lepton mixing originates from the breaking of modular symmetry and couplings responsible for lepton masses are modular forms. The latter are decomposed into irreducible multiplets of the finite modular group $\Gamma_7$. At weight 2, there are 26 linearly independent modular forms, organised into a triplet, a septet and two octets of $\Gamma_7$. A full list of modular forms up to weight 8 are provided. Assuming the absence of flavons, the simplest modular-invariant models are constructed in which neutrinos gain masses via either the Weinberg operator or the type-I seesaw mechanism, and their predictions compared to experiment.

We organise the rest of this paper in the following. In Section~\ref{sec:rev_mods}, we describe properties of the modular group $\Gamma$ and its finite subgroup $\Gamma_7$. A full list of modular forms of level 7 and weight up to 10 are derived in Section~\ref{sec:form}, based on the SageMath algebra system~\cite{SageMath:2018}. We construct a class of flavon-less lepton flavour models and study their experimental constraints in Section~\ref{sec:model}. Summary is given in Section~\ref{sec:conclusion}. The group theory of $\Gamma_7$ is listed in Appendix~\ref{sec:Group}. We refer to Appendices~\ref{sec:eta} and \ref{sec:theta} for alternative methods based on the Dedekind eta function method \cite{Feruglio:2017spp} and the theta function method \cite{Novichkov:2018nkm}, respectively. How to find an independent set of higher weight modular forms from hundreds of constraints is discussed in Appendix~\ref{appsec:w4_cons}.

\section{\label{sec:rev_mods}Modular symmetry and modular forms of level $N=7$}

In the following, we briefly review the modular symmetry and the its congruence subgroups. The special linear group $SL(2, Z)$ is constituted by $2\times2$ matrices with integer entries and
determinant equal to one ~\cite{Bruinier2008The,diamond2005first}:
\begin{equation}
SL(2, Z)=\left\{\left(\begin{array}{cc}a&b\\c&d\end{array}\right)\bigg|a,b,c,d\in \mathbb{Z},ad-bc=1\right\}\,.
\end{equation}
The $SL(2, Z)$ group acts on the upper half plane $\mathcal{H}=\left\{\tau\in\mathbb{C}~|~\Im\tau >0\right\}$ as the linear fractional transformation,
\begin{equation}
\label{eq:modular_trans}\gamma=\begin{pmatrix}
a  &  b  \\
c  &  d
\end{pmatrix}:\mathcal{H}\rightarrow\mathcal{H},\quad \tau \mapsto \gamma\tau=\gamma(\tau)=\frac{a\tau+b}{c\tau+d}\,.
\end{equation}
It is easy to see the following identity
\begin{equation}
\Im(\gamma(\tau))=\frac{\Im\tau}{|c\tau+d|^2},\quad \gamma=\begin{pmatrix}
a  &  b  \\
c  &  d
\end{pmatrix}\in SL(2, Z)\,.
\end{equation}
Consequently the image $\gamma(z)\in\mathcal{H}$ for any $\gamma\in SL(2, Z)$ and $\tau\in\mathcal{H}$. It is obvious that
\begin{equation}
\frac{a\tau+b}{c\tau+d}\quad~~\text{is~identical~with}\quad~~\frac{-a\tau-b}{-c\tau-d}\,,
\end{equation}
and therefore we identify
\begin{equation}
\label{eq:gamma_fgamma}\left(
\begin{array}{cc}
a  &  b  \\
c  &  d
\end{array}
\right)\quad~~\text{is~identical~with}\quad~~\left(
\begin{array}{cc}
-a  &  -b  \\
-c  & -d
\end{array}
\right)\,.
\end{equation}
Hence the action of $\gamma$ and $-\gamma$ act on the complex modulus $\tau$ is exactly the same, and it is sufficient to consider the projective special linear group $PSL(2, Z)=SL(2, Z)/\{I, -I\}$, which is the quotient of $SL(2, Z)$ by $\pm I$. The group $PSL(2, Z)$ is also called the modular group in the literature, it is a discrete group with infinite elements and it can be generated by two transformations $S$ and $T$~\cite{Bruinier2008The}
\begin{equation}
S=\left(
\begin{array}{cc}
0 ~&~ 1\\
-1 ~&~ 0
\end{array}
\right),\quad  T=\left(
\begin{array}{cc}
1 ~&~  1\\
0 ~&~ 1
\end{array}
\right)\,,
\end{equation}
which fulfill the relations
\begin{equation}
S^2=(ST)^3=\mathds{1}\,.
\end{equation}
The actions of $S$ and $T$ on $\mathcal{H}$ are given by
\begin{equation}
S: \tau \mapsto -\frac{1}{\tau},\qquad T: \tau \mapsto \tau+1\,.
\end{equation}
For a positive integer $N$, the principal congruence subgroup of level $N$ is defined as
\begin{equation}
\Gamma(N)=\left\{\left(\begin{array}{cc}a&b\\c&d\end{array}\right)\in SL(2,\mathbb{Z}),~~a\equiv d\equiv1~({\tt mod}~N),~~ b\equiv c\equiv0~({\tt mod}~N) \right\}\,,
\end{equation}
which is a normal subgroup of the special linear group $SL(2, Z)$.  Obviously $\Gamma(1)\cong SL(2, \mathbb{Z})$ is the special linear group. It is easy to obtain
\begin{equation}
T^{N}=\begin{pmatrix}
1  ~&~  N \\
0   ~&~  1
\end{pmatrix}\,,
\end{equation}
which implies $T^N\in\Gamma(N)$, i.e., $T^N$ is an element of $\Gamma(N)$. Taking the quotient of $\Gamma(1)$ and $\Gamma(2)$ by $\{I, -I\}$, we obtain the projective principal congruence subgroups $\overline{\Gamma}(N)=\Gamma(N)/\{I, -I\}$ for $N=1, 2$, and $\overline{\Gamma}(N>2)=\Gamma(N)$ since the element $-I$ does not belong to $\Gamma(N)$ for $N>2$. The quotient groups $\Gamma_{N}=\overline{\Gamma}(1)/\overline{\Gamma}(N)$ are usually called inhomogeneous finite modular groups, and the homogeneous finite modular group is defined as $\Gamma'_N=SL(2,Z)/\Gamma(N)$ which is the double covering of $\Gamma_N$~\cite{Liu:2019khw}. The finite modular group $\Gamma_N$ for $N\leq5$ can be obtained from $\overline{\Gamma}(1)$ by imposing the condition $T^N=\mathds{1}$. Consequently the generators $S$ and $T$ of $\Gamma_N$ obey the relations
\begin{equation}
S^2=(ST)^3=T^{N}=\mathds{1}\,.
\end{equation}
The groups $\Gamma_N$ with $N=2$, $3$, $4$, $5$ are isomorphic to the permutation groups $S_3$, $A_4$, $S_4$ and $A_5$ respectively~\cite{deAdelhartToorop:2011re}. Note again that for this group, as for all groups with $N>5$, at least one additional relation is necessary in order to render the group finite. It is easy to calculate
\begin{equation}
(ST^{3})^{4}=\begin{pmatrix}
-8  & -21 \\
21  & 55
\end{pmatrix}\,,
\end{equation}
which implies
\begin{equation}
-(ST^{3})^{4}=\begin{pmatrix}
1   &  0 \\
0   &  1
\end{pmatrix}~~(\text{mod}~7)\,.
\end{equation}
Hence the element $-(ST^{3})^{4}$ is belong to $\overline{\Gamma}(7)$. Notice that $-(ST^{3})^{4}$ and $(ST^{3})^{4}$ are identified as the same element of the $\overline{\Gamma}(1)$ group, since they lead to the same linear fraction transformations. Therefore the finite modular group $\Gamma_{7}$ of level $N=7$ be generated by two generators $S$ and $T$ which satisfy the following multiplication rules\footnote{The multiplication rule of $\Gamma_7$ is $S^2=(ST)^{3}=T^7=(ST^{-1}ST)^{4}=\mathds{1}$ in \cite{deAdelhartToorop:2011re}.}
\begin{equation}
S^2=(ST)^{3}=T^7=(ST^{3})^{4}=\mathds{1}\,.
\end{equation}
The crucial element of the modular invariance approach is the modular forms $f(\tau)$ of weight $k$ and level $N$. They are holomorphic functions of the complex
modulus $\tau$ with well-defined transformation properties under the group $\Gamma(N)$,
\begin{equation}
f\left(\gamma\tau\right)=(c\tau+d)^kf(\tau),~~~~\gamma=\left(\begin{array}{cc}
a  &  b \\
c  &  d
\end{array}
\right)\in\Gamma(N)\,,
\end{equation}
The modular forms of weight $k$ and level $N$ form a linear space $\mathcal{M}_{k}(\Gamma(N))$ of finite dimension. In the present work, we shall focus on even weight modular forms, i.e., $k$ being an even number. Then it is always possible to choose a basis of $\mathcal{M}_{k}(\Gamma(N))$ such that the modular forms transform according to a unitary irreducible representation $\mathbf{r}$ of $\Gamma_{N}$~\cite{Feruglio:2017spp},
\begin{equation}
\label{eq:irr_MFD}f_{i}(\gamma\tau)=(c\tau+d)\rho_{ij}(\gamma)f_j(\tau)\,,
\end{equation}
where $\gamma$ is a representative element of $\Gamma_N$, and $\rho_{\mathbf{r}}(\gamma)$ is the representation matrix of $\gamma$ in the irreducible representation $\mathbf{r}$. If the modular weight $k$ is odd, the modular forms can be decomposed into irreducible representations of the homogeneous finite modular group $\Gamma'_N$~\cite{Liu:2019khw}. In order to determine the proper basis, it is sufficient to apply Eq.~\eqref{eq:irr_MFD} to the generators $S$ and $T$ which can generate all elements of $\gamma_N$.

\section{\label{sec:form}Modular forms of level $N=7$}

The general dimension formula for the linear space of modular forms of level $N$ and weight $k$ is given by
\begin{equation}
\texttt{dim}\mathcal{M}_{k}(\Gamma(N))=\dfrac{(k-1)N+6}{24}N^2 \prod_{p|N}(1-\dfrac{1}{p^2}), \quad~N>2,\,k\geq 1\,.
\end{equation}
For $N=7$, we have
\begin{equation}
\texttt{dim}\mathcal{M}_{k}(\Gamma(7))=\frac{(k-1)\times7+6}{24}
\times7^2\times\left(1-\frac{1}{7^2}\right)=14k-2\,.
\end{equation}
Hence the linear space of modular forms of level 7 and weight 2 has dimension $14\times2-2=26$.

One can obtain $q$-expansions for a basis $b_i$ of the space of lowest weight modular forms for $\Gamma_7$ from the SageMath algebra system~\cite{SageMath:2018}. They are given by
\begin{eqnarray}
\nonumber b_1(\tau) &=& q^{1/7}(1 - 3q + 4q^{3} + 2q^{4} + 3q^{5} - 12q^{6} - 5q^{7} + 7q^{9})+\ldots\,,\\
\nonumber b_2(\tau) &=&  q^{2/7}(1 - 3q - q^{2} + 8q^{3} - 6q^{5} - 4q^{6} + 2q^{8}) +\ldots\,, \\
\nonumber b_3(\tau) &=& q^{4/7}(1 - 4q + 3q^{2} + 5q^{3} - 5q^{4} - 8q^{6} + 10q^{7} - 4q^{9})+\ldots\,, \\
\nonumber b_4(\tau) &=&  1 + 252q^{5} - 840q^{6} + 1344q^{7} - 840q^{8} - 420q^{9} + 588q^{10}+\ldots\,, \\
\nonumber b_5(\tau) &=&  q^{1/7}(1 + \frac{15}{2}q^{3} + 30q^{4} - 41q^{5} + 44q^{6} + 5q^{8} - 8q^{9}) +\ldots\,, \\
\nonumber b_6(\tau) &=&   q^{2/7}(1 + \frac{73}{4}q^{3} - 13q^{4} + \frac{119}{4}q^{5} - \frac{43}{4}q^{6} - \frac{15}{4}q^{7} + 30q^{8} + \frac{97}{2}q^{9})+\ldots\,,\\
\nonumber b_7(\tau) &=&   q^{3/7}(1 + 15q^{3} - q^{4} - 3q^{5} + 15q^{6} + 11q^{7} - 3q^{8} + 27q^{9})+\ldots\,, \\
\nonumber b_8(\tau) &=&   q^{4/7}(1 + \frac{62}{5}q^{3} + 9q^{4} - \frac{176}{5}q^{5} + 51q^{6} + \frac{66}{5}q^{7} - \frac{213}{5}q^{8} + \frac{88}{5}q^{9}) +\ldots\,,\\
\nonumber b_9(\tau) &=&   q^{5/7}(1 + 17q^{3} - 18q^{4} + 19q^{5} - 2q^{6} + 16q^{7} - 3q^{8} + 17q^{9})+\ldots\,,\\
\nonumber b_{10}(\tau) &=&  q^{6/7}(1 + 15q^{3} - 13q^{4} + 29q^{6} - q^{7} - 27q^{8} + 43q^{9})+\ldots\,, \\
\nonumber b_{11}(\tau) &=&  q + 69q^{5} - 234q^{6} + 385q^{7} - 231q^{8} - 130q^{9} + 168q^{10}+\ldots\,, \\
\nonumber b_{12}(\tau) &=&  q^{8/7}(1 + \frac{7}{2}q^{2} + 4q^{4} + 11q^{6} - 3q^{7} + 9q^{8})+\ldots\,,\\
\nonumber b_{13}(\tau) &=&   q^{9/7}(1 + 3q^{2} - q^{3} + 5q^{4} + 7q^{6} + 6q^{8})+\ldots\,, \\
\nonumber b_{14}(\tau) &=&  q^{10/7}(1 + 2q^{2} + 3q^{4} + 2q^{5} + 4q^{6} + 5q^{8}) +\ldots\,,\\
\nonumber b_{15}(\tau) &=&  q^{11/7}(1 + \frac{6}{5}q^{2} + \frac{22}{5}q^{4} - q^{5} + \frac{23}{5}q^{6} - \frac{4}{5}q^{7} + \frac{29}{5}q^{8})+\ldots\,, \\
\nonumber b_{16}(\tau) &=&  q^{12/7}(1 + 2q^{3} + 2q^{4} + 3q^{6} + 3q^{8})+\ldots\,,\\
\nonumber b_{17}(\tau) &=&  q^{13/7}(1 - q^{2} + 3q^{3} + 3q^{4} - 4q^{5} + 3q^{6} + 6q^{7} - 3q^{8})+\ldots\,, \\
\nonumber b_{18}(\tau) &=&  q^{2} - 8q^{5} + 39q^{6} - 62q^{7} + 38q^{8} + 28q^{9} - 28q^{10}+\ldots\,,\\
\nonumber b_{19}(\tau) &=& q^{15/7}(1 - q + 3q^{3} - 3q^{5} + 5q^{6})+\ldots\,, \\
\nonumber b_{20}(\tau) &=&  q^{16/7}(1 - \frac{9}{4}q + 4q^{2} - \frac{15}{4}q^{3} + \frac{15}{4}q^{4} - \frac{1}{4}q^{5} - \frac{9}{2}q^{7})+\ldots\,,\\
\nonumber b_{21}(\tau) &=&  q^{17/7}(1 - 2q + 2q^{2} + q^{3} - q^{4} - q^{5} + 4q^{6} - 3q^{7})+\ldots\,, \\
\nonumber b_{22}(\tau) &=&  q^{18/7}(1 - \frac{9}{5}q + \frac{32}{5}q^{3} - 7q^{4} - \frac{12}{5}q^{5} + \frac{61}{5}q^{6} - \frac{16}{5}q^{7})+\ldots\,,\\
\nonumber b_{23}(\tau) &=& q^{19/7}(1 - 3q + 5q^{2} - 4q^{3} + 3q^{4} - 3q^{5} + 4q^{6} - 3q^{7})+\ldots\,, \\
\nonumber b_{24}(\tau) &=&  q^{20/7}(1 - 3q + 4q^{2} - 4q^{4} + q^{5} + 8q^{6} - 9q^{7}) +\ldots\,,\\
\nonumber b_{25}(\tau) &=&  q^{3} - 15q^{5} + 55q^{6} - 88q^{7} + 54q^{8} + 34q^{9} - 41q^{10} +\ldots\,, \\
\label{eq:miller-like_basis} b_{26}(\tau) &=&  q^{4} - 6q^{5} + 17q^{6} - 26q^{7} + 18q^{8} + 4q^{9} - 7q^{10}+\ldots\,,
\end{eqnarray}
with $q = e^{2\pi i\,\tau}$ and where fractional powers $q^{n/7}$ should be read as $q^{n/7} = e^{2 n \pi i\,\tau/7}$. The above lowest weight modular forms can be organized into a triplet transforming in the representation $\mathbf{3}$ of $\Gamma_7$, a septet transforming in the representation $\mathbf{7}$ of $\Gamma_7$, and two octets in the $\mathbf{8}$ of $\Gamma_7$. To be more explicit, we have
\begin{eqnarray}
\label{eq:Y2_3}&&Y^{(2)}_{\mathbf{3}}(\tau)\equiv
 \left(\begin{array}{c}
Y_{1}(\tau)\\
Y_{2}(\tau)\\
Y_{3}(\tau)
\end{array}\right)
=\begin{pmatrix}
b_1(\tau)\\
b_2(\tau)\\
-b_3(\tau)
\end{pmatrix},\\
\label{eq:Y2_7} &&Y^{(2)}_{\mathbf{7}}(\tau)\equiv \left(\begin{array}{c}
Y_{4}(\tau)\\
Y_{5}(\tau)\\
Y_{6}(\tau)\\
Y_{7}(\tau)\\
Y_{8}(\tau) \\
Y_{9}(\tau)\\
Y_{10}(\tau)
\end{array}\right)
=\begin{pmatrix}
b_4(\tau)+4b_{11}(\tau)+12b_{18}(\tau)+16b_{25}(\tau)+28b_{26}(\tau)\\
-\sqrt{2}\left[b_5(\tau)+15b_{12}(\tau)+24b_{19}(\tau)\right] \\
-\sqrt{2}\left[3b_6(\tau)+13b_{13}(\tau)+31b_{20}(\tau)\right] \\
-2\sqrt{2}\left[2b_7(\tau)+9b_{14}(\tau)+9b_{21}(\tau)\right]\\
-\sqrt{2}\left[7b_8(\tau)+12b_{15}(\tau)+39b_{22}(\tau)\right]\\
-2\sqrt{2}\left[3b_9(\tau)+14b_{16}(\tau)+10b_{23}(\tau)\right]\\
-2\sqrt{2}\left[6b_{10}(\tau)+7b_{17}(\tau)+21b_{24}(\tau)\right]
\end{pmatrix}\,,\\
\label{eq:Y2_8a}&&Y^{(2)}_{\mathbf{8}a}(\tau)\equiv \left(\begin{array}{c}
Y_{11}(\tau)\\
Y_{12}(\tau)\\
Y_{13}(\tau) \\
Y_{14}(\tau)\\
Y_{15}(\tau)\\
Y_{16}(\tau) \\
Y_{17}(\tau)\\
Y_{18}(\tau)
\end{array}\right)
=\frac{1}{2}\begin{pmatrix}
2b_4(\tau)+2b_{11}(\tau)-24b_{18}(\tau)+56b_{25}(\tau)-b_{26}(\tau)\\
-\sqrt{3}\left[22b_{11}(\tau)+30b_{18}(\tau)+56b_{25}(\tau)+59b_{26}(\tau)\right]\\
-2\sqrt{2}\left[2b_5(\tau)+15b_{12}(\tau)+21b_{19}(\tau)\right] \\
\sqrt{2}\left[3b_{6}(\tau)-16b_{13}(\tau)+11b_{20}(\tau)\right] \\
7\sqrt{2}\left[2b_{7}(\tau)+3b_{14}(\tau)+12b_{21}(\tau)\right]\\
-\sqrt{2}\left[11b_{8}(\tau)+54b_{15}(\tau)+54b_{22}(\tau)\right]\\
7\sqrt{2}\left[b_{16}(\tau)+2b_{23}(\tau)\right]\\
7\sqrt{2}\left[3b_{10}(\tau)+8b_{17}(\tau)+9b_{24}(\tau)\right]
\end{pmatrix}\,,\\
\label{eq:mod}&&Y^{(2)}_{\mathbf{8}b}(\tau)\equiv \left(\begin{array}{c}
Y_{19}(\tau)\\
Y_{20}(\tau)\\
Y_{21}(\tau)\\
Y_{22}(\tau)\\
Y_{23}(\tau)\\
Y_{24}(\tau)\\
Y_{25}(\tau)\\
Y_{26}(\tau)
\end{array}\right)
=\frac{1}{60}\begin{pmatrix}
3\left[20b_{11}(\tau)+40b_{18}(\tau)+42b_{25}(\tau)+53b_{26}(\tau)\right]\\
-\sqrt{3}\left[2b_4(\tau)-12b_{11}(\tau)+18b_{18}(\tau)-42b_{25}(\tau)+27b_{26}(\tau)\right]\\
6\sqrt{2}\left[b_5(\tau)+4b_{12}(\tau)+14b_{19}(\tau)\right] \\
-3\sqrt{2}\left[5b_{6}(\tau)+20b_{13}(\tau)+37b_{20}(\tau)\right]\\
21\sqrt{2}\left[3b_{14}(\tau)+2b_{21}(\tau)\right]\\
-3\sqrt{2}\left[5b_{8}(\tau)+8b_{15}(\tau)+22b_{22}(\tau)\right]\\
21\sqrt{2}\left[2b_{9}(\tau)+3b_{16}(\tau)+8b_{23}(\tau)\right]\\
-21\sqrt{2}\left[b_{10}(\tau)+4b_{17}(\tau)+b_{24}(\tau)\right]
\end{pmatrix}\,.
\end{eqnarray}
Notice that the weight two modular multiplets $Y^{(2)}_{\mathbf{8}a}(\tau)$ and $Y^{(2)}_{\mathbf{8}b}(\tau)$ are not unique, and in principle they can taken to be any two linearly independent combinations of $Y^{(2)}_{\mathbf{8}a}(\tau)$ and $Y^{(2)}_{\mathbf{8}b}(\tau)$. Higher weight modular multiplets can be obtained from tensor products of the lowest weight multiplets
$Y^{(2)}_\mathbf{3}$, $Y^{(2)}_\mathbf{7}$, $Y^{(2)}_{\mathbf{8}a}$ and $Y^{(2)}_{\mathbf{8}b}$.
The missing $\mathbf{1}$ and $\mathbf{6}$ representations
arise at weight 4. Even though one can form 351 products $Y_i Y_j$ where some vanishing modular forms easily seen from the Clebsch-Gordan coefficients are not counted, the space of modular forms of weight 4 (and level 7) has dimension $14k-2=54$. Therefore, there are 297 constraints between the $Y_i Y_j$, which we list in Appendix~\ref{appsec:w4_cons}. The 54 linearly independent modular forms of weight 4 can be arranged into the following multiplets of $\Gamma_7$:
\begin{equation}\label{eq:Yw4l}
 Y^{(4)}_{\mathbf{1}a}    =\left(Y^{(2)}_{\mathbf{7}}Y^{(2)}_{\mathbf{7}}\right)_{\mathbf{1}} = 2 Y_{10} Y_{5}+Y_{4}^2+2 Y_{6} Y_{9}+2 Y_{7} Y_{8}\equiv Y^{(4)}_{1}(\tau)     \,.
\end{equation}
\begin{equation}
{Y^{(4)}_{\mathbf{3}a}}  =\left(Y^{(2)}_{\mathbf{3}}Y^{(2)}_{\mathbf{8}a}\right)_{\mathbf{3}} =
\left(
\begin{array}{c}
\sqrt{3} Y_{1} Y_{11}+Y_{1} Y_{12}-\sqrt{6}  Y_{3}Y_{16}-\sqrt{6} Y_{2}Y_{18} \\
\sqrt{6} Y_{1} Y_{13}-\sqrt{3} Y_{2}Y_{11} +Y_{2}Y_{12} -\sqrt{6}Y_{3} Y_{17} \\
\sqrt{6} Y_{1} Y_{15}-2 Y_{3}Y_{12} +\sqrt{6}Y_{2} Y_{14}
\end{array}
\right)\equiv \left(\begin{array}{c}
Y^{(4)}_{2}(\tau)\\
Y^{(4)}_{3}(\tau)\\
Y^{(4)}_{4}(\tau)
\end{array}\right) \,.
\end{equation}
\begin{eqnarray}
{Y^{(4)}_{\mathbf{6}a}}  &&=\left(Y^{(2)}_{\mathbf{3}}Y^{(2)}_{\mathbf{3}}\right)_{\mathbf{6}} =
\left(
\begin{array}{c}
\sqrt{2} Y_{3}^2\\
\sqrt{2} Y_{1}^2\\
2 Y_{1} Y_{2}\\
\sqrt{2} Y_{2}^2\\
2 Y_{1} Y_{3}\\
2 Y_{2} Y_{3}
\end{array}
\right)\equiv \left(\begin{array}{c}
Y^{(4)}_{5}(\tau)\\
Y^{(4)}_{6}(\tau)\\
Y^{(4)}_{7}(\tau)\\
Y^{(4)}_{8}(\tau)\\
Y^{(4)}_{9}(\tau)\\
Y^{(4)}_{10}(\tau)
\end{array}\right) \,,\\[2mm]
 {Y^{(4)}_{\mathbf{6}b}}  &&=
\left(Y^{(2)}_{\mathbf{3}}Y^{(2)}_{\mathbf{7}}\right)_{\mathbf{6}} =
\left(
\begin{array}{c}
2 Y_{1} Y_{4}-2 \sqrt{2} Y_{10} Y_{2}-\sqrt{2} Y_{3} Y_{8}\\
-\sqrt{2} Y_{1} Y_{5}+2 Y_{2} Y_{4}-2 \sqrt{2} Y_{3} Y_{9}\\
-Y_{1} Y_{6}-2 Y_{10} Y_{3}+3 Y_{2} Y_{5}\\
-2 \sqrt{2} Y_{1} Y_{7}-\sqrt{2} Y_{2} Y_{6}+2 Y_{3} Y_{4}\\
3 Y_{1} Y_{8}-2 Y_{2} Y_{7}-Y_{3} Y_{5}\\
-2 Y_{1} Y_{9}-Y_{2} Y_{8}+3 Y_{3} Y_{6}
\end{array}
\right)\equiv \left(\begin{array}{c}
Y^{(4)}_{11}(\tau)\\
Y^{(4)}_{12}(\tau)\\
Y^{(4)}_{13}(\tau)\\
Y^{(4)}_{14}(\tau)\\
Y^{(4)}_{15}(\tau)\\
Y^{(4)}_{16}(\tau)
\end{array}\right) \,.
\end{eqnarray}
\begin{eqnarray}
Y^{(4)}_{\mathbf{7}a}  &&=
\left(Y^{(2)}_{\mathbf{3}}Y^{(2)}_{\mathbf{7}}\right)_{\mathbf{7}} =
\left(
\begin{array}{c}
\sqrt{2} Y_{1} Y_{10}+\sqrt{2} Y_{2} Y_{9}+\sqrt{2} Y_{3} Y_{7}\\
-\sqrt{2} Y_{1} Y_{4}-2 Y_{3} Y_{8}\\
-\sqrt{2} Y_{2} Y_{4}-2 Y_{1} Y_{5}\\
-Y_{1} Y_{6}+2 Y_{10} Y_{3}+Y_{2} Y_{5}\\
-\sqrt{2} Y_{3} Y_{4}-2 Y_{2} Y_{6}\\
Y_{1} Y_{8}+2 Y_{2} Y_{7}-Y_{3} Y_{5}\\
2 Y_{1} Y_{9}-Y_{2} Y_{8}+Y_{3} Y_{6}
\end{array}
\right)\equiv \left(\begin{array}{c}
Y^{(4)}_{17}(\tau)\\
Y^{(4)}_{18}(\tau)\\
Y^{(4)}_{19}(\tau)\\
Y^{(4)}_{20}(\tau)\\
Y^{(4)}_{21}(\tau)\\
Y^{(4)}_{22}(\tau)\\
Y^{(4)}_{23}(\tau)
\end{array}\right) \,,\\[2mm]
 {Y^{(4)}_{\mathbf{7}b}}  &&=
\left(Y^{(2)}_{\mathbf{7}}Y^{(2)}_{\mathbf{7}}\right)_{\mathbf{7_S}} =
\left(
\begin{array}{c}
-2 Y_{10} Y_{5}+6 Y_{4}^2-2 Y_{6} Y_{9}-2 Y_{7} Y_{8}\\
4 \sqrt{2} Y_{10} Y_{6}-2 Y_{4} Y_{5}+4 \sqrt{2} Y_{7} Y_{9}+2 \sqrt{2} Y_{8}^2\\
4 \sqrt{2} Y_{10} Y_{7}-2 Y_{4} Y_{6}+2 \sqrt{2} Y_{5}^2+4 \sqrt{2} Y_{8} Y_{9}\\
4 \sqrt{2} Y_{10} Y_{8}-2 Y_{4} Y_{7}+4 \sqrt{2} Y_{5} Y_{6}+2 \sqrt{2} Y_{9}^2\\
4 \sqrt{2} Y_{10} Y_{9}-2 Y_{4} Y_{8}+4 \sqrt{2} Y_{5} Y_{7}+2 \sqrt{2} Y_{6}^2\\
2 \sqrt{2} Y_{10}^2-2 Y_{4} Y_{9}+4 \sqrt{2} Y_{5} Y_{8}+4 \sqrt{2} Y_{6} Y_{7}\\
-2 Y_{10} Y_{4}+4 \sqrt{2} Y_{5} Y_{9}+4 \sqrt{2} Y_{6} Y_{8}+2 \sqrt{2} Y_{7}^2
\end{array}
\right)\equiv \left(\begin{array}{c}
Y^{(4)}_{24}(\tau)\\
Y^{(4)}_{25}(\tau)\\
Y^{(4)}_{26}(\tau)\\
Y^{(4)}_{27}(\tau)\\
Y^{(4)}_{28}(\tau)\\
Y^{(4)}_{29}(\tau)\\
Y^{(4)}_{30}(\tau)
\end{array}\right)
\,.
\end{eqnarray}
\begin{eqnarray}
{Y^{(4)}_{\mathbf{8}a}}  &&=
\left(Y^{(2)}_{\mathbf{3}}Y^{(2)}_{\mathbf{7}}\right)_{\mathbf{8}} =
\left(
\begin{array}{c}
5 Y_{1} Y_{10}-Y_{2} Y_{9}-4 Y_{3} Y_{7}\\
-\sqrt{3} Y_{1} Y_{10}+3 \sqrt{3} Y_{2} Y_{9}-2 \sqrt{3} Y_{3} Y_{7}\\
-4 Y_{1} Y_{4}-3 \sqrt{2} Y_{10} Y_{2}+2 \sqrt{2} Y_{3} Y_{8}\\
2 \sqrt{2} Y_{1} Y_{5}-4 Y_{2} Y_{4}-3 \sqrt{2} Y_{3} Y_{9}\\
4 \sqrt{2} Y_{1} Y_{6}+\sqrt{2} Y_{10} Y_{3}+2 \sqrt{2} Y_{2} Y_{5}\\
3 \sqrt{2} Y_{1} Y_{7}-2 \sqrt{2} Y_{2} Y_{6}+4 Y_{3} Y_{4}\\
-2 \sqrt{2} Y_{1} Y_{8}-\sqrt{2} Y_{2} Y_{7}-4 \sqrt{2} Y_{3} Y_{5}\\
-\sqrt{2} Y_{1} Y_{9}-4 \sqrt{2} Y_{2} Y_{8}-2 \sqrt{2} Y_{3} Y_{6}
\end{array}
\right)\equiv \left(\begin{array}{c}
Y^{(4)}_{31}(\tau)\\
Y^{(4)}_{32}(\tau)\\
Y^{(4)}_{33}(\tau)\\
Y^{(4)}_{34}(\tau)\\
Y^{(4)}_{35}(\tau)\\
Y^{(4)}_{36}(\tau)\\
Y^{(4)}_{37}(\tau)\\
Y^{(4)}_{38}(\tau)
\end{array}\right)
\,,\\[2mm]
\nonumber {Y^{(4)}_{\mathbf{8}b}}  &&=
\left(Y^{(2)}_{\mathbf{8}a}Y^{(2)}_{\mathbf{8}a}\right)_{\mathbf{8_{S,1}}} \\
&&=
\left(
\begin{array}{c}
3 \sqrt{3} Y_{11}^2-2 Y_{11} Y_{12}-3 \sqrt{3} Y_{12}^2+4 \sqrt{3} Y_{13} Y_{18}-4 \sqrt{3} Y_{15} Y_{16}\\
-Y_{11}^2-6 \sqrt{3} Y_{11} Y_{12}+Y_{12}^2-4 Y_{13} Y_{18}+8 Y_{14} Y_{17}-4 Y_{15} Y_{16}\\
-4 \sqrt{3} Y_{11} Y_{13}+4 Y_{12} Y_{13}-4 \sqrt{6} Y_{14} Y_{18}+2 \sqrt{6} Y_{16}^2\\
-8 Y_{12} Y_{14}+2 \sqrt{6} Y_{13}^2+4 \sqrt{6} Y_{16} Y_{17}\\
4 \sqrt{3} Y_{11} Y_{15}+4 Y_{12} Y_{15}+4 \sqrt{6} Y_{16} Y_{18}+2 \sqrt{6} Y_{17}^2\\
4 \sqrt{3} Y_{11} Y_{16}+4 Y_{12} Y_{16}-4 \sqrt{6} Y_{13} Y_{15}-2 \sqrt{6} Y_{14}^2\\
-8 Y_{12} Y_{17}-4 \sqrt{6} Y_{14} Y_{15}-2 \sqrt{6} Y_{18}^2\\
-4 \sqrt{3} Y_{11} Y_{18}+4 Y_{12} Y_{18}+4 \sqrt{6} Y_{13} Y_{17}-2 \sqrt{6} Y_{15}^2
\end{array}
\right)\equiv \left(\begin{array}{c}
Y^{(4)}_{39}(\tau)\\
Y^{(4)}_{40}(\tau)\\
Y^{(4)}_{41}(\tau)\\
Y^{(4)}_{42}(\tau)\\
Y^{(4)}_{43}(\tau)\\
Y^{(4)}_{44}(\tau)\\
Y^{(4)}_{45}(\tau)\\
Y^{(4)}_{46}(\tau)
\end{array}\right)
 \,, \\[2mm]
\label{eq:w4_54forms_p3} Y^{(4)}_{\mathbf{8}c} &&=
\left(Y^{(2)}_{\mathbf{8}b}Y^{(2)}_{\mathbf{8}b}\right)_{\mathbf{8_{S,2}}} =
\left(
\begin{array}{c}
4 Y_{19} Y_{20}+2 \sqrt{3} Y_{22} Y_{25}-2 \sqrt{3} Y_{23} Y_{24}\\
2 Y_{19}^2-2 Y_{20}^2-4 Y_{21} Y_{26}+2 Y_{22} Y_{25}+2 Y_{23} Y_{24}\\
4 Y_{20} Y_{21}-2 \sqrt{6} Y_{23} Y_{25}\\
-2 \sqrt{3} Y_{19} Y_{22}-2 Y_{20} Y_{22}-2 \sqrt{6} Y_{23} Y_{26}\\
2 \sqrt{3} Y_{19} Y_{23}-2 Y_{20} Y_{23}+2 \sqrt{6} Y_{21} Y_{22}\\
2 \sqrt{3} Y_{19} Y_{24}-2 Y_{20} Y_{24}-2 \sqrt{6} Y_{25} Y_{26}\\
-2 \sqrt{3} Y_{19} Y_{25}-2 Y_{20} Y_{25}+2 \sqrt{6} Y_{21} Y_{24}\\
4 Y_{20} Y_{26}+2 \sqrt{6} Y_{22} Y_{24}
\end{array}
\right)\equiv \left(\begin{array}{c}
Y^{(4)}_{47}(\tau)\\
Y^{(4)}_{48}(\tau)\\
Y^{(4)}_{49}(\tau)\\
Y^{(4)}_{50}(\tau)\\
Y^{(4)}_{51}(\tau)\\
Y^{(4)}_{52}(\tau)\\
Y^{(4)}_{53}(\tau)\\
Y^{(4)}_{54}(\tau)
\end{array}\right)\,.
\end{eqnarray}

\begin{table}[t!]
\centering
\begin{tabular}{|c|c|}
\hline  \hline

& Modular form $Y^{(k)}_{\mathbf{r}}$ \\  \hline

$k=2$ & $Y^{(2)}_{\mathbf{3}}$, $Y^{(2)}_{\mathbf{7}}$, $Y^{(2)}_{\mathbf{8}a}$, $Y^{(2)}_{\mathbf{8}b}$\\ \hline

$k=4$ & $Y^{(4)}_{\mathbf{1}a}$, $Y^{(4)}_{\mathbf{3}a}$, $Y^{(4)}_{\mathbf{6}a}$, $Y^{(4)}_{\mathbf{6}b}$, $Y^{(4)}_{\mathbf{7}a}$, $Y^{(4)}_{\mathbf{7}b}$, $Y^{(4)}_{\mathbf{8}a}$, $Y^{(4)}_{\mathbf{8}b}$, $Y^{(4)}_{\mathbf{8}c} $ \\ \hline

$k=6$ & $Y^{(6)}_{\mathbf{1}}$, $Y^{(6)}_{\mathbf{3}a}$, $Y^{(6)}_{\mathbf{3}b}$, $Y^{(6)}_{\mathbf{\bar{3}}}$, $Y^{(6)}_{\mathbf{6}a}$, $Y^{(6)}_{\mathbf{6}b}$, $Y^{(6)}_{\mathbf{7}a}$, $Y^{(6)}_{\mathbf{7}b}$, $Y^{(6)}_{\mathbf{7}c}$, $Y^{(6)}_{\mathbf{7}d}$, $Y^{(6)}_{\mathbf{8}a}$, $Y^{(6)}_{\mathbf{8}b}$, $Y^{(6)}_{\mathbf{8}c} $, $Y^{(6)}_{\mathbf{8}d}$  \\ \hline

\multirow{2}{*}{$k=8$} & $Y^{(8)}_{\mathbf{1}}$, $Y^{(8)}_{\mathbf{3}a}$, $Y^{(8)}_{\mathbf{3}b}$, $Y^{(8)}_{\mathbf{\bar{3}}}$, $Y^{(8)}_{\mathbf{6}a}$, $Y^{(8)}_{\mathbf{6}b}$, $Y^{(8)}_{\mathbf{6}c}$, $Y^{(8)}_{\mathbf{6}d}$, \\
& $Y^{(8)}_{\mathbf{7}a}$, $Y^{(8)}_{\mathbf{7}b}$, $Y^{(8)}_{\mathbf{7}c}$, $Y^{(8)}_{\mathbf{7}d}$, $Y^{(8)}_{\mathbf{8}a}$, $Y^{(8)}_{\mathbf{8}b}$, $Y^{(8)}_{\mathbf{8}c} $, $Y^{(8)}_{\mathbf{8}d}$, $Y^{(8)}_{\mathbf{8}e} $, $Y^{(8)}_{\mathbf{8}f}$ \\ \hline

\multirow{2}{*}{$k=10$} & $Y^{(10)}_{\mathbf{1}}$, $Y^{(10)}_{\mathbf{3}a}$, $Y^{(10)}_{\mathbf{3}b}$, $Y^{(10)}_{\mathbf{3}c}$, $Y^{(10)}_{\mathbf{\bar{3}}a}$, $Y^{(10)}_{\mathbf{\bar{3}}b}$, $Y^{(10)}_{\mathbf{6}a}$, $Y^{(10)}_{\mathbf{6}b}$, $Y^{(10)}_{\mathbf{6}c}$, $Y^{(10)}_{\mathbf{6}d}$, \\
& $Y^{(10)}_{\mathbf{7}a}$, $Y^{(10)}_{\mathbf{7}b}$, $Y^{(10)}_{\mathbf{7}c}$, $Y^{(10)}_{\mathbf{7}d}$, $Y^{(10)}_{\mathbf{7}e}$, $Y^{(10)}_{\mathbf{7}f}$, $Y^{(10)}_{\mathbf{8}a}$, $Y^{(10)}_{\mathbf{8}b}$, $Y^{(10)}_{\mathbf{8}c}$, $Y^{(10)}_{\mathbf{8}d}$, $Y^{(10)}_{\mathbf{8}e}$, $Y^{(10)}_{\mathbf{8}f}$ , $Y^{(10)}_{\mathbf{8}g}$ \\ \hline \hline
\end{tabular}
\caption{\label{tab:MF_summary}Summary of modular forms of level 7 up to weight 8, the subscript $\mathbf{r}$ denotes the transformation property under $\Gamma_7$ modular symmetry. Here $Y^{(2)}_{\mathbf{8}a}$ and $Y^{(2)}_{\mathbf{8}b}$ stand for two weight 2 modular forms transforming in the representation $\mathbf{8}$ of $\Gamma_7$. The same convention is adopted for other modular forms. }
\end{table}

\section{\label{sec:model}Lepton models based on $\Gamma_7$ modular symmetry }

In this section, we shall construct some typical models for neutrino masses and mixing based on the $\Gamma_7$ modular symmetry. We will not introduce any flavon field, the flavor symmetry is broken when the complex modulus $\tau$ obtains a vacuum expectation value. The Higgs doublets fields $H_{u,d}$ are assumed to be singlets of $\Gamma_7$ with vanishing modular weights. The three right-handed (RH) charged leptons $E^{c}_{1,2,3}$ transform as singlet $\mathbf{1}$ under $\Gamma_7$ modular group nevertheless they are distinguished by the different modular weights $k_{1,2,3}$. We assign the three generations of left-handed (LH) lepton doublets $L$ and the three right-handed neutrinos $N^c$ to two triplets $\mathbf{3}$ and $\mathbf{\bar{3}}$ with the weights $k_L$ and $k_N$. We shall employ potentially the lowest weight  modular forms as much as possible in order to reduce free parameters.

\subsection{Charged lepton sector}

If the left-handed lepton fields $L$ are embedded into the triplet $\mathbf{3}$, modular forms in the representation $\mathbf{\bar{3}}$ should be invoked in the charged lepton mass terms. The superpotential for the charged lepton Yukawa coupling reads as
\begin{equation}
\label{eq:We_1st}\mathcal{W}_e=\alpha\left(E^{c}_1LY^{(6)}_{\mathbf{\bar{3}}}H_{d}\right)_{\mathbf{1}}+
\beta \left(E^{c}_2LY^{(8)}_{\mathbf{\bar{3}}}\right)_{\mathbf{1}}H_{d}+\gamma_1 \left(E^{c}_3LY^{(10)}_{\mathbf{\bar{3}}a}H_{d}\right)_{\mathbf{1}}+\gamma_2 \left(E^{c}_3LY^{(10)}_{\mathbf{\bar{3}}b}H_{d}\right)_{\mathbf{1}}\,.
\end{equation}
Notice that there are two linearly independent weight ten modular forms $Y^{(10)}_{\mathbf{\bar{3}}a}$ and $Y^{(10)}_{\mathbf{\bar{3}}b}$ transforming as $\mathbf{\bar{3}}$ at level $N=7$. The phases of the coupling constants $\alpha$, $\beta$ and $\gamma_1$ can be absorbed into the lepton fields while $\gamma_2$ is generally a complex parameter. Modular invariance of the superpotential $\mathcal{W}_e$ in Eq.~\eqref{eq:We_1st} requires the modular weights should fulfill the conditions
\begin{equation}
k_1=k_2-2=k_3-4=6-k_L\,.
\end{equation}
After the value of $\tau$ is fixed by certain modulus stabilization mechanism, the charged lepton mass matrix takes the following form
\begin{equation}
M_e=\begin{pmatrix}
\alpha Y^{(6)}_{\mathbf{\bar{3}}, 1}  ~&~ \alpha Y^{(6)}_{\mathbf{\bar{3}}, 2} ~&~ \alpha Y^{(6)}_{\mathbf{\bar{3}}, 3}  \\
\beta Y^{(8)}_{\mathbf{\bar{3}}, 1}  ~&~ \beta Y^{(8)}_{\mathbf{\bar{3}}, 2} ~&~ \beta Y^{(8)}_{\mathbf{\bar{3}}, 3}  \\
\gamma_1 Y^{(10)}_{\mathbf{\bar{3}}a, 1} +\gamma_2 Y^{(10)}_{\mathbf{\bar{3}}b, 1}  ~&~ \gamma_1 Y^{(10)}_{\mathbf{\bar{3}}a, 2} +\gamma_2 Y^{(10)}_{\mathbf{\bar{3}}b, 2}  ~&~ \gamma_1 Y^{(10)}_{\mathbf{\bar{3}}a, 3} +\gamma_2 Y^{(10)}_{\mathbf{\bar{3}}b, 3}
\end{pmatrix}v_{d}\,,
\end{equation}
where we denote $Y^{(6)}_{\mathbf{\bar{3}}}\equiv(Y^{(6)}_{\mathbf{\bar{3}}, 1}, Y^{(6)}_{\mathbf{\bar{3}}, 2}, Y^{(6)}_{\mathbf{\bar{3}}, 3})^{T}$, and the expressions of $Y^{(6)}_{\mathbf{\bar{3}}, 1}$, $Y^{(6)}_{\mathbf{\bar{3}}, 2}$ and $Y^{(6)}_{\mathbf{\bar{3}}, 3}$ are given in Appendix~\ref{appsec:w4_cons}. Similar notations are adopted for $Y^{(8)}_{\mathbf{\bar{3}}}$, $Y^{(10)}_{\mathbf{\bar{3}}a}$, $Y^{(10)}_{\mathbf{\bar{3}}b}$ and other modular forms in the following. Similarly if the LH leptons $L$ are assigned to the triplet $\mathbf{\bar{3}}$, the charged lepton mass terms are
\begin{equation}
\label{eq:We_2nd}\mathcal{W}_e=\alpha\left(E^{c}_1LY^{(2)}_{\mathbf{3}}H_{d}\right)_{\mathbf{1}}+
\beta \left(E^{c}_2LY^{(4)}_{\mathbf{3}a}H_{d}\right)_{\mathbf{1}}+\gamma_1 \left(E^{c}_3LY^{(6)}_{\mathbf{3}a}H_{d}\right)_{\mathbf{1}}+\gamma_2 \left(E^{c}_3LY^{(6)}_{\mathbf{3}b}H_{d}\right)_{\mathbf{1}}\,,
\end{equation}
for $k_1=k_2-2=k_3-4=2-k_L$, and the charged lepton mass matrix is given by
\begin{equation}
M_e=\begin{pmatrix}
\alpha Y^{(2)}_{\mathbf{3}, 1}  ~&~ \alpha Y^{(2)}_{\mathbf{3}, 2} ~&~ \alpha Y^{(2)}_{\mathbf{3}, 3}  \\
\beta Y^{(4)}_{\mathbf{3}a, 1}  ~&~ \beta Y^{(4)}_{\mathbf{3}a, 2} ~&~ \beta Y^{(4)}_{\mathbf{3}a, 3}  \\
\gamma_1 Y^{(6)}_{\mathbf{3}a, 1} +\gamma_2 Y^{(6)}_{\mathbf{3}b, 1}  ~&~ \gamma_1 Y^{(6)}_{\mathbf{3}a, 2} +\gamma_2 Y^{(6)}_{\mathbf{3}b, 2}  ~&~ \gamma_1 Y^{(6)}_{\mathbf{3}a, 3} +\gamma_2 Y^{(6)}_{\mathbf{3}b, 3}
\end{pmatrix}v_{d}\,.
\end{equation}
Since there are two triplet modular forms $Y^{(8)}_{\mathbf{3}a}$ and $Y^{(8)}_{\mathbf{3}b}$ at weight 8, the superpotential $\mathcal{W}_e$ also comprises four independent terms for the weight assignment $k_1=k_2-2=k_3-6=2-k_L$,
\begin{equation}
\label{eq:We_3rd}\mathcal{W}_e=\alpha\left(E^{c}_1LY^{(2)}_{\mathbf{3}}H_{d}\right)_{\mathbf{1}}+
\beta \left(E^{c}_2LY^{(4)}_{\mathbf{3}a}H_{d}\right)_{\mathbf{1}}+\gamma_1 \left(E^{c}_3LY^{(8)}_{\mathbf{3}a}H_{d}\right)_{\mathbf{1}}+\gamma_2 \left(E^{c}_3LY^{(8)}_{\mathbf{3}b}H_{d}\right)_{\mathbf{1}}\,,
\end{equation}
which leads to
\begin{equation}\label{eq:ch_mass_C3}
M_e=\begin{pmatrix}
\alpha Y^{(2)}_{\mathbf{3}, 1}  ~&~ \alpha Y^{(2)}_{\mathbf{3}, 2} ~&~ \alpha Y^{(2)}_{\mathbf{3}, 3}  \\
\beta Y^{(4)}_{\mathbf{3}a, 1}  ~&~ \beta Y^{(4)}_{\mathbf{3}a, 2} ~&~ \beta Y^{(4)}_{\mathbf{3}a, 3}  \\
\gamma_1 Y^{(8)}_{\mathbf{3}a, 1} +\gamma_2 Y^{(8)}_{\mathbf{3}b, 1}  ~&~ \gamma_1 Y^{(8)}_{\mathbf{3}a, 2} +\gamma_2 Y^{(8)}_{\mathbf{3}b, 2}  ~&~ \gamma_1 Y^{(8)}_{\mathbf{3}a, 3} +\gamma_2 Y^{(8)}_{\mathbf{3}b, 3}
\end{pmatrix}v_{d}\,.
\end{equation}
We find that the triplet modular forms $Y^{(2)}_{\mathbf{3}}$, $Y^{(4)}_{\mathbf{3}}$, $Y^{(8)}_{\mathbf{3}a}$ and $Y^{(8)}_{\mathbf{3}b}$ satisfy the following identities
\begin{equation}
Y^{(8)}_{\mathbf{3}a}(\tau)=Y^{(4)}_{\mathbf{1}}(\tau)Y^{(4)}_{\mathbf{3}a}(\tau), \qquad
Y^{(8)}_{\mathbf{3}b}(\tau)=-\frac{\sqrt{2}}{3}Y^{(6)}_{\mathbf{1}}(\tau)Y^{(2)}_{\mathbf{3}}(\tau)+2 \sqrt{\frac{2}{3}}Y^{(4)}_{\mathbf{1}}(\tau)Y^{(4)}_{\mathbf{3}a}(\tau)\,.
\end{equation}
Therefore the third row of the above mass matrix can be written as a linear combination of the first and the second rows, and the rank of this mass matrix is 2. As a consequence, the charged lepton mass matrix in Eq.~\eqref{eq:ch_mass_C3} would lead to massless electron. This is obviously not compatible with the present observation, therefore we shall not discuss this case in the following. The resulting charged lepton mass matrices for the rest possible models considered above are summarized in table~\ref{tab:sum_ch}.

\begin{table}[t!]
\renewcommand{\tabcolsep}{0.58mm}
\centering
\begin{tabular}{|c|c|c|c|} \hline\hline
 & $\rho_{L}$ & $k_{1,2,3}+k_L$ & Charged lepton mass matrices	\\ \hline
 & & & \\[-0.15in]
$C_1$  &  $\mathbf{3}$ & 6, 8, 10 &  $ M_e=\begin{pmatrix}
\alpha Y^{(6)}_{\mathbf{\bar{3}}, 1}  ~&~ \alpha Y^{(6)}_{\mathbf{\bar{3}}, 2} ~&~ \alpha Y^{(6)}_{\mathbf{\bar{3}}, 3}  \\
 ~~&~~     ~~&~~    ~~&~~ \\[-0.15in]
\beta Y^{(8)}_{\mathbf{\bar{3}}, 1}  ~&~ \beta Y^{(8)}_{\mathbf{\bar{3}}, 2} ~&~ \beta Y^{(8)}_{\mathbf{\bar{3}}, 3}  \\
 ~~&~~     ~~&~~    ~~&~~ \\[-0.15in]
\gamma_1 Y^{(10)}_{\mathbf{\bar{3}}a, 1} +\gamma_2 Y^{(10)}_{\mathbf{\bar{3}}b, 1}  ~&~ \gamma_1 Y^{(10)}_{\mathbf{\bar{3}}a, 2} +\gamma_2 Y^{(10)}_{\mathbf{\bar{3}}b, 2}  ~&~ \gamma_1 Y^{(10)}_{\mathbf{\bar{3}}a, 3} +\gamma_2 Y^{(10)}_{\mathbf{\bar{3}}b, 3} \\
 ~~&~~     ~~&~~    ~~&~~ \\[-0.15in]
\end{pmatrix}v_{d} $\\
 & & & \\[-0.15in]  \hline
  & & & \\[-0.15in]
$C_2$  & $\mathbf{\bar{3}}$  & 2, 4, 6  &   $ M_e=\begin{pmatrix}
\alpha Y^{(2)}_{\mathbf{3}, 1}  ~&~ \alpha Y^{(2)}_{\mathbf{3}, 2} ~&~ \alpha Y^{(2)}_{\mathbf{3}, 3}  \\
 ~~&~~     ~~&~~    ~~&~~ \\[-0.15in]
\beta Y^{(4)}_{\mathbf{3}a, 1}  ~&~ \beta Y^{(4)}_{\mathbf{3}a, 2} ~&~ \beta Y^{(4)}_{\mathbf{3}a, 3}  \\
 ~~&~~     ~~&~~    ~~&~~ \\[-0.15in]
\gamma_1 Y^{(6)}_{\mathbf{3}a, 1} +\gamma_2 Y^{(6)}_{\mathbf{3}b, 1}  ~&~ \gamma_1 Y^{(6)}_{\mathbf{3}a, 2} +\gamma_2 Y^{(6)}_{\mathbf{3}b, 2}  ~&~ \gamma_1 Y^{(6)}_{\mathbf{3}a, 3} +\gamma_2 Y^{(6)}_{\mathbf{3}b, 3} \\
 ~~&~~     ~~&~~    ~~&~~ \\[-0.15in]
\end{pmatrix}v_{d}$\\ \hline	\hline	
			 	 		 		
\end{tabular}
\caption{\label{tab:sum_ch}The charged lepton mass matrices for different possible assignments of the left-handed lepton fields $L$, where the charged lepton mass matrix $M_e$ is given in the right-left basis $E^c\,M_e\,L$ with $ v_d =\langle H^0_d\rangle$.  }
\end{table}

\subsection{Neutrino sector}

In the present paper, we assume neutrinos are Majorana particles, and the neutrino masses are described by the effective Weinberg operator or the type I seesaw mechanism. From the multiplication rules $\mathbf{3}\otimes\mathbf{3}=\mathbf{\bar{3}_A}\oplus\mathbf{6_S}$ and $\mathbf{\bar{3}}\otimes\mathbf{\bar{3}}=\mathbf{3_A}\oplus\mathbf{6_S}$, we see that the sextet modular forms $Y_{\mathbf{6}}$ are necessary when neutrino masses originate from the Weinberg operator $\mathcal{W}_{W}$. We can uniquely determine the form of $\mathcal{W}_{W}$ as follow,
\begin{equation}
\label{eq:weinberg}\mathcal{W}_W=\left\{\begin{array}{cc}\frac{g_1}{2\Lambda}\left(LLY^{(4)}_{\mathbf{6}a}\right)_{\mathbf{1}}H_uH_u+\frac{g_2}{2\Lambda}\left(LLY^{(4)}_{\mathbf{6}b}\right)_{\mathbf{1}}H_uH_u, &~~   k_L=2\,, \\ [0.1in]
\frac{g_1}{2\Lambda}\left(LLY^{(6)}_{\mathbf{6}a}\right)_{\mathbf{1}}H_uH_u+\frac{g_2}{2\Lambda}\left(LLY^{(6)}_{\mathbf{6}b}\right)_{\mathbf{1}}H_uH_u, &~~   k_L=3 \,.
\end{array}
\right.
\end{equation}
Applying the decomposition rules of $\Gamma_7$, we obtain the light neutrino mass matrix as
\begin{small}
\begin{eqnarray}
\nonumber \hskip-0.4in&& m_{\nu}=
\frac{v^2_u}{\Lambda}\begin{pmatrix}
\sqrt{2}(g_1Y^{(2k_L)}_{\mathbf{6}a, 5}+g_2Y^{(2k_L)}_{\mathbf{6}b, 5})  &  g_1Y^{(2k_L)}_{\mathbf{6}a, 4}+g_2Y^{(2k_L)}_{\mathbf{6}b, 4}  & g_1Y^{(2k_L)}_{\mathbf{6}a, 2}+g_2Y^{(2k_L)}_{\mathbf{6}b, 2} \\
g_1Y^{(2k_L)}_{\mathbf{6}a, 4}+g_2Y^{(2k_L)}_{\mathbf{6}b, 4} &  \sqrt{2}(g_1Y^{(2k_L)}_{\mathbf{6}a, 3}+g_2Y^{(2k_L)}_{\mathbf{6}b, 3}) & g_1Y^{(2k_L)}_{\mathbf{6}a, 1}+g_2Y^{(2k_L)}_{\mathbf{6}b, 1}  \\
g_1Y^{(2k_L)}_{\mathbf{6}a, 2}+g_2Y^{(2k_L)}_{\mathbf{6}b, 2}  & g_1Y^{(2k_L)}_{\mathbf{6}a, 1}+g_2Y^{(2k_L)}_{\mathbf{6}b, 1}  &
\sqrt{2}(g_1Y^{(2k_L)}_{\mathbf{6}a, 6}+g_2Y^{(2k_L)}_{\mathbf{6}b, 6})
\end{pmatrix},~\text{for}~L\sim\mathbf{3}\,,\\
\hskip-0.4in&& m_{\nu}=
\frac{v^2_u}{\Lambda}\begin{pmatrix}
\sqrt{2}(g_1Y^{(2k_L)}_{\mathbf{6}a, 2}+g_2Y^{(2k_L)}_{\mathbf{6}b, 2})&  g_1Y^{(2k_L)}_{\mathbf{6}a, 3}+g_2Y^{(2k_L)}_{\mathbf{6}b, 3}  &  g_1Y^{(2k_L)}_{\mathbf{6}a, 5}+g_2Y^{(2k_L)}_{\mathbf{6}b, 5}   \\
g_1Y^{(2k_L)}_{\mathbf{6}a, 3}+g_2Y^{(2k_L)}_{\mathbf{6}b, 3}  & \sqrt{2}(g_1Y^{(2k_L)}_{\mathbf{6}a, 4}+g_2Y^{(2k_L)}_{\mathbf{6}b, 4})  & g_1Y^{(2k_L)}_{\mathbf{6}a, 6}+g_2Y^{(2k_L)}_{\mathbf{6}b, 6}    \\
 g_1Y^{(2k_L)}_{\mathbf{6}a, 5}+g_2Y^{(2k_L)}_{\mathbf{6}b, 5}  &  g_1Y^{(2k_L)}_{\mathbf{6}a, 6}+g_2Y^{(2k_L)}_{\mathbf{6}b, 6} & \sqrt{2}(g_1Y^{(2k_L)}_{\mathbf{6}a, 1}+g_2Y^{(2k_L)}_{\mathbf{6}b, 1})
\end{pmatrix},~\text{for}~L\sim\mathbf{\bar{3}}\,,
\end{eqnarray}
\end{small}
where $Y^{(2k_L)}_{\mathbf{6}a, \mathbf{6}b}$ stands for $Y^{(4)}_{\mathbf{6}a, \mathbf{6}b}$ and  $Y^{(6)}_{\mathbf{6}a, \mathbf{6}b}$ for $k_L=2$ and $k_L=3$, respectively.

If the light neutrino masses are generated by the type I seesaw mechanism, the superpotential for the neutrino masses can be generally written as
\begin{equation}
\mathcal{W}_{\nu}=g(N^cLY_DH_u)_{\mathbf{1}}+\frac{1}{2}\Lambda(N^cN^cY_{N})_{\mathbf{1}}\,,
\end{equation}
where $Y_D$ and $Y_{N}$ denote the modular form multiplets, and they are required to ensure modular invariance. The explicit forms of $Y_D$ and $Y_{N}$ are determined by the weight and representation assignments for $L$ and $N^c$. The Majorana mass term for the heavy neutrinos $N^c$ is similar to the Weinberg operator in Eq.~\eqref{eq:weinberg}, and $Y_{N}$ should be modular form multiplets transforming as $\mathbf{6}$ under $\Gamma_7$. The mass matrix for the Majorana neutrinos $N^c$ reads as
\begin{small}
\begin{eqnarray}
\nonumber \hskip-0.4in&& m_{N}=
\Lambda\begin{pmatrix}
\sqrt{2}(g_1Y^{(2k_{N})}_{\mathbf{6}a, 5}+g_2Y^{(2k_{N})}_{\mathbf{6}b, 5})  &  g_1Y^{(2k_{N})}_{\mathbf{6}a, 4}+g_2Y^{(2k_{N})}_{\mathbf{6}b, 4}  & g_1Y^{(2k_{N})}_{\mathbf{6}a, 2}+g_2Y^{(2k_{N})}_{\mathbf{6}b, 2} \\
g_1Y^{(2k_{N})}_{\mathbf{6}a, 4}+g_2Y^{(2k_{N})}_{\mathbf{6}b, 4} &  \sqrt{2}(g_1Y^{(2k_{N})}_{\mathbf{6}a, 3}+g_2Y^{(2k_{N})}_{\mathbf{6}b, 3}) & g_1Y^{(2k_{N})}_{\mathbf{6}a, 1}+g_2Y^{(2k_{N})}_{\mathbf{6}b, 1}  \\
g_1Y^{(2k_{N})}_{\mathbf{6}a, 2}+g_2Y^{(2k_{N})}_{\mathbf{6}b, 2}  & g_1Y^{(2k_{N})}_{\mathbf{6}a, 1}+g_2Y^{(2k_{N})}_{\mathbf{6}b, 1}  &
\sqrt{2}(g_1Y^{(2k_{N})}_{\mathbf{6}a, 6}+g_2Y^{(2k_{N})}_{\mathbf{6}b, 6})
\end{pmatrix},~\text{for}~N^c\sim\mathbf{3}\,,\\
\hskip-0.4in&& m_{N}=\Lambda\begin{pmatrix}
\sqrt{2}(g_1Y^{(2k_{N})}_{\mathbf{6}a, 2}+g_2Y^{(2k_{N})}_{\mathbf{6}b, 2})&  g_1Y^{(2k_{N})}_{\mathbf{6}a, 3}+g_2Y^{(2k_{N})}_{\mathbf{6}b, 3}  &  g_1Y^{(2k_{N})}_{\mathbf{6}a, 5}+g_2Y^{(2k_{N})}_{\mathbf{6}b, 5}   \\
g_1Y^{(2k_{N})}_{\mathbf{6}a, 3}+g_2Y^{(2k_{N})}_{\mathbf{6}b, 3}  & \sqrt{2}(g_1Y^{(2k_{N})}_{\mathbf{6}a, 4}+g_2Y^{(2k_{N})}_{\mathbf{6}b, 4})  & g_1Y^{(2k_{N})}_{\mathbf{6}a, 6}+g_2Y^{(2k_{N})}_{\mathbf{6}b, 6}    \\
 g_1Y^{(2k_{N})}_{\mathbf{6}a, 5}+g_2Y^{(2k_{N})}_{\mathbf{6}b, 5}  &  g_1Y^{(2k_{N})}_{\mathbf{6}a, 6}+g_2Y^{(2k_{N})}_{\mathbf{6}b, 6} & \sqrt{2}(g_1Y^{(2k_{N})}_{\mathbf{6}a, 1}+g_2Y^{(2k_{N})}_{\mathbf{6}b, 1})
\end{pmatrix},~\text{for}~N^c\sim\mathbf{\bar{3}}\,,
\end{eqnarray}
\end{small}
where $Y^{(2k_{N})}_{\mathbf{6}a, \mathbf{6}b}=Y^{(4)}_{\mathbf{6}a, \mathbf{6}b}$ for $k_{N}=2$ and $Y^{(2k_{N})}_{\mathbf{6}a, \mathbf{6}b}=Y^{(6)}_{\mathbf{6}a, \mathbf{6}b}$ for $k_{N}=3$. Now we analyze the neutrino Yukawa couplings.

\begin{itemize}[labelindent=-0.8em, leftmargin=1.2em]
\item{$L\sim\mathbf{3}$, $N^c\sim\mathbf{3}$  }

In this case, modular invariance requires $Y_D$ should be transform in $\mathbf{3}$ or $\mathbf{6}$. In the case of $k_L+k_N=2$, and the Dirac
neutrino mass matrix take the following form
\begin{equation}
m_D=g\begin{pmatrix}
0   &  Y^{(2)}_{\mathbf{3}, 3}  & -Y^{(2)}_{\mathbf{3}, 2} \\
-Y^{(2)}_{\mathbf{3}, 3} &  0  & Y^{(2)}_{\mathbf{3}, 1}  \\
Y^{(2)}_{\mathbf{3}, 2}  &  -Y^{(2)}_{\mathbf{3}, 1}  & 0
\end{pmatrix}v_u\,.
\end{equation}
Since this $m_D$ is a $3\times3$ anti-symmetric matrix with zero determinant, the light neutrino mass matrix given by seesaw formula is at most of rank 2, such that at least one light neutrino is massless. If $k_L+k_N=4$, the neutrino Yukawa couplings would involve three independent terms, and we have
\begin{equation}
m_D=
\begin{pmatrix}
\sqrt{2}(g_2Y^{(4)}_{\mathbf{6}a, 5}+g_3Y^{(4)}_{\mathbf{6}b, 5})  &  g_1Y^{(4)}_{\mathbf{3}, 3}+g_2Y^{(4)}_{\mathbf{6}a, 4}+g_3Y^{(4)}_{\mathbf{6}b, 4}  & -g_1Y^{(4)}_{\mathbf{3}, 2} +g_2Y^{(4)}_{\mathbf{6}a, 2}+g_3Y^{(4)}_{\mathbf{6}b, 2} \\
-g_1Y^{(4)}_{\mathbf{3}, 3}+g_2Y^{(4)}_{\mathbf{6}a, 4}+g_3Y^{(4)}_{\mathbf{6}b, 4} &  \sqrt{2}(g_2Y^{(4)}_{\mathbf{6}a, 3}+g_3Y^{(4)}_{\mathbf{6}b, 3}) & g_1Y^{(4)}_{\mathbf{3}, 1}+g_2Y^{(4)}_{\mathbf{6}a, 1}+g_3Y^{(4)}_{\mathbf{6}b, 1}  \\
g_1Y^{(4)}_{\mathbf{3}, 2}+g_2Y^{(4)}_{\mathbf{6}a, 2}+g_3Y^{(4)}_{\mathbf{6}b, 2}  & -g_1Y^{(4)}_{\mathbf{3}, 1}+g_2Y^{(4)}_{\mathbf{6}a, 1}+g_3Y^{(4)}_{\mathbf{6}b, 1}  &
\sqrt{2}(g_2Y^{(4)}_{\mathbf{6}a, 6}+g_3Y^{(4)}_{\mathbf{6}b, 6})
\end{pmatrix}v_u\,.
\end{equation}
The Dirac neutrino mass matrices for $k_L+k_N\geq4$ contain more free parameters than $k_L+k_N=2$. We shall not consider these cases in the present work.

\item{$L\sim\mathbf{\bar{3}}$, $N^c\sim\mathbf{\bar{3}}$  }

In this case, the Dirac neutrino mass matrix for the modular weights $k_L+k_N=4$ contains a minimum number of input parameters. Here the modular form $Y_D$ can be $Y^{(4)}_{\mathbf{6}a}$ and $Y^{(4)}_{\mathbf{6}b}$. We can read out the Dirac neutrino mass matrix as follow
\begin{equation}
m_D=\begin{pmatrix}
\sqrt{2}(h_1Y^{(4)}_{\mathbf{6}a, 2}+h_2Y^{(4)}_{\mathbf{6}b, 2})&  h_1Y^{(4)}_{\mathbf{6}a, 3}+h_2Y^{(4)}_{\mathbf{6}b, 3}  &  h_1Y^{(4)}_{\mathbf{6}a, 5}+h_2Y^{(4)}_{\mathbf{6}b, 5}   \\
h_1Y^{(4)}_{\mathbf{6}a, 3}+h_2Y^{(4)}_{\mathbf{6}b, 3}  & \sqrt{2}(h_1Y^{(4)}_{\mathbf{6}a, 4}+h_2Y^{(4)}_{\mathbf{6}b, 4})  & h_1Y^{(4)}_{\mathbf{6}a, 6}+h_2Y^{(4)}_{\mathbf{6}b, 6}    \\
 h_1Y^{(4)}_{\mathbf{6}a, 5}+h_2Y^{(4)}_{\mathbf{6}b, 5}  &  h_1Y^{(4)}_{\mathbf{6}a, 6}+h_2Y^{(4)}_{\mathbf{6}b, 6} & \sqrt{2}(h_1Y^{(4)}_{\mathbf{6}a, 1}+h_2Y^{(4)}_{\mathbf{6}b, 1})
\end{pmatrix}v_u\,.
\end{equation}
Similar in the case of $k_L+k_N=6$, we have
\begin{equation}
m_D=\begin{pmatrix}
\sqrt{2}(g_2Y^{(6)}_{\mathbf{6}a, 2}+g_3Y^{(6)}_{\mathbf{6}b, 2})&  g_1Y^{(6)}_{\mathbf{\bar{3}}, 3}+ g_2Y^{(6)}_{\mathbf{6}a, 3}+g_3Y^{(6)}_{\mathbf{6}b, 3}  &  -g_1Y^{(6)}_{\mathbf{\bar{3}}, 2}+g_2Y^{(6)}_{\mathbf{6}a, 5}+g_3Y^{(6)}_{\mathbf{6}b, 5}   \\
-g_1Y^{(6)}_{\mathbf{\bar{3}}, 3}+g_2Y^{(6)}_{\mathbf{6}a, 3}+g_3Y^{(6)}_{\mathbf{6}b, 3}  & \sqrt{2}(g_2Y^{(6)}_{\mathbf{6}a, 4}+g_3Y^{(6)}_{\mathbf{6}b, 4})  & g_1Y^{(6)}_{\mathbf{\bar{3}}, 1}+g_2Y^{(6)}_{\mathbf{6}a, 6}+g_3Y^{(6)}_{\mathbf{6}b, 6}    \\
g_1Y^{(6)}_{\mathbf{\bar{3}}, 2}+ g_2Y^{(6)}_{\mathbf{6}a, 5}+g_3Y^{(6)}_{\mathbf{6}b, 5}  &  -g_1Y^{(6)}_{\mathbf{\bar{3}}, 1}+g_2Y^{(6)}_{\mathbf{6}a, 6}+g_3Y^{(6)}_{\mathbf{6}b, 6} & \sqrt{2}(g_2Y^{(6)}_{\mathbf{6}a, 1}+g_3Y^{(6)}_{\mathbf{6}b, 1})
\end{pmatrix}v_u\,.
\end{equation}
It contains three complex input parameters.

\item{$L\sim\mathbf{\bar{3}}$, $N^c\sim\mathbf{3}$  }

For this type of assignment, $L$ and $N^c$ can form an invariant singlet for $k_L+k_N=0$, and consequently $m_D$ has a simple structure,
\begin{equation}
\label{eq:mD_1}m_D=g\begin{pmatrix}
1  ~&~  0  ~&~ 0 \\
0  ~&~  1  ~&~  0 \\
0  ~&~ 0  ~&~ 1
\end{pmatrix}v_u\,.
\end{equation}
The lowest non-vanishing weight $k_L+k_N=2$ gives rise to the following neutrino Yukawa coupling
\begin{equation}
h_1(N^cLY^{(2)}_{\mathbf{8}a}H_u)_{\mathbf{1}}+h_2(N^cLY^{(2)}_{\mathbf{8}b}H_u)_{\mathbf{1}}\,.
\end{equation}
The Dirac neutrino mass matrix is given by
\begin{equation}
\label{eq:mD_2}m_D=\begin{pmatrix}
\left.\begin{array}{c}\sqrt{3}(g_1Y^{(2)}_{\mathbf{8}a,1}+g_2Y^{(2)}_{\mathbf{8}b,1})\\
+g_1Y^{(2)}_{\mathbf{8}a,2}+g_2Y^{(2)}_{\mathbf{8}b,2}
\end{array}\right.   ~&~  \sqrt{6}(g_1Y^{(2)}_{\mathbf{8}a,3}+g_2Y^{(2)}_{\mathbf{8}b,3})  ~&~ \sqrt{6}(g_1Y^{(2)}_{\mathbf{8}a,5}+g_2Y^{(2)}_{\mathbf{8}b,5})  \\
-\sqrt{6}(g_1Y^{(2)}_{\mathbf{8}a,8}+g_2Y^{(2)}_{\mathbf{8}b,8})  ~&~ \left.\begin{array}{c}-\sqrt{3}(g_1Y^{(2)}_{\mathbf{8}a,1}+g_2Y^{(2)}_{\mathbf{8}b,1})\\
  +g_1Y^{(2)}_{\mathbf{8}a,2}+g_2Y^{(2)}_{\mathbf{8}b,2}\end{array}\right.  ~&~ \sqrt{6}(g_1Y^{(2)}_{\mathbf{8}a,4}+g_2Y^{(2)}_{\mathbf{8}b,4}) \\
-\sqrt{6}(g_1Y^{(2)}_{\mathbf{8}a,6}+g_2Y^{(2)}_{\mathbf{8}b,6})  ~&~ -\sqrt{6}(g_1Y^{(2)}_{\mathbf{8}a,7}+g_2Y^{(2)}_{\mathbf{8}b,7}) ~&~ -2(g_1Y^{(2)}_{\mathbf{8}a,2}+g_2Y^{(2)}_{\mathbf{8}b,2})
\end{pmatrix}v_u\,,
\end{equation}
which contains one more complex parameter than the above case of $k_L+k_N=0$. A number of free coupling constants are introduced for higher modular weights, the expression of $m_D$ would be too lengthy to display. For the assignment $L\sim\mathbf{3}$ and $N^c\sim\mathbf{\bar{3}}$, the matrix $m_D$ can be obtained from Eq.~\eqref{eq:mD_1} by performing transposition. The different possible forms of the neutrino mass matrices for Weinberg operator and type I seesaw mechanism are summarized in table~\ref{tab:sum_nu}. We see that the light neutrino mass matrix of $S_2$ contains more free parameters than other possible cases. Hence we will not perform numerical analysis for the $S_2$ model in the following.

\end{itemize}

\begin{table}[t!]
\renewcommand{\tabcolsep}{0.6mm}
\centering
\begin{tabular}{|c|c|c|c|c|} \hline\hline

& $\rho_{L}$, $\rho_{N^c}$ & $k_L$, $k_{N}$ &	Neutrino mass matrices  \\ \hline
   &   &   &  \\ [-0.15in]	
$W_1$ & $\mathbf{3}$, --- & 2(3), --- & {\footnotesize $ m_{\nu}=
\frac{v^2_u}{\Lambda}\begin{pmatrix}
\sqrt{2}(g_1Y^{(2k_L)}_{\mathbf{6}a, 5}+g_2Y^{(2k_L)}_{\mathbf{6}b, 5})  &  g_1Y^{(2k_L)}_{\mathbf{6}a, 4}+g_2Y^{(2k_L)}_{\mathbf{6}b, 4}  & g_1Y^{(2k_L)}_{\mathbf{6}a, 2}+g_2Y^{(2k_L)}_{\mathbf{6}b, 2} \\
g_1Y^{(2k_L)}_{\mathbf{6}a, 4}+g_2Y^{(2k_L)}_{\mathbf{6}b, 4} &  \sqrt{2}(g_1Y^{(2k_L)}_{\mathbf{6}a, 3}+g_2Y^{(2k_L)}_{\mathbf{6}b, 3}) & g_1Y^{(2k_L)}_{\mathbf{6}a, 1}+g_2Y^{(2k_L)}_{\mathbf{6}b, 1}  \\
g_1Y^{(2k_L)}_{\mathbf{6}a, 2}+g_2Y^{(2k_L)}_{\mathbf{6}b, 2}  & g_1Y^{(2k_L)}_{\mathbf{6}a, 1}+g_2Y^{(2k_L)}_{\mathbf{6}b, 1}  &
\sqrt{2}(g_1Y^{(2k_L)}_{\mathbf{6}a, 6}+g_2Y^{(2k_L)}_{\mathbf{6}b, 6})
\end{pmatrix}$}  \\
   &   &   &   \\ [-0.15in] \hline
   &   &   &   \\ [-0.15in]
   $W_2$ & $\mathbf{\bar{3}}$, --- & 2(3), ---  & {\footnotesize $ m_{\nu}=\frac{v^2_u}{\Lambda}\begin{pmatrix}
\sqrt{2}(g_1Y^{(2k_L)}_{\mathbf{6}a, 2}+g_2Y^{(2k_L)}_{\mathbf{6}b, 2})&  g_1Y^{(2k_L)}_{\mathbf{6}a, 3}+g_2Y^{(2k_L)}_{\mathbf{6}b, 3}  &  g_1Y^{(2k_L)}_{\mathbf{6}a, 5}+g_2Y^{(2k_L)}_{\mathbf{6}b, 5}   \\
g_1Y^{(2k_L)}_{\mathbf{6}a, 3}+g_2Y^{(2k_L)}_{\mathbf{6}b, 3}  & \sqrt{2}(g_1Y^{(2k_L)}_{\mathbf{6}a, 4}+g_2Y^{(2k_L)}_{\mathbf{6}b, 4})  & g_1Y^{(2k_L)}_{\mathbf{6}a, 6}+g_2Y^{(2k_L)}_{\mathbf{6}b, 6}    \\
 g_1Y^{(2k_L)}_{\mathbf{6}a, 5}+g_2Y^{(2k_L)}_{\mathbf{6}b, 5}  &  g_1Y^{(2k_L)}_{\mathbf{6}a, 6}+g_2Y^{(2k_L)}_{\mathbf{6}b, 6} & \sqrt{2}(g_1Y^{(2k_L)}_{\mathbf{6}a, 1}+g_2Y^{(2k_L)}_{\mathbf{6}b, 1})
\end{pmatrix}$}  \\
   &   &  &   \\ [-0.15in] \hline
  &   &  &    \\ [-0.15in]
\multirow{4}{*}{$S_1$} &  \multirow{4}{*}{$\mathbf{3}$, $\mathbf{3}$} & \multirow{4}{*}{$0(-1)$,} \multirow{4}{*}{2(3)} &  {\footnotesize $ m_D=g\begin{pmatrix}
0   &  Y^{(2)}_{\mathbf{3}, 3}  & -Y^{(2)}_{\mathbf{3}, 2} \\
-Y^{(2)}_{\mathbf{3}, 3} &  0  & Y^{(2)}_{\mathbf{3}, 1}  \\
Y^{(2)}_{\mathbf{3}, 2}  &  -Y^{(2)}_{\mathbf{3}, 1}  & 0
\end{pmatrix}v_u $}, \\
 &   &  &    \\ [-0.18in]
    &   &  &     \\ [-0.18in]
 &   &  &
{\footnotesize	$m_{N}=\Lambda\begin{pmatrix}
\sqrt{2}(g_1Y^{(2k_{N})}_{\mathbf{6}a, 5}+g_2Y^{(2k_{N})}_{\mathbf{6}b, 5})  &  g_1Y^{(2k_{N})}_{\mathbf{6}a, 4}+g_2Y^{(2k_{N})}_{\mathbf{6}b, 4}  & g_1Y^{(2k_{N})}_{\mathbf{6}a, 2}+g_2Y^{(2k_{N})}_{\mathbf{6}b, 2} \\
g_1Y^{(2k_{N})}_{\mathbf{6}a, 4}+g_2Y^{(2k_{N})}_{\mathbf{6}b, 4} &  \sqrt{2}(g_1Y^{(2k_{N})}_{\mathbf{6}a, 3}+g_2Y^{(2k_{N})}_{\mathbf{6}b, 3}) & g_1Y^{(2k_{N})}_{\mathbf{6}a, 1}+g_2Y^{(2k_{N})}_{\mathbf{6}b, 1}  \\
g_1Y^{(2k_{N})}_{\mathbf{6}a, 2}+g_2Y^{(2k_{N})}_{\mathbf{6}b, 2}  & g_1Y^{(2k_{N})}_{\mathbf{6}a, 1}+g_2Y^{(2k_{N})}_{\mathbf{6}b, 1}  &
\sqrt{2}(g_1Y^{(2k_{N})}_{\mathbf{6}a, 6}+g_2Y^{(2k_{N})}_{\mathbf{6}b, 6})
\end{pmatrix}$} \\
 &   &  &   \\ [-0.15in] \hline
  &   &  &     \\ [-0.15in]
 &   &  &     \\ [-0.15in]
\multirow{4}{*}{$S_2$} &  \multirow{4}{*}{$\mathbf{\bar{3}}$, $\mathbf{\bar{3}}$} & \multirow{4}{*}{2(1),} \multirow{4}{*}{2(3)} &  {\footnotesize $ m_D=\begin{pmatrix}
\sqrt{2}(h_1Y^{(4)}_{\mathbf{6}a, 2}+h_2Y^{(4)}_{\mathbf{6}b, 2})&  h_1Y^{(4)}_{\mathbf{6}a, 3}+h_2Y^{(4)}_{\mathbf{6}b, 3}  &  h_1Y^{(4)}_{\mathbf{6}a, 5}+h_2Y^{(4)}_{\mathbf{6}b, 5}   \\
h_1Y^{(4)}_{\mathbf{6}a, 3}+h_2Y^{(4)}_{\mathbf{6}b, 3}  & \sqrt{2}(h_1Y^{(4)}_{\mathbf{6}a, 4}+h_2Y^{(4)}_{\mathbf{6}b, 4})  & h_1Y^{(4)}_{\mathbf{6}a, 6}+h_2Y^{(4)}_{\mathbf{6}b, 6}    \\
 h_1Y^{(4)}_{\mathbf{6}a, 5}+h_2Y^{(4)}_{\mathbf{6}b, 5}  &  h_1Y^{(4)}_{\mathbf{6}a, 6}+h_2Y^{(4)}_{\mathbf{6}b, 6} & \sqrt{2}(h_1Y^{(4)}_{\mathbf{6}a, 1}+h_2Y^{(4)}_{\mathbf{6}b, 1})
\end{pmatrix}v_u $},\\
 &   &  &    \\ [-0.18in]
  &   &  &     \\ [-0.18in]
 &   &  &
{\footnotesize	$m_{N}=\Lambda\begin{pmatrix}
\sqrt{2}(g_1Y^{(2k_{N})}_{\mathbf{6}a, 2}+g_2Y^{(2k_{N})}_{\mathbf{6}b, 2})&  g_1Y^{(2k_{N})}_{\mathbf{6}a, 3}+g_2Y^{(2k_{N})}_{\mathbf{6}b, 3}  &  g_1Y^{(2k_{N})}_{\mathbf{6}a, 5}+g_2Y^{(2k_{N})}_{\mathbf{6}b, 5}   \\
g_1Y^{(2k_{N})}_{\mathbf{6}a, 3}+g_2Y^{(2k_{N})}_{\mathbf{6}b, 3}  & \sqrt{2}(g_1Y^{(2k_{N})}_{\mathbf{6}a, 4}+g_2Y^{(2k_{N})}_{\mathbf{6}b, 4})  & g_1Y^{(2k_{N})}_{\mathbf{6}a, 6}+g_2Y^{(2k_{N})}_{\mathbf{6}b, 6}    \\
 g_1Y^{(2k_{N})}_{\mathbf{6}a, 5}+g_2Y^{(2k_{N})}_{\mathbf{6}b, 5}  &  g_1Y^{(2k_{N})}_{\mathbf{6}a, 6}+g_2Y^{(2k_{N})}_{\mathbf{6}b, 6} & \sqrt{2}(g_1Y^{(2k_{N})}_{\mathbf{6}a, 1}+g_2Y^{(2k_{N})}_{\mathbf{6}b, 1})
\end{pmatrix}$} \\
 &   &  &    \\ [-0.15in] \hline
  &   &  &     \\ [-0.15in]
\multirow{4}{*}{$S_3$} &  \multirow{4}{*}{$\mathbf{\bar{3}}$, $\mathbf{3}$} & \multirow{4}{*}{$-2(-3)$,} \multirow{4}{*}{2(3)} &  {\small $ m_D=g\begin{pmatrix}
1  ~&~  0  ~&~ 0 \\
0  ~&~  1  ~&~  0 \\
0  ~&~ 0  ~&~ 1
\end{pmatrix}v_u$},\\
 &   &  &    \\ [-0.18in]
 &   &  &    \\ [-0.18in]
 &   &  &
{\footnotesize	$m_{N}=\Lambda\begin{pmatrix}
\sqrt{2}(g_1Y^{(2k_{N})}_{\mathbf{6}a, 5}+g_2Y^{(2k_{N})}_{\mathbf{6}b, 5})  &  g_1Y^{(2k_{N})}_{\mathbf{6}a, 4}+g_2Y^{(2k_{N})}_{\mathbf{6}b, 4}  & g_1Y^{(2k_{N})}_{\mathbf{6}a, 2}+g_2Y^{(2k_{N})}_{\mathbf{6}b, 2} \\
g_1Y^{(2k_{N})}_{\mathbf{6}a, 4}+g_2Y^{(2k_{N})}_{\mathbf{6}b, 4} &  \sqrt{2}(g_1Y^{(2k_{N})}_{\mathbf{6}a, 3}+g_2Y^{(2k_{N})}_{\mathbf{6}b, 3}) & g_1Y^{(2k_{N})}_{\mathbf{6}a, 1}+g_2Y^{(2k_{N})}_{\mathbf{6}b, 1}  \\
g_1Y^{(2k_{N})}_{\mathbf{6}a, 2}+g_2Y^{(2k_{N})}_{\mathbf{6}b, 2}  & g_1Y^{(2k_{N})}_{\mathbf{6}a, 1}+g_2Y^{(2k_{N})}_{\mathbf{6}b, 1}  &
\sqrt{2}(g_1Y^{(2k_{N})}_{\mathbf{6}a, 6}+g_2Y^{(2k_{N})}_{\mathbf{6}b, 6})
\end{pmatrix}$}      \\
 &   &  &    \\ [-0.15in] \hline
  &   &  &     \\ [-0.15in]
\multirow{4}{*}{$S_4$} &  \multirow{4}{*}{$\mathbf{3}$, $\mathbf{\bar{3}}$} & \multirow{4}{*}{$-2(-3)$,} \multirow{4}{*}{2(3)} &  {\small $ m_D=g\begin{pmatrix}
1  ~&~  0  ~&~ 0 \\
0  ~&~  1  ~&~  0 \\
0  ~&~ 0  ~&~ 1
\end{pmatrix}v_u$},\\
 &   &  &    \\ [-0.18in]
 &   &  &     \\ [-0.18in]
 &   &  &
{\footnotesize	$m_{N}=\Lambda\begin{pmatrix}
\sqrt{2}(g_1Y^{(2k_{N})}_{\mathbf{6}a, 2}+g_2Y^{(2k_{N})}_{\mathbf{6}b, 2})&  g_1Y^{(2k_{N})}_{\mathbf{6}a, 3}+g_2Y^{(2k_{N})}_{\mathbf{6}b, 3}  &  g_1Y^{(2k_{N})}_{\mathbf{6}a, 5}+g_2Y^{(2k_{N})}_{\mathbf{6}b, 5}   \\
g_1Y^{(2k_{N})}_{\mathbf{6}a, 3}+g_2Y^{(2k_{N})}_{\mathbf{6}b, 3}  & \sqrt{2}(g_1Y^{(2k_{N})}_{\mathbf{6}a, 4}+g_2Y^{(2k_{N})}_{\mathbf{6}b, 4})  & g_1Y^{(2k_{N})}_{\mathbf{6}a, 6}+g_2Y^{(2k_{N})}_{\mathbf{6}b, 6}    \\
 g_1Y^{(2k_{N})}_{\mathbf{6}a, 5}+g_2Y^{(2k_{N})}_{\mathbf{6}b, 5}  &  g_1Y^{(2k_{N})}_{\mathbf{6}a, 6}+g_2Y^{(2k_{N})}_{\mathbf{6}b, 6} & \sqrt{2}(g_1Y^{(2k_{N})}_{\mathbf{6}a, 1}+g_2Y^{(2k_{N})}_{\mathbf{6}b, 1})
\end{pmatrix}$}      \\ \hline \hline				 		
\end{tabular} 
\caption{\label{tab:sum_nu}The predictions for the neutrino mass matrices, where only lower modular weight assignments are displayed. The models in which neutrino masses are generated through Weinberg operator and seesaw mechanism are denoted as $W_{1,2}$ and $S_{1,2,3,4}$ respectively.}
\end{table}

\subsection{Benchmark models}

In the following, we consider two scenarios: the modular symmetry only acts on the neutrino sector and the charged lepton matrix is diagonal in the first scenario and the modular symmetry acts on both charged lepton and neutrino sector in the second scenario.

\subsubsection{Modular symmetry on neutrino sector}

The possibility that the dynamics of flavor in the charged lepton and neutrino sectors are different can not be excluded~\cite{Criado:2019tzk}. For instance, certain flavon may be involved in the charged lepton sector and neutrino sector is dictated by modular symmetry~\cite{Criado:2019tzk}. For simplicity, we assume that charged lepton sector is diagonal. In this case, there are six different models which are shown in table~\ref{tab:sum_nu}. We find that the model $S_1$ can not accommodate the experimental data, the model $S_2$ contains more input parameters than other models, consequently we will not discuss  it. Then the rest four models in table~\ref{tab:sum_nu} only depend on the following four inputs
\begin{equation}\label{eq:four_inputs}
\Re\tau, ~~~ \Im\tau, ~~~ |g_2/g_1|,~~~ \arg{(g_2/g_1)}\,,
\end{equation}
and an overall parameter which can be fixed by the mass squared difference $\Delta m^2_{21}=m_2^2- m_1^2$. The five dimensionless observable quantities:
\begin{equation}\label{eq:obs_qua}
\sin^2\theta_{12},~~ \sin^2\theta_{13},~~ \sin^2\theta_{23}, ~~\delta_{CP},~~\Delta m^2_{21}/\Delta m^2_{3\ell}\,,
\end{equation}
only depend on the four input parameters in Eq.~\eqref{eq:four_inputs}, where $\Delta m^2_{3\ell}=m_3^2-m_1^2>0$ for normal ordering (NO) and $\Delta m^2_{3\ell}=m_3^2-m_2^2 < 0$ for inverted ordering (IO)~\cite{Esteban:2018azc}.
In order to quantitatively assess how well a model can describe the experimental data on the five dimensionless observable quantities in Eq.~\eqref{eq:obs_qua}. We define a $\chi^2$ function to estimate the goodness-of-fit of a set of chosen values of the input parameters,
\begin{equation}\label{eq:chisq}
\chi^2 = \sum_{i=1}^5 \left( \frac{P_i-O_i}{\sigma_i}\right)^2\,,
\end{equation}
where $O_{i}$ denote the global best fit values of the five observable quantities in Eq.~\eqref{eq:obs_qua}, and $\sigma_i$ refer to the $1\sigma$ deviations of the corresponding quantities, and $P_i$ are the theoretical predictions for the five physical observable quantities for the input parameters taking certain values. Here the contribution of the Dirac phase $\delta_{CP}$ is also included in the $\chi^2$ function. For each value of the input parameters, one can obtain the predicted values $P_i$ and the corresponding $\chi^2$, then one can find out the lowest $\chi^2$. After performing a detailed numerical analysis for the three mixing angles, Dirac CP  phase and $\Delta m^2_{21}/\Delta m^2_{3\ell}$, we find that only models $W_1$ with $k_L=3$ and $S_4$ with $k_N=3$ for NO case and models $W_2$ with $k_L=3$ and $S_3$ with $k_N=3$ for IO case can give results in agreement with the experimental data. As an example, we only show the results of NO case. For model $W_1$ with $k_L=3$, we find the minimum of $\chi^2$ is $\chi^2_{\text{min}}=1.937$, and the best fit values of the
free parameters are
\begin{equation}
\Re\tau=0.448, ~~ \Im\tau=0.915, ~~ |g_2/g_1|=0.129, ~~ \arg{(g_2/g_1)}=0.566\pi, ~~ g^2_1v^2_u/\Lambda=56.981\,\text{meV}\,.
\end{equation}
The predictions for various observable quantities obtained at the best fit point are
\begin{eqnarray}
\nonumber &&\sin^2\theta_{13}=0.02236, \quad \sin^2\theta_{12}=0.311, \quad \sin^2\theta_{23}=0.556, \quad  \delta_{CP}=-0.984\pi\,, \\
\nonumber&&\alpha_{21}=-0.594\pi,\quad \alpha_{31}=0.0814\pi, \quad  m_1=63.167\,\text{meV} ,\quad m_2=63.749\,\text{meV}\,, \\
&&  m_3=80.733\,\text{meV}, \quad
m_{\beta}=63.789\,\text{meV}\,,~~
m_{\beta\beta}=42.787\,\text{meV}\,,
\end{eqnarray}
where $ m_{\beta}$ is the effective mass probed by direct kinematic search in beta decay and $m_{\beta\beta}$ refers to the effective Majorana mass in neutrinoless double beta decay. The latest result from KATRIN is $m_{\beta}<1.1$ eV at $90\%$ CL~\cite{Aker:2019uuj}. The combined results from KamLAND-Zen and EXO-200 give a Majorana neutrino mass limit of $m_{\beta\beta}<(120-250)$ meV~\cite{Gando:2012zm}. Our above predictions for both $m_{\beta}$ and $m_{\beta\beta}$ are compatible with these latest experimental bounds. The experimentally measured values of lepton mixing angles, Dirac CP phase and neutrino masses can also be accommodated well in model $S_4$ with $k_N=3$. The values of input parameters and predictions for mixing parameters and neutrino masses at the best fit point are given by
\begin{eqnarray}
\nonumber &&\Re\tau=-0.491, \quad \Im\tau=1.125, \quad |g_2/g_1|=0.103, \quad  \arg{(g_2/g_1)}=0.0267\pi, \quad g^2_1v^2_u/\Lambda=101.093\,\text{meV}\,, \\
\nonumber &&\sin^2\theta_{13}=0.02237, \quad \sin^2\theta_{12}=0.310, \quad \sin^2\theta_{23}=0.563, \quad  \delta_{CP}=-0.814\pi\,, \\
\nonumber&&\alpha_{21}=0.466\pi,\quad \alpha_{31}=0.352\pi, \quad  m_1=122.204\,\text{meV} ,\quad m_2=122.506\,\text{meV}\,, \\
\label{eq:bf_S4}&&  m_3=132.143\,\text{meV}, \quad
m_{\beta}=122.527\,\text{meV}\,,~~m_{\beta\beta}=96.646\,\text{meV}\,.
\end{eqnarray}
Accordingly the global minimum of the $\chi^2$ function is $\chi^2_{\text{min}}=0.0732$. The predicted values of $m_{\beta}$ and $m_{\beta\beta}$ are compatible with the latest results of KATRIN~\cite{Aker:2019uuj} and KamLAND-Zen and EXO-200~\cite{Gando:2012zm}, and they would potentially be tested in next generation experiments. As regards the experimental bound on neutrino mass sum, the result sensitively depends on the cosmological model and the experimental data considered. Combining the Planck TT, TE, EE, lowE  polarization spectra, baryon acoustic oscillation (BAO) data with the CMB lensing reconstruction power spectrum, the Planck collaboration gives  $\sum_i m_i<120$ meV at $95\%$ confidence level~\cite{Aghanim:2018eyx}. However, if only the BAO data and the CMB lensing reconstruction power spectrum are taken into account in the data analysis, this bound becomes $\sum_i m_i<600$ meV~\cite{Aghanim:2018eyx}. For the above two models, we find the neutrino mass sum $\sum_i m_i$ is $207.649\,\text{meV}$ and $376.853\,\text{meV}$ respectively which are consistent with the Planck's looser constraint $\sum_i m_i<600$ meV.

\begin{figure}[t!]
\centering
\begin{tabular}{ccc}
\includegraphics[width=0.48\linewidth]{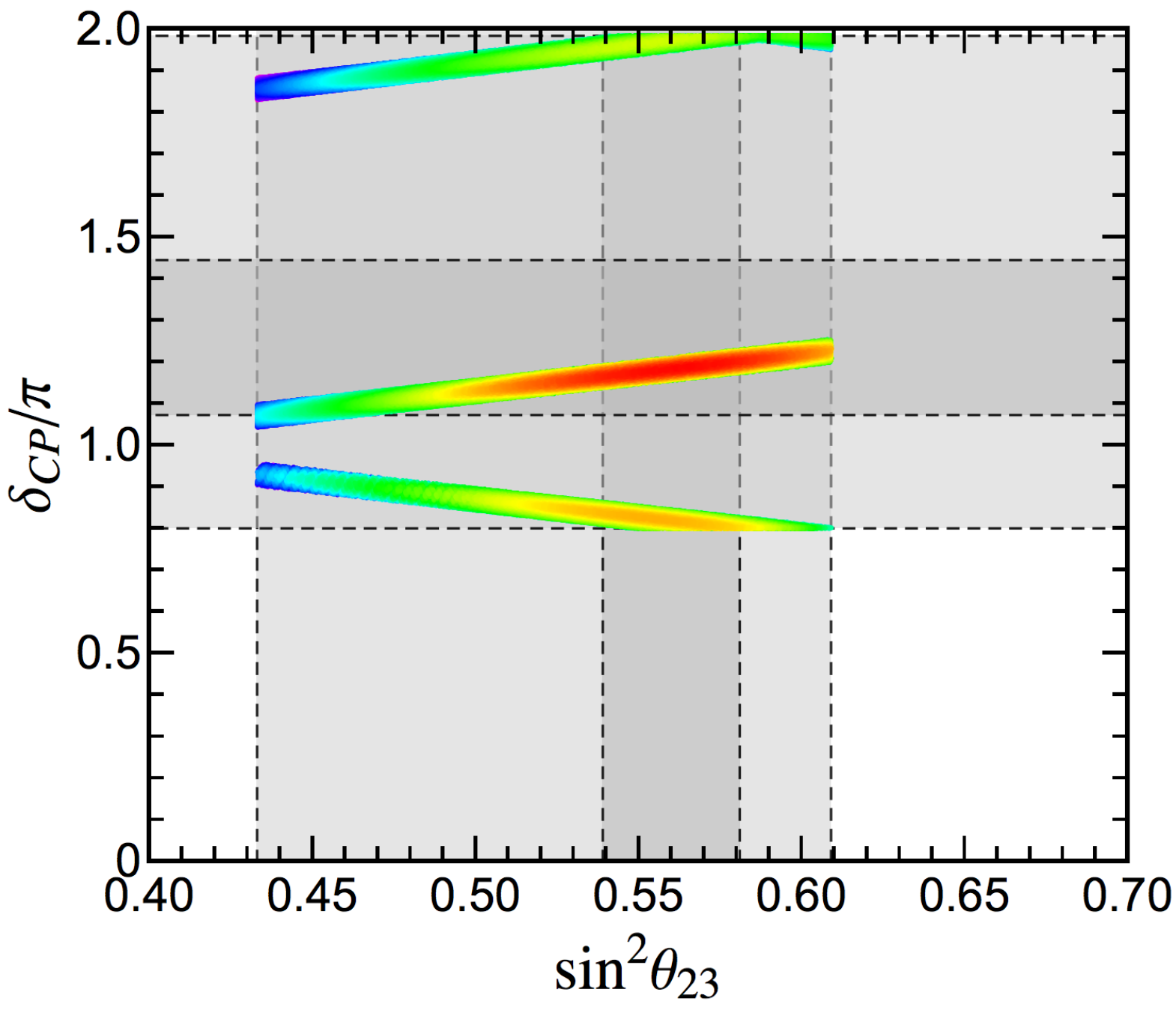}&
\includegraphics[width=0.48\linewidth]{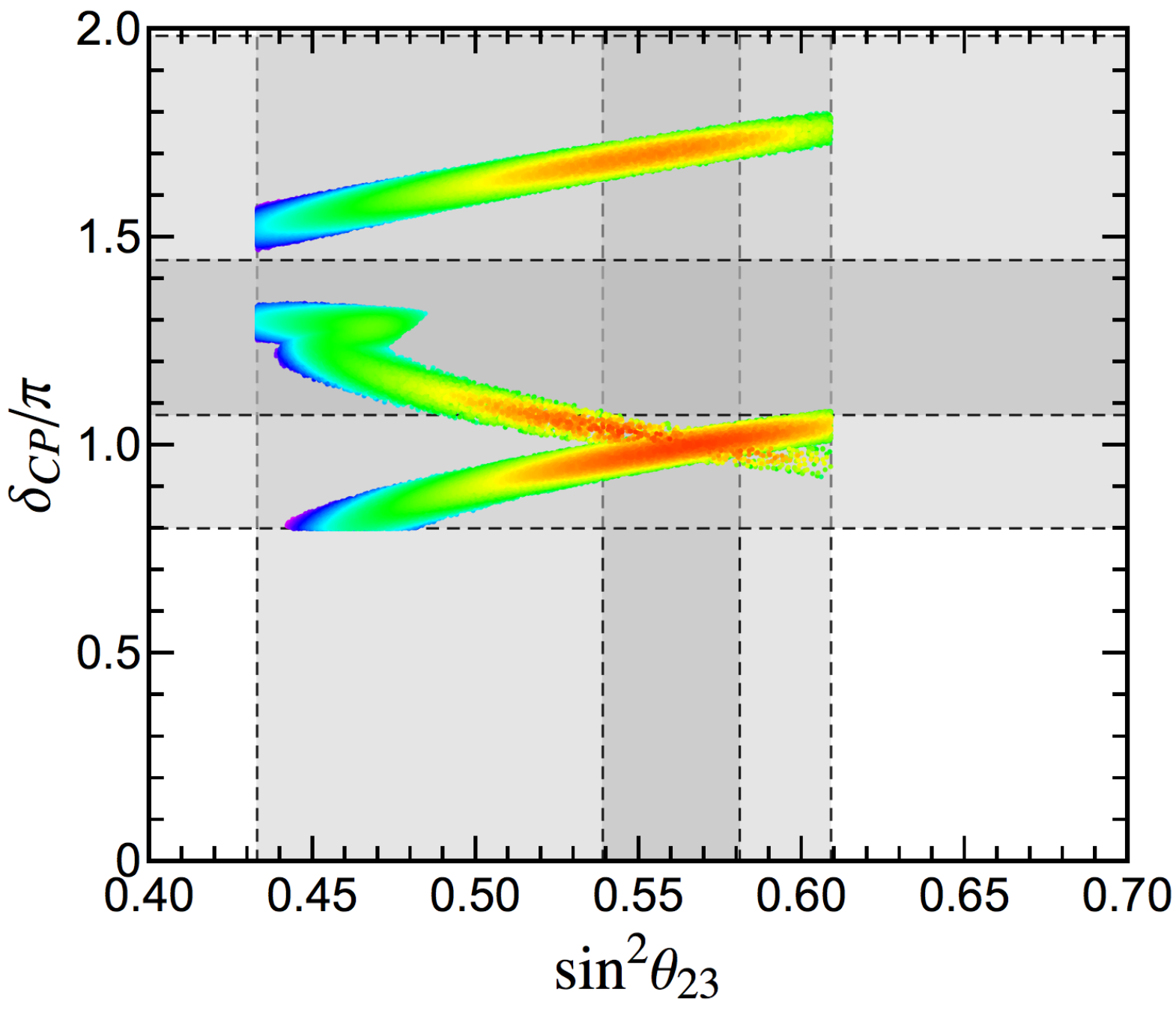}&
\multirow{1}{*}[6.7cm]{\includegraphics[height=6cm]{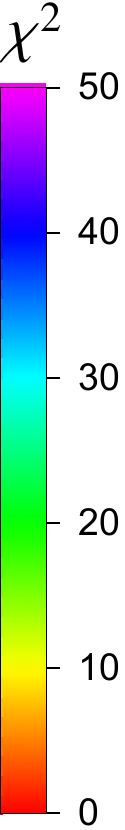}}
\end{tabular}
\caption{\label{fig:corr-S4-W1} The correlation between $\delta_{CP}$ and $\sin^2\theta_{23}$, the left and right panels are for the models $S_4$ with $k_N=3$ and  $W_1$ with $k_L=3$ respectively in the charged lepton diagonal basis. The gray bands denote the experimentally preferred $1\sigma$ and $3\sigma$ ranges of $\delta_{CP}$ and $\sin^2\theta_{23}$ adapted from~\cite{Esteban:2018azc}.}
\end{figure}

\begin{figure}[hptb!]
\centering
\includegraphics[width=0.6\linewidth]{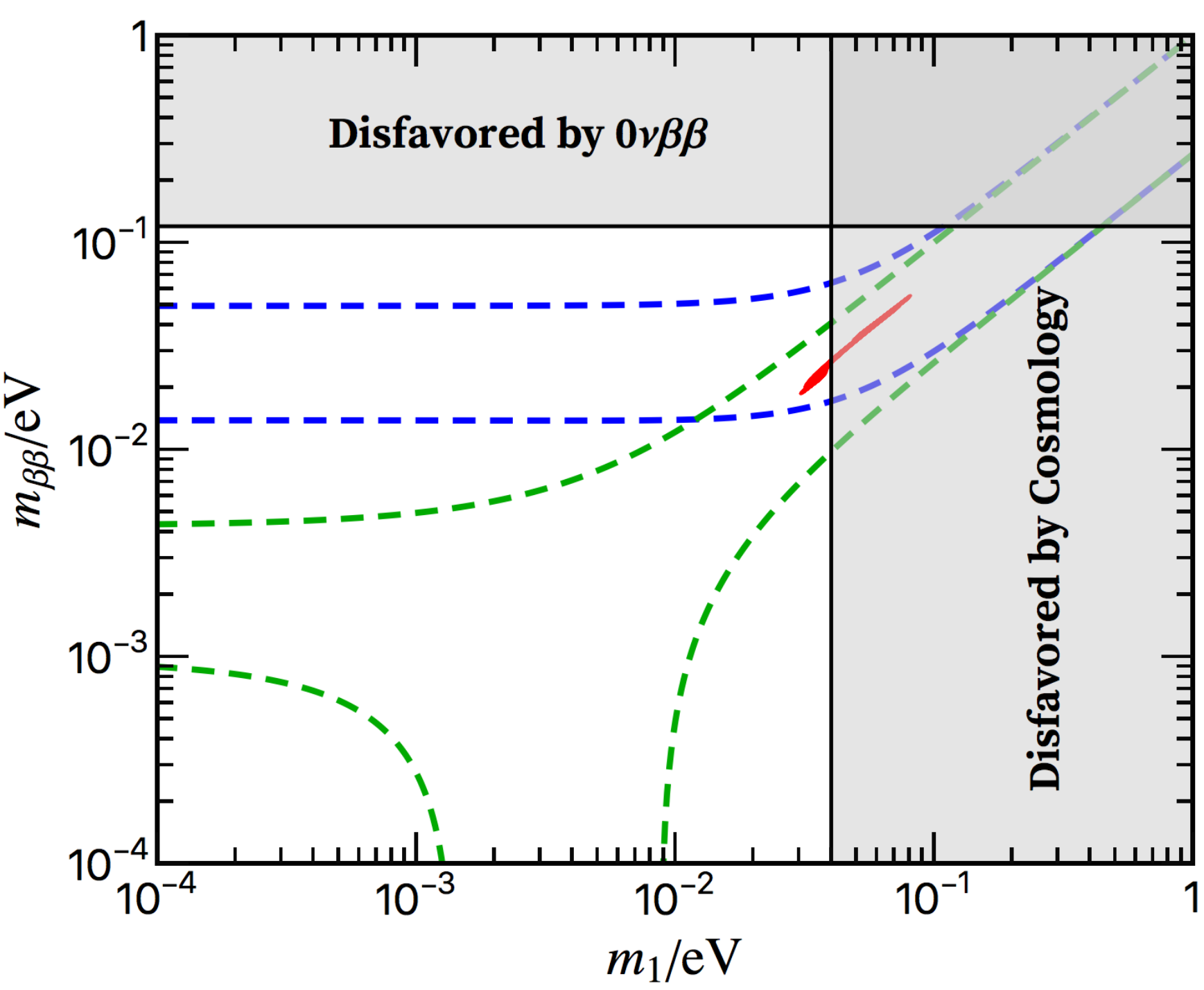}
\caption{\label{fig:mee-W1} The allowed values of the effective Majorana neutrino mass $m_{\beta\beta}$ with respect to the lightest neutrino mass $m_1$ in the model $W_1$ with $k_L=3$, where we consider the case of NO neutrino masses. The blue (green) dashed lines represent the most general allowed regions for IO (NO) where the neutrino oscillation parameters are varied within their $3\sigma$ ranges~\cite{Esteban:2018azc}. The horizontal grey band denote the current experimental bound $m_{\beta\beta}<(120-250)$ meV from KamLAND-Zen~\cite{Gando:2012zm}. The vertical grey exclusion band denotes the bound on the lightest neutrino mass extracted from $\sum_i m_i<120$ meV by the Planck collaboration~\cite{Aghanim:2018eyx}. }
\end{figure}

We perform a comprehensive numerical scan over the free parameters of the above two models. We find that the three mixing angles can take any values in their $3\sigma$ ranges. The two Majorana CP phases are restricted to the ranges $\alpha_{21}/\pi\in[0.562,0.638]\cup[1.353,1.446]$ and $\alpha_{31}/\pi\in[0,0.159]\cup[1.353,1.118]\cup[1.495,2)$ in the model $W_1$ with $k_L=3$, and they are $\alpha_{21}/\pi\in[0.442,0.497]\cup[1.503,1.557]$ and $\alpha_{31}/\pi\in[0.278,0.389]\cup[1.611,1.709)$ for the model $S_4$ with $k_N=3$. The Dirac CP phase $\delta_{CP}$ and $\theta_{23}$ are strongly correlated in the two models, as shown in figure~\ref{fig:corr-S4-W1}. The allowed values of the effective Majorana mass $m_{\beta\beta}$ for the model $W_1$ with $k_L=3$ are displayed in figure~\ref{fig:mee-W1}. We see that there is portion of parameter space where all the bounds from neutrino oscillation experiments and neutrino mass bound $\sum_i m_i<120$ meV from Planck are fulfilled. For the model $S_4$ with $k_N=3$, the neutrino masses and $m_{\beta\beta}$ lie in quite small regions around the best fit values in Eq.~\eqref{eq:bf_S4}, consequently the corresponding figure is not shown here.

\subsubsection{Modular symmetry on both neutrino and charged lepton sector sectors }

\begin{table*}[hptb!]
\renewcommand{\tabcolsep}{1.0mm}
\centering
\begin{tabular}{|c|c|c||c|c|} \hline\hline
\multirow{2}{*}{Models} & \multirow{2}{*}{mass matrices} &\multirow{2}{*}{$\rho_{L}$, $\rho_{N^c}$} & \multicolumn{2}{c|}{modular weights}

\\ \cline{4-5}
 & & &  $k_{{1,2,3}}$+$k_L $ & $k_L $, $k_{N}$  \\ \hline
$\mathcal{M}_1$ & $ C_1,W_1$ &$\mathbf{3}$, --- & $6,8,10$  & $2$ ($3$), ---
\\ \cline{1-5}

 $\mathcal{M}_2$  & $ C_2,W_2$ & $\mathbf{\bar{3}}$, --- & $2,4,6$  & $2$ ($3$), ---

 \\ \cline{1-5}

$\mathcal{M}_3$   & $C_2 ,S_3$ & $\mathbf{\bar{3}}$, $\mathbf{3}$ & $2,4,6$  & $-2(-3)$, 2(3)

 \\ \cline{1-5}

 $\mathcal{M}_{4}$ & $ C_1,S_4$ & $\mathbf{3}$, $\mathbf{\bar{3}}$ & $6,8,10$  & $-2(-3)$, 2(3)

 \\ \hline \hline	      	      	
\end{tabular}
\caption{\label{tab:sum_models}Summary of the ``minimal'' neutrino mass models with the $\Gamma_7$ modular symmetry. }
\end{table*}

Combining the different possible constructions in the charged lepton and neutrino sectors, we can easily get all possible models based on $\Gamma_7$ modular symmetry. Focusing on the cases with lower weight modular forms and free parameters as few as possible, we find four different types of models named as $\mathcal{M}_{1,2,3,4}$. The assignments of the weights and representations for the leptonic fields are listed in table~\ref{tab:sum_models}. From the superpotential of $C_1$ and $C_2$, we see that the phases of the parameters $\alpha$, $\beta$ and $\gamma_1$ can be absorbed into the right-handed lepton fields. Then the coupling constants $\alpha$, $\beta$ and $\gamma_1$ in the charged lepton mass matrix can be taken to be real and positive and $\gamma_2$ is a complex parameters for both $C_1$ and $C_2$. As a consequence, the charged lepton sector only contain four real dimensionless parameters $\beta/\alpha$, $\gamma_1/\alpha$, $|\gamma_2/\alpha|$, $\arg{(\gamma_2/\alpha)}$ and an overall scale $\alpha v_d$. In the neutrino sector, it is easy to check that light neutrino mass matrices for all the four cases $W_1$, $W_2$, $S_1$ and $S_2$ depend on two positive dimensionless parameters $|g_2/g_1|$, $\arg{(g_2/g_1)}$ and an overall neutrino mass $g^2_1v^2_u/\Lambda$ or $g^2v^2_u/(g_1\Lambda)$. We perform a numerical analysis for each model, the complex modulus $\tau$ is restricted to lie in the fundamental domain $\{\tau|\Im\tau>0, |\Re\tau|\leq\frac{1}{2}, |\tau|\geq1\}$ of the modular group. The other dimensionless parameters randomly vary in the following regions
\begin{equation}
\arg{(\gamma_2/\alpha)}, \, \arg{(g_2/g_1)}\in [0,2\pi)\,, \qquad
\beta/\alpha, \,\gamma/\alpha, \, \gamma_1/\alpha, \, |\gamma_2/\alpha|, \, |g_2/g_1|\in [0,10^4]\,.
\end{equation}
The overall parameters of charged lepton and neutrino mass matrices are fixed by the electron mass and the solar neutrino mass squared difference $\Delta m^2_{21}$. Then one can obtained all the predictions for six lepton masses and three lepton mixing angles as well as three CP violating phases. We find that good agreement with experimental data can be achieved for certain values of input parameters for all these four models in both NO and IO cases. Since NO is slightly preferred by present data, we only present the numerical results of the four models for NO. The predictions for lepton mixing parameters and neutrino masses are listed in table~\ref{tab:bf_NO}. The charged lepton mass matrix depends on four parameters $\alpha$, $\beta$, $\gamma_1$ and $\gamma_2$, the measured values of the three charged lepton masses can be reproduced exactly, hence we do not show the results of charged lepton masses in table~\ref{tab:bf_NO}. We find that the global best fit values of neutrino mixing angles and mass squared differences $\Delta m^2_{21}$, $\Delta m^2_{31}$ from NuFIT v4.1 can be obtained, as shown in table~\ref{tab:bf_NO}. Moreover, the most stringent neutrino mass bound $\sum_i m_i<120$ meV~\cite{Aghanim:2018eyx} is saturated in models $\mathcal{M}_2$ and $\mathcal{M}_3$, and the less stringent bound $\sum_i m_i<600$ meV~\cite{Aghanim:2018eyx} is fulfilled in models $\mathcal{M}_1$ and $\mathcal{M}_4$. The effective Majorana mass $m_{\beta\beta}$ is predicted to be around 10 meV or few tens of meV which is within the reach of forthcoming $0\nu\beta\beta$ decay experiments such as nEXO~\cite{Albert:2017hjq}.

\begin{table}[t!]
\renewcommand{\tabcolsep}{1.7mm}
\centering
\renewcommand{\arraystretch}{1.1}
\begin{tabular}{|c|cc|cc|cc|cc|c|c|} \hline\hline
&  \multicolumn{8}{c|}{Best fit values for NO }               \\ \cline{2-9}
& \multicolumn{2}{c|}{  $\mathcal{M}_{1}$} & \multicolumn{2}{c|}{ $\mathcal{M}_{2}$}  & \multicolumn{2}{c|}{ $\mathcal{M}_{3}$} & \multicolumn{2}{c|}{ $\mathcal{M}_{4}$} \\ \cline{2-11}
& $k_L=2$  &  $k_L=3$  & $k_L=2$   &  $k_L=3$ & $k_N=2$   &  $k_N=3$& $k_N=2$   &  $k_N=3$  \\  \hline

$\Re\langle\tau\rangle$  & $-0.219$ & $-0.299$ & $0.248$ & $0.235$ & $-0.206$ & $-0.0712$ & $0.446$ & $-0.437$ \\
$\Im\langle\tau\rangle$  & $0 .962$ & $1 .435$ & $1 .315$ & $1 .881$ & $1 .343$ & $1 .962$ & $0 .836$ & $1 .285$ \\
$\beta/\alpha$ & $0 .00579$ & $337 .346$ & $75 .905$ & $29 .100$ & $49 .078$ & $51 .055$ & $197 .311$ & $392 .758$ \\
$\gamma_1/\alpha$  & $5 .641$ & $1760 .614$ & $1702 .628$ & $692 .861$ & $1706 .880$ & $326 .593$ & $3858 .297$ & $2119 .228$ \\
$|\gamma_2/\alpha|$   & $0 .632$ & $328 .957$ & $898 .143$ & $363 .564$ & $477 .413$ & $817 .142$ & $84 .991$ & $389 .175$ \\
$\arg{(\gamma_2/\alpha)}/\pi$  & $0 .213$ & $1 .022$ & $0 .0762$ & $0 .530$ & $1 .611$ & $0 .251$ & $0 .945$ & $0 .819$ \\
$|g_2/g_1|$   & $0 .394$ & $0 .0344$ & $0 .274$ & $0 .00880$ & $0 .353$ & $0 .0410$ & $1 .780$ & $0 .0781$ \\
$\arg{(g_2/g_1)}/\pi$  & $0 .538$ & $1 .404$ & $1 .050$ & $1 .474$ & $0 .290$ & $0 .304$ & $0 .326$ & $0 .584$ \\
$(g^2_1v^2_u/\Lambda)/$eV  & $0 .104$ & $0 .225$ & $0 .221$ & $0 .392$ & $0 .00384$ & $0 .00125$ & $0 .212$ & $0 .0143$ \\ \hline

$\sin^2\theta_{13}$  & $0 .0224$ & $0 .0224$ & $0 .0224$ & $0 .0224$ & $0 .0224$ & $0 .0224$ & $0 .0224$ & $0 .0224$ \\
$\sin^2\theta_{12}$  & $0 .310$ & $0 .310$ & $0 .310$ & $0 .310$ & $0 .310$ & $0 .310$ & $0 .310$ & $0 .310$ \\
$\sin^2 \theta_{23}$  & $0 .563$ & $0 .563$ & $0 .563$ & $0 .563$ & $0 .563$ & $0 .563$ & $0 .563$ & $0 .563$ \\
$\delta_{CP}/\pi$    & $-0.772$ & $-0.772$ & $-0.771$ & $-0.773$ & $-0.773$ & $-0.772$ & $-0.772$ & $-0.770$ \\
$\alpha_{21}/\pi$   & $0 .0188$ & $-0.937$ & $-0.225$ & $0 .304$ & $0 .0732$ & $-0.591$ & $-0.490$ & $0 .415$ \\
$\alpha_{31}/\pi$  & $0.381$ & $-0.279$ & $0.360$ & $0.0342$ & $0.0629$ & $-0.375$ & $0.615$ & $-0.321$ \\ \hline

$m_1$/meV   & $39 .606$ & $40 .078$ & $15 .843$ & $10 .324$ & $10 .841$ & $9 .072$ & $77 .446$ & $31 .159$ \\

$m_2$/meV   & $40 .528$ & $40 .989$ & $18 .025$ & $13 .434$ & $13 .836$ & $12 .498$ & $77 .922$ & $32 .323$ \\

$m_3$/meV   & $64 .005$ & $64 .298$ & $52 .717$ & $51 .331$ & $51 .434$ & $51 .091$ & $92 .336$ & $59 .151$ \\

$m_{\beta}$/meV  & 40.591& 41.052& 18.167  & 13.624 & 14.020  & 12.701  & 77.955 & 32.402   \\

$m_{\beta\beta}$/meV   & $40.373$ & $14.317$ & $16.434$ & $9.852$ & $11.782$ & $5.856$ & $59.281$ & $24.172$ \\ \hline \hline
\end{tabular} 
\caption{\label{tab:bf_NO}The predictions for lepton mixing parameters and neutrino masses for the four models given in table~\ref{tab:sum_models}, where we assume neutrino mass spectrum is NO, and similar results can be obtained for IO.}
\end{table}

\section{\label{sec:conclusion}Conclusion}

We have considered the finite modular group $\Gamma_7 \simeq PSL(2,Z_7)$ in the framework of the modular invariance approach to lepton flavour. $\Gamma_7$ is a quotient group of the infinite modular group $\Gamma$ which is achieved by imposing the generator condition $T^7=\mathds{1}$. An additional condition $(S T^3)^4=\mathds{1}$ should also be satisfied, which is essential to make $\Gamma_7$ finite.

One crucial ingredient of modular-invariant theories is the introduction of modular forms, which involves the modulus field
$\tau$. Given level 7 and an even weight $k$, there are $14k-2$ linearly independent modular forms, which can all be decomposed into irreducible representations (irreps) of $\Gamma_7$.
At weight $k=2$, we constructed all 26 modular forms with the help of the SageMath algebra system~\cite{SageMath:2018}. They are decomposed into a triplet $\mathbf{3}$, a septet $\mathbf{7}$ and two octets $\mathbf{8}$ of $\Gamma_7$. A full list of linearly independent modular forms of weight up to 8 is provided in Appendix~\ref{appsec:w4_cons}.
We have also considered two alternative ways to derive modular forms of level 7, the Dedekind eta function method proposed in \cite{Feruglio:2017spp} and the theta function method proposed in \cite{Novichkov:2018nkm}, discussed in Appendices~\ref{sec:eta} and \ref{sec:theta}, respectively. Results from these methods are consistent with the former one, but incomplete: the Dedekind eta function method gives only 7 modular forms of weight 2, forming the septet of $\Gamma_7$; and the theta function method gives 23 modular forms of weight 2, decomposed as $\mathbf{7}+\mathbf{8}+\mathbf{8}$.

We also considered flavon-free lepton flavour models, constructed by assigning couplings and matter superfields transforming in irreps of $\Gamma_7$. In the considered models, charged leptons gain masses via renormalisable Yukawa couplings, neutrinos gain masses via either the Weinberg operator or the type-I seesaw mechanism. Flavour textures arise after the modulus field $\tau$ gains a vacuum expectation value. Lists of charged lepton ($C_1$, $C_2$) and neutrino mass matrices ($W_1$, $W_2$ and $S_1$, ..., $S_4$) involving $\tau$ are given in Tables~\ref{tab:sum_ch} and \ref{tab:sum_nu}, respectively.

In the numerical studies, we have considered two scenarios: the modular symmetry acting on {\em only} the neutrino sector, or {\em both} charged lepton and neutrino sectors. We have performed a $\chi^2$ analysis with experimental data of neutrino oscillation parameters in the $1\sigma$ range taken into account. In the {\em first} scenario, we found that, in the normal neutrino mass ordering, only models $W_1$ with lepton doublet weight $k_L=3$ and $S_4$ with right-handed neutrino weight $k_N=3$, give results that agree with the experimental data. Similarly, in the inverted mass ordering, only models $W_2$ with $k_L=3$ and $S_3$ with $k_N=3$ are allowed by data. The effective neutrino masses $m_\beta$ and $m_{\beta\beta}$, which control the beta decay and neutrinoless double beta decay rates, respectively, are predicted to be compatible with current data. However, the prediction of the sum of neutrino masses is disfavoured by the current cosmological constraint $\sum_i m_i<120$ meV~\cite{Aghanim:2018eyx} although the less stringent bound $\sum_i m_i<600$ meV~\cite{Aghanim:2018eyx} is satisfied. In the {\em second} scenario, eight benchmark models ($\mathcal{M}_1$, ..., $\mathcal{M}_4$ with two sets of modular weights), listed in Table~\ref{tab:sum_models}, have been studied. While all of them are compatible with oscillation data, the models $\mathcal{M}_2$ and $\mathcal{M}_3$ are also favoured by the cosmological constraint on the sum of neutrino masses $\sum_i m_i<120$ meV, while the prediction of $m_{\beta\beta}$ is within the sensitivity of the next generation of neutrinoless double beta decay experiments.

\section*{Acknowledgements}

GJD and CCL are grateful to Dr. Chang-Yuan Yao and Jun-Nan Lu for their kind help on group theory and numerical analysis. YLZ would like to thank A. Titov for the useful discussion. GJD is supported by the National Natural Science Foundation of China under Grant Nos 11975224, 11835013, 11947301. SFK and YLZ acknowledge the STFC Consolidated Grant ST/L000296/1 and the European Union's Horizon 2020 Research and Innovation programme under Marie Sk\l{}odowska-Curie grant agreements Elusives ITN No.\ 674896 and InvisiblesPlus RISE No.\ 690575. CCL is supported by  the Anhui Province Natural Science Foundation Grant No. 1908085QA24 and the National Natural Science Foundation of China under Grant  No. 11947301.

\section*{Appendix}

\setcounter{equation}{0}
\renewcommand{\theequation}{\thesection.\arabic{equation}}

\begin{appendix}

\section{\label{sec:Group}Group Theory of $\Gamma_{7}\cong PSL(2,Z_{7})$ }
The group $\Gamma_{7}\cong PSL(2,Z_{7})$ is a non-Abelian finite subgroup of $SU(3)$ of order $168$. $\Gamma_{7}$ group can be generated by two generators $S$ and $T$ which satisfy the multiplication rules:
\begin{equation}
S^2=(ST)^{3}= T^7= (ST^{3})^{4}=1.
\end{equation}
The 168 elements of $\Gamma_{7}$ group are divided into 6 conjugacy classes:
\begin{eqnarray}
\nonumber 1C_{1} :&&\{1\}\,,\\
21C_{2} :&&
\{S,T^3ST^5ST^3,T^2ST^4ST^2,T^6ST,ST^3ST^4S,ST^5ST^2S,ST^4ST^3S,TST^6, \\
\nonumber && ST^2ST^5S,T^5ST^2,TST^5ST^5,(ST^4)^2,T^2ST^5,(ST^3)^2, T^5ST^5ST,T^4ST^3, \\
\nonumber && ST^5ST^6, T^2ST^3ST,ST^2ST,T^3ST^4,TST^3ST^2 \}\,,\\
\nonumber 56C_{3} : &&
\{ST,TS,ST^3ST^5ST^3,ST^2ST^4ST^2,ST^2ST^4ST^3,ST^3ST^5ST^4,ST^3ST^5ST^2,\\
\nonumber &&ST^2ST^4ST,ST^5ST^3S,ST^3ST^5ST^5,ST^3ST^5ST,ST^2ST^4S,
ST^2ST^4ST^5,\\
\nonumber && ST^4ST^2S,ST^2ST^4ST^6,ST^3ST^5S,T^6S,ST^4ST^2,
ST^3ST^5,ST^5ST^3,\\
\nonumber && ST^2ST^4,T^5ST^5ST^4,T^2ST^3ST^6,T^5ST,ST^2ST^5,
T^3ST^4ST^3,ST^6,\\
\nonumber &&TST^5ST^2,TST^3ST^4,ST^5ST^2,T^2ST^6,TST^4ST,T^5ST^3S,
T^4ST^5ST^5,\\
\nonumber &&T^4ST^2,T^5ST^5ST^5,T^6ST^2,TST^3S,T^2ST^4S,T^2ST^5ST,T^3ST^5,
TST^5ST,\\
\nonumber &&T^6ST^4S,TST^5,T^4ST^2S,TST^4ST^5,T^5ST^3,T^5ST^2S,ST^3ST,
T^3ST^3,T^3ST^5S,\\
\nonumber && T^2ST^4,T^3ST^4ST^6,T^2ST^5S,ST^4ST^6,T^4ST^4 \}\,,\\
\nonumber 42C_{4}: && \{ST^3,T^3S,TST^5ST^3,T^5ST^5ST^3,T^4S,T^3ST^4ST^2,TST^4ST^2,T^2ST^4ST^3,\\
\nonumber &&T^4ST^5ST^4, T^3ST,ST^4ST^3,ST^5ST^4,T^2ST,T^2ST^4ST,T^2ST^5ST^2,T^5ST^6,\\
\nonumber && ST^2ST^3, ST^3ST^4,T^4ST^6,T^3ST^5ST^5,TST^2,T^2ST^3S,T^4ST^3S,ST^2ST^6,T^2ST^2,\\
\nonumber && T^3ST^5ST, ST^5ST,T^3ST^4S,T^6ST^5,T^5ST^4S,T^5ST^5,T^6ST^2S,TST^3,T^3ST^2S, \\
\nonumber && ST^4ST^5,ST^5ST^5S,ST^4,ST^3ST^2,T^4ST^5S,T^6ST^4,TST^5S,ST^2ST^2S \}\,, \\
\nonumber 24C_{7}: && \{T,T^2,T^4,TST^3ST^5,T^2ST^5ST^3,T^5S,ST^3ST^6,
T^4ST,ST^2ST^3S,T^3ST^5ST^2,\\
\nonumber && STS,T^2ST^3ST^4,T^6ST^3S,(ST^5)^2,T^3ST^2,ST^5,ST^4S,
ST^3ST^2S,T^2ST^4ST^5,\\
\nonumber && ST^2S,T^2ST^3,TST^4ST^6,TST^4,(T^5S)^2\}\,, \\
\nonumber 24C^{\prime}_{7}: && \{T^3,T^5,T^6,T^2ST^3ST^5,T^4ST^5ST^3,T^2S,
T^3ST^5ST^4,ST^6S,TST^3ST^6,ST^4ST,\\
\nonumber && T^3ST^6,ST^5ST^4S,ST^2,ST^3S,ST^4ST^5S,TST^4S,(ST^2)^2,
T^4ST^5,T^3ST^4ST^5,\\
\nonumber && T^6ST^3,(T^2S)^2,T^2ST^4ST^6,ST^5S,T^5ST^4\}\,,
\end{eqnarray}
where $nC_k$ denotes a class with $n$ elements which is of order $k$. The character table of $\Gamma_{7}$ group is given in Table~\ref{tab:character}.
\begin{table}[t]
\begin{center}
\begin{tabular}{|c|c|c|c|c|c|c|}\hline\hline
 &\multicolumn{6}{c|}{\texttt{Conjugacy Classes}}\\\cline{2-7}
 &   &  & & & &    \\ [-0.16in]
&$1C_{1}$&$21C_{2}$&$56C_{3}$&$42C_{4}$&$24C_{7}$&$24C^{\prime}_{7}$ \\ \hline
&   &  & & & &    \\ [-0.16in]
$\bf{1}$ & 1 & 1 & 1 & 1 & 1 & 1\\ \hline

&   &  & & & &    \\ [-0.16in]
$\bf{3}$ & 3 & $-1$ & 0 & 1 & $b_{7}$ & $\bar{b}_7$ \\ \hline
 &   &  & & & &    \\ [-0.16in]

$\bf{\bar{3}}$ & 3 & $-1$ & 0 &$1$ & $\bar{b}_7$ & $b_{7}$ \\ \hline
&   &  & & & &    \\ [-0.16in]

$\bf{6}$ & 6 & 2 & 0 & $0$ & $-1$ & $-1$ \\ \hline
&   &  & & & &    \\ [-0.16in]

$\bf{7}$& 7 & $-1$ & $1$ & $-1$  & 0 & 0 \\ \hline
&   &  & & & &    \\ [-0.16in]

$\bf{8}$ & 8 & 0 & $-1$ & $0$ & $1$ &1 \\ \hline\hline

\end{tabular}
\caption{\label{tab:character} The character table of the $\Gamma_7$ group with $b_7=(-1+i\sqrt{7})/2$ and $\bar{b}_7=b^*_7=-(1+i\sqrt{7})/2$.
}
\end{center}
\end{table}
Following the convention of Ref.~\cite{deAdelhartToorop:2011re}, we find that $\Gamma_{7}$ group has ninety-two abelian subgroups in total:  twenty-one $Z_{2}$ subgroups, twenty-eight $Z_{3}$ subgroups, fourteen $K_{4}$ subgroups, twenty-one $Z_{4}$ subgroups and eight $Z_{7}$ subgroups. In terms of the generators $S$ and $T$, these abelian subgroups are given as follows:
\begin{itemize}[leftmargin=1.5em]
\item{$Z_{2}$ subgroups}
\begin{equation*}
\begin{array}{lllllll}
 Z^{S}_{2}=\{1,S\}, ~&~  Z^{T^3ST^5ST^3}_{2}=\{1,T^3ST^5ST^3\}, ~&~ Z^{T^2ST^4ST^2}_{2}=\{1,T^2ST^4ST^2\}, \\
  Z^{T^6ST}_{2}=\{1,T^6ST\}, ~&~  Z^{ST^3ST^4S}_{2}=\{1,ST^3ST^4S\}, ~&~  Z^{ST^5ST^2S}_{2}=\{1,ST^5ST^2S\},\\
   Z^{ST^4ST^3S}_{2}=\{1,ST^4ST^3S\}, ~&~  Z^{TST^6}_{2}=\{1,TST^6\}, ~&~  Z^{ST^2ST^5S}_{2}=\{1,ST^2ST^5S\}, \\
 Z^{T^5ST^2}_{2}=\{1,T^5ST^2\}, ~&~ Z^{TST^5ST^5}_{2}=\{1,TST^5ST^5\}, ~&~  Z^{(ST^4)^2}_{2}=\{1,(ST^4)^2\},\\
  Z^{T^2ST^5}_{2}=\{1,T^2ST^5\}, ~&~ Z^{(ST^3)^2}_{2}=\{1,(ST^3)^2\}, ~&~ Z^{T^5ST^5ST}_{2}=\{1,T^5ST^5ST\}, \\
 Z^{T^4ST^3}_{2}=\{1,T^4ST^3\},~&~  Z^{ST^5ST^6}_{2}=\{1,ST^5ST^6\}, ~&~ Z^{T^2ST^3ST}_{2}=\{1,T^2ST^3ST\}, \\
  Z^{ST^2ST}_{2}=\{1,ST^2ST\},~&~  Z^{T^3ST^4}_{2}=\{1,T^3ST^4\}, ~&~ Z^{TST^3ST^2}_{2}=\{1,TST^3ST^2\}.
\end{array}
\end{equation*}
All the above twenty-one $Z_{2}$ subgroups are conjugate to each other.

\item{$Z_{3}$ subgroups}
\begin{equation*}
\begin{array}{lllllll}
Z^{ST}_{3}=\{1,ST,T^6S\}, ~&~    Z^{TS}_{3}=\{1,TS,ST^6\}, \\ Z^{ST^2ST^4ST^2}_{3}=\{1,ST^2ST^4ST^2,ST^3ST^5ST^3\}, ~&~ Z^{ST^2ST^4ST^3}_{3}=\{1,ST^2ST^4ST^3,ST^3ST^5ST\}, \\ Z^{ST^3ST^5ST^4}_{3}=\{1,ST^3ST^5ST^4,ST^2ST^4ST^5\}, ~&~ Z^{ST^2ST^4ST^6}_{3}=\{1,ST^2ST^4ST^6,ST^3ST^5ST^2\},   \\ Z^{ST^2ST^4ST}_{3}=\{1,ST^2ST^4ST,ST^3ST^5ST^5\}  ~&~
 Z^{ST^4ST^2S}_{3}=\{1,ST^4ST^2S,ST^5ST^3S\}, \\
Z^{ST^2ST^4S}_{3}=\{1,ST^2ST^4S,ST^3ST^5S\},~&~
Z^{ST^4ST^2}_{3}=\{1,ST^4ST^2,T^5ST^3S\}, \\
Z^{ST^3ST^5}_{3}=\{1,ST^3ST^5,T^2ST^4S\}, ~&~
Z^{ST^5ST^3}_{3}=\{1,ST^5ST^3,T^4ST^2S\}, \\
Z^{ST^2ST^4}_{3}=\{1,ST^2ST^4,T^3ST^5S\}, ~&~
Z^{T^2ST^5ST}_{3}=\{1,T^2ST^5ST,T^5ST^5ST^4\}, \\
Z^{TST^4ST^5}_{3}=\{1,TST^4ST^5,T^2ST^3ST^6\}, ~&~
  Z^{T^5ST}_{3}=\{1,T^5ST,T^6ST^2\}, \\
  Z^{T^2ST^5S}_{3}=\{1,T^2ST^5S,ST^2ST^5\}, ~&~
 Z^{TST^4ST}_{3}=\{1,TST^4ST,T^3ST^4ST^3\}, \\
 Z^{TST^5ST^2}_{3}=\{1,TST^5ST^2,T^4ST^5ST^5\}, ~&~
 Z^{TST^3ST^4}_{3}=\{1,TST^3ST^4,T^3ST^4ST^6\}, \\
 Z^{T^5ST^2S}_{3}=\{1,T^5ST^2S,ST^5ST^2\},~&~
 Z^{TST^5}_{3}=\{1,TST^5,T^2ST^6\}, \\
Z^{T^4ST^2}_{3}=\{1,T^4ST^2,T^5ST^3\}, ~&~
Z^{TST^5ST}_{3}=\{1,TST^5ST,T^5ST^5ST^5\}, \\
 Z^{TST^3S}_{3}=\{1,TST^3S,ST^4ST^6\}, ~&~
Z^{T^2ST^4}_{3}=\{1,T^2ST^4,T^3ST^5\}, \\
 Z^{ST^3ST}_{3}=\{1,ST^3ST,T^6ST^4S\}, ~&~
  Z^{T^3ST^3}_{3}=\{1,T^3ST^3,T^4ST^4\}.
 \end{array}
\end{equation*}
The twenty-eight $Z_{3}$ subgroups are related with each other by group conjugation.
\item{$K_{4}$ subgroups}
\begin{eqnarray*}
&& K^{(S, TST^5ST^5)}_{4}\equiv Z^{S}_{2}\times Z^{TST^5ST^5}_{2}=\{1,S,TST^5ST^5,T^2ST^3ST\}, \\
 && K^{(T^2ST^5, ST^3ST^4S)}_{4}\equiv Z^{T^2ST^5}_{2}\times Z^{ST^3ST^4S}_{2}=\{1,T^2ST^5,ST^3ST^4S,T^3ST^5ST^3\}, \\
&& K^{(ST^2ST^5S, T^3ST^4)}_{4}\equiv Z^{ST^2ST^5S}_{2}\times Z^{T^3ST^4}_{2}=\{1,ST^2ST^5S, T^3ST^4,T^2ST^4ST^2\}, \\
 &&  K^{(T^6ST,ST^5ST^6)}_{4}\equiv Z^{T^6ST}_{2}\times Z^{ST^5ST^6}_{2}=\{1,T^6ST,ST^5ST^6, TST^3ST^2\}, \\
&&K^{(TST^6, (ST^4)^2)}_{4}\equiv Z^{TST^6}_{2}\times Z^{(ST^4)^2}_{2}=\{1,TST^6,(ST^4)^2,ST^5ST^2S\}, \\
&& K^{(T^5ST^5ST, T^4ST^3)}_{4}\equiv Z^{T^5ST^5ST}_{2}\times Z^{T^4ST^3}_{2}=\{1,T^5ST^5ST, T^4ST^3,ST^4ST^3S\}, \\
&& K^{(ST^2ST, T^5ST^2)}_{4}\equiv Z^{ST^2ST}_{2}\times Z^{T^5ST^2}_{2}=\{1,ST^2ST, T^5ST^2,(ST^3)^2\} \\
 &&  K^{(S, T^5ST^5ST)}_{4}\equiv Z^{S}_{2}\times Z^{T^5ST^5ST}_{2}=\{1,S, T^5ST^5ST,TST^3ST^2\}, \\
&& K^{(T^5ST^2, ST^4ST^3S)}_{4}\equiv Z^{T^5ST^2}_{2}\times Z^{ST^4ST^3S}_{2}=\{1,T^5ST^2, ST^4ST^3S,T^3ST^5ST^3\}\\
&& K^{(ST^5ST^2S, T^4ST^3)}_{4}\equiv Z^{ST^5ST^2S}_{2}\times Z^{T^4ST^3}_{2}=\{1,ST^5ST^2S, T^4ST^3,T^2ST^4ST^2\}, \\
&& K^{(T^6ST , ST^2ST^5S)}_{4}\equiv Z^{T^6ST}_{2}\times Z^{ST^2ST^5S}_{2}=\{1,T^6ST , ST^2ST^5S,(ST^3)^2\}\\
 && K^{(TST^5ST^5 , T^3ST^4)}_{4}\equiv Z^{TST^5ST^5}_{2}\times Z^{T^3ST^4}_{2}=\{1,TST^5ST^5 , T^3ST^4,ST^3ST^4S\}, \\
 &&K^{(TST^6 ,ST^2ST)}_{4}\equiv Z^{TST^6}_{2}\times Z^{ST^2ST}_{2}=\{1,TST^6 ,ST^2ST,T^2ST^3ST\}\\
&&  K^{(ST^5ST^6 ,T^2ST^5)}_{4}\equiv Z^{ST^5ST^6}_{2}\times Z^{T^2ST^5}_{2}=\{1,ST^5ST^6 ,T^2ST^5,(ST^4)^2\}. \\
\end{eqnarray*}
All the fourteen $K_{4}$ subgroups are conjugate as well.
\item{$Z_{4}$ subgroups}
\begin{equation*}
\begin{array}{lllllll}
Z^{ST^3}_{4}=\{1,ST^3,(ST^3)^2,T^4S\}, ~&~
Z^{T^3S}_{4}=\{1,T^3S,(ST^4)^2,ST^4\},  \\
Z^{T^4ST^5ST^4}_{4}=\{1,T^4ST^5ST^4,S,T^2ST^5ST^2\},~&~
 Z^{ST^2ST^2S}_{4}=\{1,ST^2ST^2S,T^3ST^5ST^3,ST^5ST^5S\}, \\
Z^{T^2ST^2}_{4}=\{1,T^2ST^2,T^2ST^4ST^2,T^5ST^5\},~&~
 Z^{TST^5ST^3}_{4}=\{1,TST^5ST^3,T^6ST,T^3ST^5ST^5\},\\
Z^{T^3ST^5ST}_{4}=\{1,T^3ST^5ST,TST^6,T^5ST^5ST^3\}, ~&~
Z^{T^2ST^4ST}_{4}=\{1,T^2ST^4ST,ST^5ST^6,T^3ST^4ST^2\},   \\
Z^{TST^4ST^2}_{4}=\{1,TST^4ST^2,ST^2ST,T^2ST^4ST^3\}, ~&~
Z^{TST^3}_{4}=\{1,TST^3,ST^3ST^4S,T^4ST^6\},\\
Z^{ST^3ST^2}_{4}=\{1,ST^3ST^2,ST^5ST^2S,T^5ST^4S\}, ~&~
Z^{T^3ST}_{4}=\{1,T^3ST,ST^4ST^3S,T^6ST^4\}, \\
Z^{T^4ST^3S}_{4}=\{1,ST^4ST^3,T^5ST^5ST,T^4ST^3S\},~&~
Z^{ST^5ST^4}_{4}=\{1,ST^5ST^4,T^5ST^2,T^3ST^2S\},\\
Z^{T^2ST}_{4}=\{1,T^2ST,T^2ST^3ST,T^6ST^5\}, ~&~
Z^{TST^2}_{4}=\{1,TST^2,TST^3ST^2,T^5ST^6\},\\
Z^{ST^2ST^3}_{4}=\{1,ST^2ST^3,T^2ST^5,T^4ST^5S\}, ~&~
Z^{T^3ST^4S}_{4}=\{1,T^3ST^4S,TST^5ST^5,ST^3ST^4\}, \\
Z^{ST^4ST^5}_{4}=\{1,ST^4ST^5,ST^2ST^5S,T^2ST^3S\},~&~
 Z^{TST^5S}_{4}=\{1,ST^2ST^6,T^4ST^3,TST^5S\}, \\
Z^{ST^5ST}_{4}=\{1,ST^5ST,T^3ST^4,T^6ST^2S\}.
\end{array}
\end{equation*}
All the twenty-one $Z_{4}$ subgroups are related to each other under group conjugation.
\item{$Z_{7}$ subgroups}
\begin{eqnarray*}
&&Z^{T}_{7}=\{1,T,T^2,T^3,T^4,T^5,T^6\},\\
&&Z^{T^4ST}_{7}=\{1,T^4ST,ST^3ST^2S,T^2ST^4ST^6,
TST^3ST^5,ST^5ST^4S,T^6ST^3\}, \\
&&Z^{TST^4}_{7}=\{1,TST^4,ST^2ST^3S,T^2ST^3ST^5,
T^2ST^4ST^5,ST^4ST^5S,T^3ST^6\},\\
&&Z^{T^2ST^3}_{7}=\{1,T^2ST^3,T^6ST^3S,T^3ST^5ST^4,
T^2ST^5ST^3,T^4ST^5,ST^4ST\},\\
&&Z^{T^3ST^2}_{7}=\{1,T^3ST^2,TST^4S,T^5ST^4,T^3ST^5ST^2,
T^4ST^5ST^3,ST^3ST^6\}, \\
&&Z^{T^2S}_{7}=\{1,T^2S,ST^5,(ST^5)^2,TST^3ST^6,
TST^4ST^6,(T^2S)^2\}, \\
&&Z^{ST^2}_{7}=\{1,ST^2,(ST^2)^2,T^2ST^3ST^4,
T^3ST^4ST^5,(T^5S)^2,T^5S\}, \\
&&Z^{STS}_{7}=\{1,STS,ST^2S,ST^3S,ST^4S,ST^5S,ST^6S\}.
\end{eqnarray*}
All the eight $Z_{7}$ subgroups are related to each other as well under group conjugation.

\end{itemize}

The $\Gamma_{7}$ group has six irreducible representations: one singlet representation $\bf{1}$, two three-dimensional representations $\bf{3}$ and $\bf{\bar{3}}$, one six-dimensional representation $\mathbf{6}$, one seven-dimensional representation $\mathbf{7}$ and one eight-dimensional representation $\mathbf{8}$. The explicit forms of the generators $S$ and $T$ in the five irreducible representations are chosen as follows
\footnote{In the basis of~\cite{King:2009mk} the triplet representation was given in terms of the standard
generators $A,B$ in~\cite{Luhn:2007yr} and may be related to four
generators $S,T,U,V$, with
$\Gamma_7 \supset S_4 \supset A_4$ corresponding to the respective generators being
$S,T,U,V \supset S,T,U \supset S,T$.
However for our purposes here we shall use the representation theory for $\Gamma_7$ developed for
finite modular groups in \cite{deAdelhartToorop:2011re}. Thus the notation we use for the generators $S,T$
corresponds to \cite{deAdelhartToorop:2011re} rather than \cite{King:2009mk}.}
\begin{equation*}
  \begin{array}{ll}
 \bf{3} &   ~S=\frac{2}{\sqrt{7}}
\begin{pmatrix}
 -s_2 &~ -s_1 ~& s_3 \\
 -s_1 &~ s_3 ~& -s_2 \\
 s_3 &~ -s_2 ~& -s_1 \\
\end{pmatrix}\,,~ \qquad
{\bf\bar{3}:}  ~S=\frac{2}{\sqrt{7}}
\begin{pmatrix}
 -s_2 &~ -s_1 ~& s_3 \\
 -s_1 &~ s_3 ~& -s_2 \\
 s_3 &~ -s_2 ~& -s_1 \\
\end{pmatrix}\,,
~~\\[-14pt] \\[4pt]
\bf{6:} & ~S=\frac{2\sqrt{2}}{7}
\begin{pmatrix}
 \frac{1-c_{2}}{\sqrt{2}} & \frac{1-c_{1}}{\sqrt{2}} & c_{2}-c_{1} & \frac{1-c_{3}}{\sqrt{2}} & c_{3}-c_{2} & c_{1}-c_{3} \\
 \frac{1-c_{1}}{\sqrt{2}} & \frac{1-c_{3}}{\sqrt{2}} & c_{1}-c_{3} & \frac{1-c_{2}}{\sqrt{2}} & c_{2}-c_{1} & c_{3}-c_{2} \\
 c_{2}-c_{1} & c_{1}-c_{3} & \frac{1-c_{1}}{\sqrt{2}} & c_{3}-c_{2} & \frac{1-c_{2}}{\sqrt{2}} & \frac{1-c_{3}}{\sqrt{2}} \\
 \frac{1-c_{3}}{\sqrt{2}} & \frac{1-c_{2}}{\sqrt{2}} & c_{3}-c_{2} & \frac{1-c_{1}}{\sqrt{2}} & c_{1}-c_{3} & c_{2}-c_{1} \\
 c_{3}-c_{2} & c_{2}-c_{1} & \frac{1-c_{2}}{\sqrt{2}} & c_{1}-c_{3} & \frac{1-c_{3}}{\sqrt{2}} & \frac{1-c_{1}}{\sqrt{2}} \\
 c_{1}-c_{3} & c_{3}-c_{2} & \frac{1-c_{3}}{\sqrt{2}} & c_{2}-c_{1} & \frac{1-c_{1}}{\sqrt{2}} & \frac{1-c_{2}}{\sqrt{2}} \\
\end{pmatrix}\,, ~ \\[-14pt] \\[4pt]
{\bf7:} & ~S=\frac{2}{7}
\begin{pmatrix}
 -\frac{1}{2} & \sqrt{2} & \sqrt{2} & \sqrt{2} & \sqrt{2} & \sqrt{2} & \sqrt{2} \\
 \sqrt{2} & \frac{s2+4 s3}{\sqrt{7}} & \frac{s1-4 s_2}{\sqrt{7}} & \frac{2 s_1-2 s_2-4 s_3}{\sqrt{7}} & \frac{-4 s_1-s_3}{\sqrt{7}} & \frac{4 s_1+2 s_2+2 s_3}{\sqrt{7}} & \frac{-2 s_1+4 s_2-2 s_3}{\sqrt{7}} \\
 \sqrt{2} & \frac{s_1-4 s_2}{\sqrt{7}} & \frac{-4 s_1-s_3}{\sqrt{7}} & \frac{-2 s_1+4 s_2-2 s_3}{\sqrt{7}} & \frac{s_2+4 s_3}{\sqrt{7}} & \frac{2 s_1-2 s_2-4 s_3}{\sqrt{7}} & \frac{4 s_1+2 s_2+2 s_3}{\sqrt{7}} \\
 \sqrt{2} & \frac{2 s_1-2 s_2-4 s_3}{\sqrt{7}} & \frac{-2 s_1+4 s_2-2 s_3}{\sqrt{7}} & \frac{s_1-4 s_2}{\sqrt{7}} & \frac{4 s_1+2 s_2+2 s_3}{\sqrt{7}} & \frac{s_2+4 s_3}{\sqrt{7}} & \frac{-4 s_1-s_3}{\sqrt{7}} \\
 \sqrt{2} & \frac{-4 s_1-s_3}{\sqrt{7}} & \frac{s_2+4 s_3}{\sqrt{7}} & \frac{4 s_1+2 s_2+2 s_3}{\sqrt{7}} & \frac{s_1-4 s_2}{\sqrt{7}} & \frac{-2 s_1+4 s_2-2 s_3}{\sqrt{7}} & \frac{2 s_1-2 s_2-4 s_3}{\sqrt{7}} \\
 \sqrt{2} & \frac{4 s_1+2 s_2+2 s_3}{\sqrt{7}} & \frac{2 s_1-2 s_2-4 s_3}{\sqrt{7}} & \frac{s_2+4 s_3}{\sqrt{7}} & \frac{-2 s_1+4 s_2-2 s_3}{\sqrt{7}} & \frac{-4 s_1-s_3}{\sqrt{7}} & \frac{s_1-4 s_2}{\sqrt{7}} \\
 \sqrt{2} & \frac{-2 s_1+4 s_2-2 s_3}{\sqrt{7}} & \frac{4 s_1+2 s_2+2 s_3}{\sqrt{7}} & \frac{-4 s_1-s_3}{\sqrt{7}} & \frac{2 s_1-2 s_2-4 s_3}{\sqrt{7}} & \frac{s_1-4 s_2}{\sqrt{7}} & \frac{s_2+4 s_3}{\sqrt{7}} \\
\end{pmatrix}\,,\\[-14pt] \\[4pt]
{\bf8:} & ~S=\frac{2\sqrt{6}}{7}
\begin{pmatrix}
 \frac{2 c_{2}-c_{1}-c_{3}}{2 \sqrt{6}} & \frac{c_{1}-c_{3}}{2 \sqrt{2}} & \frac{c_{1}+c_{2}-2 c_{3}}{2 \sqrt{3}} & \frac{c_{1}-2 c_{2}+c_{3}}{2 \sqrt{3}} & \frac{c_{2}+c_{3}-2 c_{1}}{2 \sqrt{3}} & \frac{2 c_{1}-c_{2}-c_{3}}{2 \sqrt{3}} & \frac{2 c_{2}-c_{1}-c_{3}}{2 \sqrt{3}} & \frac{2 c_{3}-c_{1}-c_{2}}{2 \sqrt{3}} \\
 \frac{c_{1}-c_{3}}{2 \sqrt{2}} & \frac{c_{1}-2 c_{2}+c_{3}}{2 \sqrt{6}} & \frac{c_{1}-c_{2}}{2} & \frac{c_{3}-c_{1}}{2} & \frac{c_{2}-c_{3}}{2} & \frac{c_{3}-c_{2}}{2} & \frac{c_{1}-c_{3}}{2} & \frac{c_{2}-c_{1}}{2} \\
 \frac{c_{1}+c_{2}-2 c_{3}}{2 \sqrt{3}} & \frac{c_{1}-c_{2}}{2} & \frac{c_{2}-c_{1}}{\sqrt{6}} & \frac{c_{1}-c_{3}}{\sqrt{6}} & \frac{1-c_{3}}{\sqrt{6}} & \frac{c_{2}-c_{3}}{\sqrt{6}} & \frac{c_{1}-1}{\sqrt{6}} & \frac{c_{2}-1}{\sqrt{6}} \\
 \frac{c_{1}-2 c_{2}+c_{3}}{2 \sqrt{3}} & \frac{c_{3}-c_{1}}{2} & \frac{c_{1}-c_{3}}{\sqrt{6}} & \frac{c_{3}-c_{2}}{\sqrt{6}} & \frac{1-c_{2}}{\sqrt{6}} & \frac{c_{1}-c_{2}}{\sqrt{6}} & \frac{c_{3}-1}{\sqrt{6}} & \frac{c_{1}-1}{\sqrt{6}} \\
 \frac{c_{2}+c_{3}-2 c_{1}}{2 \sqrt{3}} & \frac{c_{2}-c_{3}}{2} & \frac{1-c_{3}}{\sqrt{6}} & \frac{1-c_{2}}{\sqrt{6}} & \frac{c_{1}-c_{3}}{\sqrt{6}} & \frac{c_{1}-1}{\sqrt{6}} & \frac{c_{1}-c_{2}}{\sqrt{6}} & \frac{c_{2}-c_{3}}{\sqrt{6}} \\
 \frac{2 c_{1}-c_{2}-c_{3}}{2 \sqrt{3}} & \frac{c_{3}-c_{2}}{2} & \frac{c_{2}-c_{3}}{\sqrt{6}} & \frac{c_{1}-c_{2}}{\sqrt{6}} & \frac{c_{1}-1}{\sqrt{6}} & \frac{c_{1}-c_{3}}{\sqrt{6}} & \frac{1-c_{2}}{\sqrt{6}} & \frac{1-c_{3}}{\sqrt{6}} \\
 \frac{2 c_{2}-c_{1}-c_{3}}{2 \sqrt{3}} & \frac{c_{1}-c_{3}}{2} & \frac{c_{1}-1}{\sqrt{6}} & \frac{c_{3}-1}{\sqrt{6}} & \frac{c_{1}-c_{2}}{\sqrt{6}} & \frac{1-c_{2}}{\sqrt{6}} & \frac{c_{3}-c_{2}}{\sqrt{6}} & \frac{c_{1}-c_{3}}{\sqrt{6}} \\
 \frac{2 c_{3}-c_{1}-c_{2}}{2 \sqrt{3}} & \frac{c_{2}-c_{1}}{2} & \frac{c_{2}-1}{\sqrt{6}} & \frac{c_{1}-1}{\sqrt{6}} & \frac{c_{2}-c_{3}}{\sqrt{6}} & \frac{1-c_{3}}{\sqrt{6}} & \frac{c_{1}-c_{3}}{\sqrt{6}} & \frac{c_{2}-c_{1}}{\sqrt{6}} \\
\end{pmatrix}\,, \\
\end{array}
\end{equation*}
\begin{eqnarray*}
&& {\bf3:}~T=\text{diag}(\rho,\rho^2,\rho^4)\,, \qquad {\bf\bar{3}:}~T=\text{diag}(\rho^6,\rho^5,\rho^3)\,, \qquad {\bf6:}~T=\text{diag}(\rho^1,\rho^2,\rho^3,\rho^4,\rho^5,\rho^6)\,, \\
&& {\bf7:}~T=\text{diag}(1,\rho,\rho^2,\rho^3,\rho^4,\rho^5,\rho^6)\,, \qquad
{\bf8:}~T=\text{diag}(1,1,\rho,\rho^2,\rho^3,\rho^4,\rho^5,\rho^6)\,,
\end{eqnarray*}
where the parameter $\rho$ is the seventh unit root $\rho=e^{2\pi i/7}$, $s_{n}=\sin{\frac{2n\pi}{7}}$ and $c_{n}=\cos{\frac{2n\pi}{7}}$ with $n=1,2,3$. We can straightforwardly obtain the Kronecker products between various representations:
\begin{eqnarray}
\nonumber&&\bf{1}\otimes \bf{r}=\bf{r}\otimes\bf{1}=\bf{r},~~~\bf{3}\otimes\bf{3}=\bf{\bar{3}_{A}}\oplus\bf{6_{S}},~~~\bf{3}\otimes\bf{\bar{3}}=\bf{1}\oplus\bf{8}, ~~~\mathbf{3}\times\bf{6}=\mathbf{\bar{3}}\oplus\mathbf{7}\oplus\mathbf{8},\\
\nonumber&&\bf{3}\otimes\bf{7}=\bf{6}\oplus\bf{7}\oplus\bf{8},~~~\mathbf{3}\otimes\mathbf{8}=\mathbf{3}\oplus\mathbf{6}\oplus\mathbf{7}\oplus\mathbf{8},~~~\mathbf{\bar{3}}\otimes\mathbf{\bar{3}}
=\mathbf{3_{A}}\oplus\mathbf{6_{S}},\\
\nonumber&&\mathbf{\bar{3}}\otimes\mathbf{6}=\mathbf{3}\oplus\mathbf{7}\oplus\mathbf{8},~~~\mathbf{\bar{3}}\otimes\mathbf{7}=\mathbf{6}\oplus\mathbf{7}\oplus\mathbf{8},
~~~\mathbf{\bar{3}}\otimes\mathbf{8}=\mathbf{\bar{3}}\oplus\mathbf{6}\oplus\mathbf{7}\oplus\mathbf{8},\\
\nonumber &&\mathbf{6}\otimes\mathbf{6}=\mathbf{1_S}\oplus\mathbf{6_{S,1}}\oplus\mathbf{6_{S,2}}\oplus\mathbf{7_{A}}\oplus\mathbf{8_{S}}\oplus\mathbf{8_{A}}, ~~~\mathbf{6}\otimes\mathbf{7}=\mathbf{3}\oplus\mathbf{\bar{3}}\oplus\mathbf{6}\oplus\mathbf{7_{1}}\oplus\mathbf{7_{2}}\oplus\mathbf{8_{1}}\oplus\mathbf{8_{2}}\,, \\
\nonumber && \mathbf{6}\otimes\mathbf{8}=\mathbf{3}\oplus\mathbf{\bar{3}}\oplus\mathbf{6_{1}}\oplus\mathbf{6_{2}}\oplus\mathbf{7_{1}}\oplus\mathbf{7_{2}}\oplus\mathbf{8_{1}}\oplus\mathbf{8_{2}}\,, \\
\nonumber && \mathbf{7}\otimes\mathbf{7}=\mathbf{1_S}\oplus\mathbf{3_{A}}\oplus\mathbf{\bar{3}_{A}}\oplus\mathbf{6_{S,1}}\oplus\mathbf{6_{S,2}}\oplus\mathbf{7_{S}}\oplus\mathbf{7_{A}}\oplus\mathbf{8_{S}}\oplus\mathbf{8_{A}}\,, \\
\nonumber && \mathbf{7}\otimes\mathbf{8}=\mathbf{3}\oplus\mathbf{\bar{3}}\oplus\mathbf{6_{1}}\oplus\mathbf{6_{2}}\oplus\mathbf{7_{1}}\oplus\mathbf{7_{2}}\oplus\mathbf{8_{1}}\oplus\mathbf{8_{2}}\oplus\mathbf{8_{3}}\,, \\
&& \mathbf{8}\otimes\mathbf{8}=\mathbf{1_S}\oplus\mathbf{3_{A}}\oplus\mathbf{\bar{3}_{A}}\oplus\mathbf{6_{S,1}}\oplus\mathbf{6_{S,2}}\oplus\mathbf{7_{S}}\oplus\mathbf{7_{A,1}}\oplus\mathbf{7_{A,2}}\oplus\mathbf{8_{S,1}}\oplus\mathbf{8_{S,2}}\oplus\mathbf{8_{A}}\,.
\end{eqnarray}
where $\bf{r}$ denotes any irreducible representation of $\Gamma_7$, and $\bf{6_{1}}$, $\bf{6_{2}}$, $\bf{7_{1}}$, $\bf{7_{2}}$, $\bf{8_{1}}$, $\bf{8_{2}}$ and $\bf{8_{3}}$ stand for the two $\bf{6}$,$\bf{7}$ and three $\bf{8}$ representations which appear in the Kronecker products. The subscript "$\bf{S}$" ("$\bf{A}$") refers to symmetric (antisymmetric) combinations. We now list the Clebsch-Gordan coefficients in our basis. We use the notation $\alpha_{i}$ ($\beta_{i}$) to denote the elements of the first (second) representation.

\begin{eqnarray*}
\begin{array}{|c|c|c|}\hline\hline
~\bf{3}\otimes\bf{3}=\bf{\bar{3}_{A}} \oplus\bf{6_{S}}~ & ~\bf{\bar{3}}\otimes\bf{\bar{3}}=\bf{3_{A}} \oplus\bf{6_{S}}~ &  ~\bf{3}\otimes\bf{\bar{3}}=\bf{1}\oplus\bf{8} \\ \hline
 & & \\[-0.08in]
& &
~\bf{1}:
~\alpha _1 \beta _1+\alpha _2 \beta _2+\alpha _3 \beta _3  \\[-0.28in]
~\bf{\bar{3}}_A:
\begin{pmatrix}
 \alpha _2 \beta _3-\alpha _3 \beta _2 \\
 \alpha _3 \beta _1-\alpha _1 \beta _3 \\
 \alpha _1 \beta _2-\alpha _2 \beta _1
\end{pmatrix}
~ &
~\bf{3_{A}}:
\begin{pmatrix}
 \alpha _2 \beta _3-\alpha _3 \beta _2 \\
 \alpha _3 \beta _1-\alpha _1 \beta _3 \\
 \alpha _1 \beta _2-\alpha _2 \beta _1
\end{pmatrix}
~ &
~ \\
& & \\[-0.4in]
~\bf{6_S}:
\begin{pmatrix}
 \sqrt{2} \alpha _3 \beta _3 \\
\sqrt{2} \alpha _1 \beta _1 \\
 \alpha _1 \beta _2+\alpha _2 \beta _1 \\
 \sqrt{2} \alpha _2 \beta _2 \\
 \alpha _1 \beta _3+\alpha _3 \beta _1 \\
 \alpha _2 \beta _3+\alpha _3 \beta _2 \\
\end{pmatrix}
~  &
~\bf{6_{S}}:
\begin{pmatrix}
 \alpha _2 \beta _3+\alpha _3 \beta _2 \\
 \alpha _1 \beta _3+\alpha _3 \beta _1 \\
  \sqrt{2} \alpha _2 \beta _2 \\
 \alpha _1 \beta _2+\alpha _2 \beta _1 \\
 \sqrt{2} \alpha _1 \beta _1 \\
 \sqrt{2} \alpha _3 \beta _3
\end{pmatrix}
~ &
~\bf{8}:
\begin{pmatrix}
 \sqrt{3} \left(\alpha _1 \beta _1-\alpha _2 \beta _2\right) \\
 \alpha _1 \beta _1+\alpha _2 \beta _2-2 \alpha _3 \beta _3 \\
 \sqrt{6} \alpha _2 \beta _1 \\
 \sqrt{6} \alpha _3 \beta _2 \\
 \sqrt{6} \alpha _3 \beta _1 \\
 -\sqrt{6} \alpha _1 \beta _3 \\
 -\sqrt{6} \alpha _2 \beta _3 \\
 -\sqrt{6} \alpha _1 \beta _2
\end{pmatrix}\\
& & \\[-0.14in] \hline \hline
\end{array}
\end{eqnarray*}

\begin{eqnarray*}
\begin{array}{|c|c|} \hline\hline
~\bf{3}\otimes\bf{6}=\bf{\bar{3}}\oplus\bf{7}\oplus\bf{8}~~ & ~\bf{\bar{3}}\otimes\bf{6}=\bf{3}\oplus\bf{7}\oplus\bf{8}~ \\ \hline
  & \\[-0.16in]
~\bf{\bar{3}}:
\begin{pmatrix}
 \alpha _2 \beta _4+\alpha _3 \beta _2+\sqrt{2} \alpha _1 \beta _5 \\
 \alpha _1 \beta _4+\alpha _3 \beta _1+\sqrt{2} \alpha _2 \beta _3 \\
 \alpha _1 \beta _2+\alpha _2 \beta _1+\sqrt{2} \alpha _3 \beta _6
\end{pmatrix}
~~ &  ~\bf{3}:
\begin{pmatrix}
  \sqrt{2} \alpha _1 \beta _2+\alpha _2 \beta _3+\alpha _3 \beta _5 \\
 \sqrt{2} \alpha _2 \beta _4+\alpha _1 \beta _3+\alpha _3 \beta _6 \\
 \sqrt{2} \alpha _3 \beta _1+\alpha _1 \beta _5+\alpha _2 \beta _6
\end{pmatrix}\\
  & \\[-0.16in]
~\bf{7}:
\begin{pmatrix}
2 \left(\alpha _1 \beta _6+\alpha _2 \beta _5+\alpha _3 \beta _3\right) \\
 -2 \left(\sqrt{2} \alpha _2 \beta _6+\alpha _3 \beta _4\right) \\
 -2 \left(\alpha _1 \beta _1+\sqrt{2} \alpha _3 \beta _5\right) \\
 3 \alpha _1 \beta _2-\alpha _2 \beta _1-\sqrt{2} \alpha _3 \beta _6 \\
 -2 \left(\sqrt{2} \alpha _1 \beta _3+\alpha _2 \beta _1\right) \\
 -\alpha _1 \beta _4-\sqrt{2} \alpha _2 \beta _3+3 \alpha _3 \beta _1 \\
 -\sqrt{2} \alpha _1 \beta _5+3 \alpha _2 \beta _4-\alpha _3 \beta _2
\end{pmatrix}
~~ &
~\bf{7}:
\begin{pmatrix}
 2 \left(\alpha_1 \beta _1+\alpha _2 \beta _2+\alpha _3 \beta _4\right) \\
 -\sqrt{2} \alpha _1 \beta _2+3 \alpha _2 \beta _3-\alpha _3 \beta _5 \\
 -\sqrt{2} \alpha _2 \beta _4-\alpha _1 \beta _3+3 \alpha _3 \beta _6 \\
 -2 \left(\sqrt{2} \alpha _1 \beta _4+\alpha _2 \beta _5\right) \\
 -\sqrt{2} \alpha _3 \beta _1+3 \alpha _1 \beta _5-\alpha _2 \beta _6 \\
 -2 \left(\sqrt{2} \alpha _3 \beta _2+\alpha _1 \beta _6\right) \\
 -2 \left(\sqrt{2} \alpha _2 \beta _1+\alpha _3 \beta _3\right)
\end{pmatrix}
~  \\
 &  \\
\hline\hline
\end{array}
\end{eqnarray*}

\begin{eqnarray*}
\begin{array}{|c|c|} \hline\hline
~\bf{3}\otimes\bf{7}=\bf{6}\oplus\bf{7}\oplus\bf{8}~~ & ~\bf{\bar{3}}\otimes\bf{7}=\bf{6}\oplus\bf{7}\oplus\bf{8}~ \\ \hline
  & \\[-0.16in]
~\bf{6}:
\begin{pmatrix}
2 \alpha _1 \beta _1-\sqrt{2} \left(\alpha _3 \beta _5+2 \alpha _2 \beta _7\right)\\
2 \alpha _2 \beta _1-\sqrt{2} \left(\alpha _1 \beta _2+2 \alpha _3 \beta _6\right)\\
3 \alpha _2 \beta _2-\alpha _1 \beta _3-2 \alpha _3 \beta _7\\
2 \alpha _3 \beta _1-\sqrt{2} \left(\alpha _2 \beta _3+2 \alpha _1 \beta _4\right)\\
-\alpha _3 \beta _2-2 \alpha _2 \beta _4+3 \alpha _1 \beta _5\\
3 \alpha _3 \beta _3-\alpha _2 \beta _5-2 \alpha _1 \beta _6
\end{pmatrix}
~~ &  ~\bf{6}:
\begin{pmatrix}
2 \alpha _1 \beta _3+\alpha _2 \beta _4-3 \alpha _3 \beta _6\\
-3 \alpha _1 \beta _4+2 \alpha _2 \beta _5+\alpha _3 \beta _7\\
\sqrt{2} \left(2 \alpha _1 \beta _5+\alpha _2 \beta _6\right)-2 \alpha _3 \beta _1\\
2 \alpha _3 \beta _2+\alpha _1 \beta _6-3 \alpha _2 \beta _7\\
\sqrt{2} \left(2 \alpha _3 \beta _3+\alpha _1 \beta _7\right)-2 \alpha _2 \beta _1\\
\sqrt{2} \left(2 \alpha _2 \beta _2+\alpha _3 \beta _4\right)-2 \alpha _1 \beta _1
\end{pmatrix}\\
  & \\
\hline\hline
\end{array}
\end{eqnarray*}

\begin{eqnarray*}
\begin{array}{|c|c|} \hline\hline
~\bf{3}\otimes\bf{8}=\bf{3}\oplus\bf{6}\oplus\bf{7}\oplus\bf{8}~~ & ~\bf{\bar{3}}\otimes\bf{8}=\bf{\bar{3}}\oplus\bf{6}\oplus\bf{7}\oplus\bf{8}~ \\ \hline
  & \\[-0.16in]
  ~\bf{3}:
\begin{pmatrix}
\sqrt{3} \alpha _1 \beta _1+\alpha _1 \beta _2-\sqrt{6} (\alpha _3 \beta _6+ \alpha _2 \beta _8)\\
-\sqrt{3} \alpha _2 \beta _1+\alpha _2 \beta _2+\sqrt{6} (\alpha _1 \beta _3- \alpha _3 \beta _7)\\
-2 \alpha _3 \beta _2+\sqrt{6} (\alpha _2 \beta _4+ \alpha _1 \beta _5)
\end{pmatrix}
~~ &  ~\bf{\bar{3}}:
\begin{pmatrix}
\sqrt{3} \alpha _1 \beta _1+\alpha _1 \beta _2+\sqrt{6} (\alpha _2 \beta _3+ \alpha _3 \beta _5)\\
-\sqrt{3} \alpha _2 \beta _1+\alpha _2 \beta _2+\sqrt{6} (\alpha _3 \beta _4- \alpha _1 \beta _8)\\
-2 \alpha _3 \beta _2-\sqrt{6}( \alpha _1 \beta _6+ \alpha _2 \beta _7)
\end{pmatrix}\\
  & \\[-0.16in]
~\bf{6}:
\begin{pmatrix}
 \alpha _1\beta _1-\sqrt{3} \alpha _1\beta _2+\sqrt{2} \left(\alpha _3 \beta _6-\alpha _2 \beta _8\right)\\
\alpha _2\beta _1+\sqrt{3} \alpha _2\beta _2-\sqrt{2} \left(\alpha _1 \beta _3+\alpha _3 \beta _7\right)\\
2 \left(\alpha _1 \beta _4+\alpha _3 \beta _8\right)\\
\sqrt{2} \left(\alpha _1 \beta _5-\alpha _2 \beta _4\right)-2 \alpha _3 \beta _1\\
2 \left(\alpha _3 \beta _3-\alpha _2 \beta _5\right)\\
2 \left(\alpha _1 \beta _7-\alpha _2 \beta _6\right)
\end{pmatrix}
~~ &  ~\bf{6}:
\begin{pmatrix}
2 \alpha _1 \beta _4-2 \alpha _2 \beta _5\\
2 \alpha _3 \beta _8-2 \alpha _2 \beta _6\\
2 \alpha _3 \beta _1+\sqrt{2} \left(\alpha _1 \beta _6-\alpha _2 \beta _7\right)\\
2 \left(\alpha _3 \beta _3+\alpha _1 \beta _7\right)\\
-\alpha _2 \beta _1+\sqrt{3} \alpha _2\beta _2-\sqrt{2} \left(\alpha _3 \beta _4+\alpha _1 \beta _8\right)\\
 \sqrt{3} \alpha _1\beta _2-\alpha _1\beta _1+\sqrt{2} \left(\alpha _3 \beta _5-\alpha _2 \beta _3\right)
\end{pmatrix}\\
  & \\
\hline\hline
\end{array}
\end{eqnarray*}

\begin{eqnarray*}
\begin{array}{|c|} \hline\hline
~\bf{6}\otimes\bf{6}=\bf{1_{S}}\oplus\bf{6_{S,1}}\oplus\bf{6_{S,2}}\oplus\bf{7_{A}}\oplus\bf{8_{S}}\oplus\bf{8}_{A}~ \\ \hline \\[-0.16in]
~\mathbf{1_{S}}: \alpha _6 \beta _1+\alpha _5 \beta _2+\alpha _4 \beta _3+\alpha _3 \beta _4+\alpha _2 \beta _5+\alpha _1 \beta _6 ~ \\
~\bf{6_{S,1}}:
\begin{pmatrix}
\sqrt{2} \left(\alpha _5 \beta _3-\alpha _4 \beta _4+\alpha _3 \beta _5\right)\\
\sqrt{2} \left(-\alpha _1 \beta _1+\alpha _6 \beta _3+\alpha _3 \beta _6\right)\\
\alpha _2 \beta _1+\alpha _1 \beta _2-\sqrt{2} \left(\alpha _6 \beta _4+\alpha _4 \beta _6\right)\\
\sqrt{2} \left(-\alpha _2 \beta _2+\alpha _6 \beta _5+\alpha _5 \beta _6\right)\\
\alpha _4 \beta _1-\sqrt{2} \alpha _3 \beta _2-\sqrt{2} \alpha _2 \beta _3+\alpha _1 \beta _4\\
-\sqrt{2} \alpha _5 \beta _1+\alpha _4 \beta _2+\alpha _2 \beta _4-\sqrt{2} \alpha _1 \beta _5
\end{pmatrix}
~ \\
~\bf{6_{S,2}}:
\begin{pmatrix}
\sqrt{2} \alpha _6 \beta _2-\alpha _5 \beta _3-\alpha _3 \beta _5+\sqrt{2} \alpha _2 \beta _6\\
-\alpha _6 \beta _3+\sqrt{2} \alpha _5 \beta _4+\sqrt{2} \alpha _4 \beta _5-\alpha _3 \beta _6\\
-\sqrt{2} \left(\alpha _2 \beta _1+\alpha _1 \beta _2-\alpha _5 \beta _5\right)\\
\sqrt{2} \alpha _3 \beta _1+\sqrt{2} \alpha _1 \beta _3-\alpha _6 \beta _5-\alpha _5 \beta _6\\
-\sqrt{2} \left(\alpha _4 \beta _1+\alpha _1 \beta _4-\alpha _6 \beta _6\right)\\
-\sqrt{2} \left(\alpha _4 \beta _2-\alpha _3 \beta _3+\alpha _2 \beta _4\right)
\end{pmatrix}
~ \\
\hline\hline
\end{array}
\end{eqnarray*}

\begin{eqnarray*}
\begin{array}{|c|} \hline\hline
~\bf{6}\otimes\bf{7}=\bf{3}\oplus\bf{\bar{3}}\oplus\bf{6}\oplus\bf{7_{1}}\oplus\bf{7_{2}}\oplus\bf{8}_{1}\oplus\bf{8}_{2}~ \\ \hline \\[-0.16in]
~\bf{3}:
\begin{pmatrix}
2 \alpha _1 \beta _1-2 \alpha _6 \beta _3+3 \alpha _5 \beta _4-2 \sqrt{2} \alpha _4 \beta _5-\alpha _3 \beta _6-\sqrt{2} \alpha _2 \beta _7\\
2 \alpha _2 \beta _1-2 \sqrt{2} \alpha _1 \beta _2-\alpha _6 \beta _4-2 \alpha _5 \beta _5-\sqrt{2} \alpha _4 \beta _6+3 \alpha _3 \beta _7\\
2 \alpha _4 \beta _1-2 \alpha _3 \beta _2-2 \sqrt{2} \alpha _2 \beta _3-\sqrt{2} \alpha _1 \beta _4+3 \alpha _6 \beta _6-\alpha _5 \beta _7
\end{pmatrix}~ \\
~\bf{\bar{3}}:
\begin{pmatrix}
2 \alpha _6 \beta _1-\sqrt{2} \alpha _5 \beta _2-\alpha _4 \beta _3-2 \sqrt{2} \alpha _3 \beta _4+3 \alpha _2 \beta _5-2 \alpha _1 \beta _6\\
2 \alpha _5 \beta _1+3 \alpha _4 \beta _2-\sqrt{2} \alpha _3 \beta _3-2 \alpha _2 \beta _4-\alpha _1 \beta _5-2 \sqrt{2} \alpha _6 \beta _7\\
2 \alpha _3 \beta _1-\alpha _2 \beta _2+3 \alpha _1 \beta _3-\sqrt{2} \alpha _6 \beta _5-2 \sqrt{2} \alpha _5 \beta _6-2 \alpha _4 \beta _7
\end{pmatrix}
~ \\
~\bf{6}:
\begin{pmatrix}
\alpha _1 \beta _1+\alpha _5 \beta _4+\sqrt{2} \alpha _4 \beta _5-\alpha _3 \beta _6+\sqrt{2} \alpha _2 \beta _7\\
\alpha _2 \beta _1+\sqrt{2} \alpha _1 \beta _2-\alpha _6 \beta _4+\sqrt{2} \alpha _4 \beta _6+\alpha _3 \beta _7\\
-\alpha _3 \beta _1+\alpha _2 \beta _2-\alpha _1 \beta _3-\sqrt{2} \alpha _6 \beta _5-\sqrt{2} \alpha _5 \beta _6\\
\alpha _4 \beta _1+\sqrt{2} \alpha _2 \beta _3+\sqrt{2} \alpha _1 \beta _4+\alpha _6 \beta _6-\alpha _5 \beta _7\\
-\alpha _5 \beta _1-\alpha _4 \beta _2-\sqrt{2} \alpha _3 \beta _3+\alpha _1 \beta _5-\sqrt{2} \alpha _6 \beta _7\\
-\alpha _6 \beta _1-\sqrt{2} \alpha _5 \beta _2+\alpha _4 \beta _3-\sqrt{2} \alpha _3 \beta _4-\alpha _2 \beta _5
\end{pmatrix}
~ \\
\hline\hline
\end{array}
\end{eqnarray*}

\begin{eqnarray*}
\begin{array}{|c|} \hline\hline
~\bf{6}\otimes\bf{8}=\bf{3}\oplus\bf{\bar{3}}\oplus\bf{6_1}\oplus\bf{6_2}\oplus\bf{7_{1}}\oplus\bf{7_{2}}\oplus\bf{8}_{1}\oplus\bf{8}_{2}~ \\ \hline \\[-0.16in]
~\bf{3}:
\begin{pmatrix}
\alpha _1 \beta _1-\sqrt{3} \alpha _1 \beta _2-2 \alpha _6 \beta _4-\sqrt{2} \alpha _4 \beta _6-2 \alpha _3 \beta _7+\sqrt{2} \alpha _2 \beta _8\\
\alpha _2 \beta _1+\sqrt{3} \alpha _2 \beta _2+\sqrt{2} \alpha _1 \beta _3+2 \alpha _6 \beta _5+2 \alpha _5 \beta _6+\sqrt{2} \alpha _4 \beta _7\\
-2 \alpha _4 \beta _1-2 \alpha _3 \beta _3+\sqrt{2} \alpha _2 \beta _4-\sqrt{2} \alpha _1 \beta _5-2 \alpha _5 \beta _8
\end{pmatrix}~ \\
~\bf{\bar{3}}:
\begin{pmatrix}
\alpha _6 \beta _1-\sqrt{3} \alpha _6 \beta _2-\sqrt{2} \alpha _5 \beta _3+2 \alpha _4 \beta _4+\sqrt{2} \alpha _3 \beta _5+2 \alpha _1 \beta _7\\
\alpha _5 \beta _1+\sqrt{3} \alpha _5 \beta _2-\sqrt{2} \alpha _3 \beta _4-2 \alpha _2 \beta _5-2 \alpha _1 \beta _6-\sqrt{2} \alpha _6 \beta _8\\
-2 \alpha _3 \beta _1+2 \alpha _2 \beta _3+\sqrt{2} \alpha _6 \beta _6-\sqrt{2} \alpha _5 \beta _7+2 \alpha _4 \beta _8
\end{pmatrix}
~ \\
~\bf{6_1}:
\begin{pmatrix}
4 \alpha _1 \beta _2-2 \sqrt{3} \alpha _6 \beta _4-2 \sqrt{3} \alpha _5 \beta _5\\
-2 \sqrt{3} \alpha _2 \beta _1-2 \alpha _2 \beta _2+2 \sqrt{3} \alpha _5 \beta _6+2 \sqrt{3} \alpha _3 \beta _8\\
-2 \alpha _3 \beta _2-2 \sqrt{3} \alpha _2 \beta _3+\sqrt{6} \alpha _6 \beta _6+\sqrt{6} \alpha _5 \beta _7+2 \sqrt{3} \alpha _4 \beta _8\\
2 \sqrt{3} \alpha _4 \beta _1-2 \alpha _4 \beta _2-2 \sqrt{3} \alpha _3 \beta _3+2 \sqrt{3} \alpha _6 \beta _7\\
-\sqrt{3} \alpha _5 \beta _1+\alpha _5 \beta _2-\sqrt{6} \alpha _3 \beta _4-2 \sqrt{3} \alpha _2 \beta _5+2 \sqrt{3} \alpha _1 \beta _6+\sqrt{6} \alpha _6 \beta _8\\
\sqrt{3} \alpha _6 \beta _1+\alpha _6 \beta _2-\sqrt{6} \alpha _5 \beta _3-2 \sqrt{3} \alpha _4 \beta _4-\sqrt{6} \alpha _3 \beta _5+2 \sqrt{3} \alpha _1 \beta _7
\end{pmatrix}
~ \\
\hline\hline
\end{array}
\end{eqnarray*}

\begin{eqnarray*}
\begin{array}{|c|} \hline\hline
~\bf{7}\otimes\bf{7}=\bf{1_S}\oplus\bf{3_A}\oplus\bf{\bar{3}_A}\oplus\bf{6_{S,1}}\oplus\bf{6_{S,2}}\oplus\bf{7_{S}}\oplus\bf{7_{A}}\oplus\bf{8}_{S}\oplus\bf{8}_{A}~ \\ \hline \\[-0.16in]
~{\bf1_S}: \alpha _1 \beta _1+\alpha _7 \beta _2+\alpha _6 \beta _3+\alpha _5 \beta_4+\alpha _4 \beta _5+\alpha _3 \beta _6+\alpha _2 \beta _7~\\
~{\bf 3_A}:
\begin{pmatrix}
 -\sqrt{2} \alpha _2 \beta _1+\sqrt{2} \alpha _1 \beta _2+2 \alpha _7 \beta_3+\alpha _6 \beta _4-\alpha _4 \beta _6-2 \alpha _3 \beta _7 \\
 -\sqrt{2} \alpha _3 \beta _1+\sqrt{2} \alpha _1 \beta _3-\alpha _7 \beta _4+2\alpha _6 \beta _5-2 \alpha _5 \beta _6+\alpha _4 \beta _7 \\
 -\sqrt{2} \alpha _5 \beta _1+2 \alpha _4 \beta _2-2 \alpha _2 \beta
   _4+\sqrt{2} \alpha _1 \beta _5+\alpha _7 \beta _6-\alpha _6 \beta _7
\end{pmatrix}~ \\
~{\bf\bar{3}_A}:
\begin{pmatrix}
 -\sqrt{2} \alpha _7 \beta _1-2 \alpha _6 \beta _2-\alpha _5 \beta _3+\alpha
   _3 \beta _5+2 \alpha _2 \beta _6+\sqrt{2} \alpha _1 \beta _7 \\
 -\sqrt{2} \alpha _6 \beta _1+\alpha _5 \beta _2-2 \alpha _4 \beta _3+2 \alpha
   _3 \beta _4-\alpha _2 \beta _5+\sqrt{2} \alpha _1 \beta _6 \\
 -\sqrt{2} \alpha _4 \beta _1-\alpha _3 \beta _2+\alpha _2 \beta _3+\sqrt{2}
   \alpha _1 \beta _4-2 \alpha _7 \beta _5+2 \alpha _5 \beta _7
\end{pmatrix}
~ \\
\hline\hline
\end{array}
\end{eqnarray*}

\begin{eqnarray*}
\begin{array}{|c|} \hline\hline
~\bf{7}\otimes\bf{8}=\bf{3}\oplus\bf{\bar{3}}\oplus\bf{6_{1}}\oplus\bf{6_{2}}\oplus\bf{7_{1}}\oplus\bf{7_{2}}\oplus\bf{8}_{1}\oplus\bf{8}_{2}\oplus\bf{8}_{3}~ \\ \hline \\[-0.16in]
~{\bf 3}:
\begin{pmatrix}
-5 \alpha _2 \beta _1+\sqrt{3} \alpha _2 \beta _2+4 \alpha _1 \beta _3-2 \sqrt{2} \alpha _7 \beta _4-4 \sqrt{2} \alpha _6 \beta _5-3 \sqrt{2} \alpha _5 \beta _6+2 \sqrt{2} \alpha _4 \beta _7+\sqrt{2} \alpha _3 \beta _8\\
\alpha _3 \beta _1-3 \sqrt{3} \alpha _3 \beta _2+3 \sqrt{2} \alpha _2 \beta _3+4 \alpha _1 \beta _4-2 \sqrt{2} \alpha _7 \beta _5+2 \sqrt{2} \alpha _6 \beta _6+\sqrt{2} \alpha _5 \beta _7+4 \sqrt{2} \alpha _4 \beta _8\\
4 \alpha _5 \beta _1+2 \sqrt{3} \alpha _5 \beta _2-2 \sqrt{2} \alpha _4 \beta _3+3 \sqrt{2} \alpha _3 \beta _4-\sqrt{2} \alpha _2 \beta _5-4 \alpha _1 \beta _6+4 \sqrt{2} \alpha _7 \beta _7+2 \sqrt{2} \alpha _6 \beta _8
\end{pmatrix}~ \\
~{\bf\bar{3}}:
\begin{pmatrix}
5 \alpha _7 \beta _1-\sqrt{3} \alpha _7 \beta _2+\sqrt{2} \alpha _6 \beta _3+2 \sqrt{2} \alpha _5 \beta _4-3 \sqrt{2} \alpha _4 \beta _5-4 \sqrt{2} \alpha _3 \beta _6-2 \sqrt{2} \alpha _2 \beta _7+4 \alpha _1 \beta _8\\
-\alpha _6 \beta _1+3 \sqrt{3} \alpha _6 \beta _2+4 \sqrt{2} \alpha _5 \beta _3+\sqrt{2} \alpha _4 \beta _4+2 \sqrt{2} \alpha _3 \beta _5-2 \sqrt{2} \alpha _2 \beta _6+4 \alpha _1 \beta _7+3 \sqrt{2} \alpha _7 \beta _8\\
-4 \alpha _4 \beta _1-2 \sqrt{3} \alpha _4 \beta _2+2 \sqrt{2} \alpha _3 \beta _3+4 \sqrt{2} \alpha _2 \beta _4-4 \alpha _1 \beta _5-\sqrt{2} \alpha _7 \beta _6+3 \sqrt{2} \alpha _6 \beta _7-2 \sqrt{2} \alpha _5 \beta _8
\end{pmatrix}
~ \\
\hline\hline
\end{array}
\end{eqnarray*}

\begin{eqnarray*}
\begin{array}{|c|}\hline
~\bf{8}_{2}:
\begin{pmatrix}
2 \sqrt{3} \alpha _1 \beta _2-2 \alpha _7 \beta _3-2 \alpha _6 \beta _4-\alpha _5 \beta _5-4 \alpha _4 \beta _6-2 \alpha _3 \beta _7+\alpha _2 \beta _8\\
-2 \sqrt{3} \alpha _1 \beta _1+2 \sqrt{3} \alpha _7 \beta _3-2 \sqrt{3} \alpha _6 \beta _4+\sqrt{3} \alpha _5 \beta _5+\sqrt{3} \alpha _2 \beta _8\\
\alpha _2 \beta _1+\sqrt{3} \alpha _2 \beta _2-2 \alpha _1 \beta _3+2 \sqrt{2} \alpha _6 \beta _5-3 \sqrt{2} \alpha _5 \beta _6+2 \sqrt{2} \alpha _4 \beta _7\\
-2 \alpha _3 \beta _1+3 \sqrt{2} \alpha _2 \beta _3-2 \alpha _1 \beta _4-2 \sqrt{2} \alpha _7 \beta _5-2 \sqrt{2} \alpha _4 \beta _8\\
-4 \alpha _4 \beta _1-\sqrt{2} \alpha _3 \beta _3+\sqrt{2} \alpha _2 \beta _4+2 \alpha _1 \beta _5+3 \sqrt{2} \alpha _5 \beta _8\\
-\alpha _5 \beta _1+\sqrt{3} \alpha _5 \beta _2-3 \sqrt{2} \alpha _3 \beta _4-2 \alpha _1 \beta _6+2 \sqrt{2} \alpha _7 \beta _7-2 \sqrt{2} \alpha _6 \beta _8\\
-2 \alpha _6 \beta _1-2 \sqrt{3} \alpha _6 \beta _2-\sqrt{2} \alpha _5 \beta _3+3 \sqrt{2} \alpha _3 \beta _5-\sqrt{2} \alpha _2 \beta _6+2 \alpha _1 \beta _7\\
-2 \alpha _7 \beta _1+2 \sqrt{3} \alpha _7 \beta _2+\sqrt{2} \alpha _5 \beta _4+\sqrt{2} \alpha _3 \beta _6-3 \sqrt{2} \alpha _2 \beta _7+2 \alpha _1 \beta _8
\end{pmatrix}
~ \\
\hline\hline
\end{array}
\end{eqnarray*}

\begin{eqnarray*}
\begin{array}{|c|} \hline\hline
~\bf{8}\otimes\bf{8}=\bf{1_S}\oplus\bf{3_A}\oplus\bf{\bar{3}_A}\oplus\bf{6_{S,1}}\oplus\bf{6_{S,2}}\oplus\bf{7_{S}}\oplus\bf{7_{A,1}}\oplus\bf{7_{A,2}}\oplus\bf{8}_{S,1}\oplus\bf{8}_{S,2}\oplus\bf{8}_{A}~ \\ \hline \\[-0.16in]
~{\bf1_S}: \alpha _1 \beta _1+\alpha _2 \beta _2-\alpha _8 \beta _3-\alpha _7 \beta _4-\alpha _6 \beta _5-\alpha _5 \beta _6-\alpha _4 \beta _7-\alpha _3 \beta _8~\\
~{\bf 3_A}:
\begin{pmatrix}
-\alpha _3 \beta _1-\sqrt{3} \alpha _3 \beta _2+\alpha _1 \beta _3+\sqrt{3} \alpha _2 \beta _3+\sqrt{2} \alpha _8 \beta _4-\sqrt{2} \alpha _7 \beta _5+\sqrt{2} \alpha _5 \beta _7-\sqrt{2} \alpha _4 \beta _8\\
2 \alpha _4 \beta _1-2 \alpha _1 \beta _4+\sqrt{2} \alpha _8 \beta _5-\sqrt{2} \alpha _7 \beta _6+\sqrt{2} \alpha _6 \beta _7-\sqrt{2} \alpha _5 \beta _8\\
\alpha _6 \beta _1-\sqrt{3} \alpha _6 \beta _2-\sqrt{2} \alpha _5 \beta _3+\sqrt{2} \alpha _3 \beta _5-\alpha _1 \beta _6+\sqrt{3} \alpha _2 \beta _6+\sqrt{2} \alpha _8 \beta _7-\sqrt{2} \alpha _7 \beta _8
\end{pmatrix}~ \\
~{\bf\bar{3}_A}:
\begin{pmatrix}
-\alpha _8 \beta _1-\sqrt{3} \alpha _8 \beta _2+\sqrt{2} \alpha _7 \beta _3-\sqrt{2} \alpha _6 \beta _4+\sqrt{2} \alpha _4 \beta _6-\sqrt{2} \alpha _3 \beta _7+\alpha _1 \beta _8+\sqrt{3} \alpha _2 \beta _8\\
2 \alpha _7 \beta _1+\sqrt{2} \alpha _6 \beta _3-\sqrt{2} \alpha _5 \beta _4+\sqrt{2} \alpha _4 \beta _5-\sqrt{2} \alpha _3 \beta _6-2 \alpha _1 \beta _7\\
\alpha _5 \beta _1-\sqrt{3} \alpha _5 \beta _2+\sqrt{2} \alpha _4 \beta _3-\sqrt{2} \alpha _3 \beta _4-\alpha _1 \beta _5+\sqrt{3} \alpha _2 \beta _5-\sqrt{2} \alpha _8 \beta _6+\sqrt{2} \alpha _6 \beta _8
\end{pmatrix}
~ \\
\hline\hline
\end{array}
\end{eqnarray*}

\begin{eqnarray*}
\begin{array}{|c|}\hline
~\bf{7_{A,1}}:
\begin{pmatrix}
4 \alpha _1 \beta _2-4 \alpha _2 \beta _1\\
\sqrt{3} \alpha _3 \beta _1-\alpha _3 \beta _2-\sqrt{3} \alpha _1 \beta _3+\alpha _2 \beta _3+\sqrt{6} \alpha _8 \beta _4+\sqrt{6} \alpha _7 \beta _5-\sqrt{6} \alpha _5 \beta _7-\sqrt{6} \alpha _4 \beta _8\\
2 \alpha _4 \beta _2-2 \alpha _2 \beta _4-\sqrt{6} \alpha _8 \beta _5-\sqrt{6} \alpha _7 \beta _6+\sqrt{6} \alpha _6 \beta _7+\sqrt{6} \alpha _5 \beta _8\\
-\sqrt{3} \alpha _5 \beta _1-\alpha _5 \beta _2+\sqrt{6} \alpha _4 \beta _3-\sqrt{6} \alpha _3 \beta _4+\sqrt{3} \alpha _1 \beta _5+\alpha _2 \beta _5+\sqrt{6} \alpha _8 \beta _6-\sqrt{6} \alpha _6 \beta _8\\
\sqrt{3} \alpha _6 \beta _1+\alpha _6 \beta _2-\sqrt{6} \alpha _5 \beta _3+\sqrt{6} \alpha _3 \beta _5-\sqrt{3} \alpha _1 \beta _6-\alpha _2 \beta _6-\sqrt{6} \alpha _8 \beta _7+\sqrt{6} \alpha _7 \beta _8\\
-2 \alpha _7 \beta _2+\sqrt{6} \alpha _6 \beta _3+\sqrt{6} \alpha _5 \beta _4-\sqrt{6} \alpha _4 \beta _5-\sqrt{6} \alpha _3 \beta _6+2 \alpha _2 \beta _7\\
-\sqrt{3} \alpha _8 \beta _1+\alpha _8 \beta _2-\sqrt{6} \alpha _7 \beta _3-\sqrt{6} \alpha _6 \beta _4+\sqrt{6} \alpha _4 \beta _6+\sqrt{6} \alpha _3 \beta _7+\sqrt{3} \alpha _1 \beta _8-\alpha _2 \beta _8
\end{pmatrix}
~  \\
\hline\hline
\end{array}
\end{eqnarray*}

\section{\label{sec:eta}Constructing weight 2 modular forms of $\Gamma(7)$ by derivative of Dedekind eta function }

For any complex number $\tau$ with $\Im\tau>0$, the Dedekind eta-function $\eta(\tau)$ is defined as
\begin{equation}
\eta(\tau)=q^{1/24}\prod_{n=1}^\infty \left(1-q^n \right),\qquad q\equiv e^{i 2 \pi\tau}\,.
\end{equation}
The $\eta$ function satisfies the following identities
\begin{equation}
\eta(\tau+1)=e^{i \pi/12}\eta(\tau),\qquad \eta(-1/\tau)=\sqrt{-i \tau}~\eta(\tau)\,,
\end{equation}
which implies $\eta(\tau)$ is a modular function of weight half. Moreover, we see that the set of functions $\eta(7\tau)$, $\eta(\tau/7)$, $\eta((\tau+1)/7)$, $\eta((\tau+2)/7)$, $\eta((\tau+3)/7)$, $\eta((\tau+4)/7)$, $\eta((\tau+5)/7)$ and $\eta((\tau+6)/7)$ are closed under the action of the generators $S$ and $T$. To be more specific, we have the following transformation rules under $T$,
\begin{eqnarray}
\nonumber&&\eta\left(7\tau\right)\rightarrow e^{i\frac{7\pi}{12}}\eta\left(7\tau\right)\,,\quad \eta\left(\frac{\tau}{7}\right)\rightarrow \eta\left(\frac{\tau+1}{7}\right)\,, \\
\nonumber&&\eta\left(\frac{\tau+1}{7}\right)\rightarrow \eta\left(\frac{\tau+2}{7}\right)\,,\quad \eta\left(\frac{\tau+2}{7}\right)\rightarrow \eta\left(\frac{\tau+3}{7}\right)\,, \\
\nonumber&&\eta\left(\frac{\tau+3}{7}\right)\rightarrow \eta\left(\frac{\tau+4}{7}\right)\,,\quad \eta\left(\frac{\tau+4}{7}\right)\rightarrow \eta\left(\frac{\tau+5}{7}\right)\,,\\
&&\eta\left(\frac{\tau+5}{7}\right)\rightarrow \eta\left(\frac{\tau+6}{7}\right)\,,\quad \eta\left(\frac{\tau+6}{7}\right)\rightarrow e^{i\frac{\pi}{12}}\eta\left(\frac{\tau}{7}\right)\,.
\end{eqnarray}
Moreover, we find the following transformation behaviors under $S$
\begin{eqnarray}
\nonumber&&\eta\left(7\tau\right)\rightarrow 
\sqrt{\frac{-i\tau}{7}}\,\eta\left(\frac{\tau}{7}\right)\,,\quad \eta\left(\frac{\tau}{7}\right)\rightarrow \sqrt{-7i\tau}\,\eta\left(7\tau\right)\,,\\
\nonumber&&\eta\left(\frac{\tau+1}{7}\right)\rightarrow e^{-5i\pi/12}\sqrt{-i\tau}\,\eta\left(\frac{\tau+6}{7}\right)\,,\quad
\eta\left(\frac{\tau+2}{7}\right)\rightarrow e^{-i\pi/12} \sqrt{-i\tau}\,\eta\left(\frac{\tau+3}{7}\right)\,,\\
\nonumber&&\eta\left(\frac{\tau+3}{7}\right)\rightarrow e^{i\pi/12} \sqrt{-i\tau}\,\eta\left(\frac{\tau+2}{7}\right)\,,\quad
\eta\left(\frac{\tau+4}{7}\right)\rightarrow e^{-i\pi/12}\sqrt{-i\tau}\,\eta\left(\frac{\tau+5}{7}\right)\,,\\
&&\eta\left(\frac{\tau+5}{7}\right)\rightarrow e^{i\pi/12}\sqrt{-i\tau}\,\eta\left(\frac{\tau+4}{7}\right)\,,\quad
\eta\left(\frac{\tau+6}{7}\right)\rightarrow e^{5i\pi/12}\sqrt{-i\tau}\,\eta\left(\frac{\tau+1}{7}\right)
\end{eqnarray}
Following the approach proposed in~\cite{Feruglio:2017spp}, we can construct the weight 2 modular form by linear combination of the logarithmic derivative of above mentioned complete set of $\eta$ functions,
\begin{eqnarray}
\nonumber &&Y(x_1, x_2, x_3, x_4, x_5, x_6, x_7, x_8|\tau)\\
\nonumber&&~~=\frac{d}{d\tau}\Bigg[x_1\ln\eta\left(7\tau\right)+x_2\ln\eta\left(\frac{\tau}{7}\right)+x_3\ln\eta\left(\frac{\tau+1}{7}\right)+x_4\ln\eta\left(\frac{\tau+2}{7}\right)
+x_5\ln\eta\left(\frac{\tau+3}{7}\right)\\
\nonumber&&\qquad
+x_6\ln\eta\left(\frac{\tau+4}{7}\right)+x_7\ln\eta\left(\frac{\tau+5}{7}\right)
+x_8\ln\eta\left(\frac{\tau+6}{7}\right)\Bigg]\,,\\
\nonumber&&~~ = 7x_1\frac{\eta'\left(7\tau\right)}{\eta\left(7\tau\right)}+\frac{1}{7}\Bigg[x_2\frac{\eta'\left(\tau/7\right)}{\eta\left(\tau/7\right)}
+x_3\frac{\eta'\left((\tau+1)/7\right)}{\eta\left((\tau+1)/7\right)}+x_4\frac{\eta'\left((\tau+2)/7\right)}{\eta\left((\tau+2)/7\right)}
+x_5\frac{\eta'\left((\tau+3)/7\right)}{\eta\left((\tau+3)/7\right)}\\
&&\qquad +x_6\frac{\eta'\left((\tau+4)/7\right)}{\eta\left((\tau+4)/7\right)}+x_7\frac{\eta'\left((\tau+5)/7\right)}{\eta\left((\tau+5)/7\right)}
+x_8\frac{\eta'\left((\tau+6)/7\right)}{\eta\left((\tau+6)/7\right)}\Bigg]\,,
\end{eqnarray}
with $x_1+x_2+x_3+x_4+x_5+x_6+x_7+x_8=0$. Notice that $12\eta'(\tau)/\eta(\tau)\equiv i\pi E_2(\tau)$, where $E_2(\tau)$ is the well-known Eisenstein series of weight 2. Under the action of the generators $S$ and $T$, this function transforms as
\begin{eqnarray}
\nonumber&&Y(x_1, x_2, x_3, x_4, x_5, x_6,x_7,x_8|\tau)\stackrel{T}{\longrightarrow}Y(x_1, x_8, x_2, x_3, x_4, x_5, x_6, x_7|\tau)\,,\\
&&Y(x_1, x_2, x_3, x_4, x_5, x_6,x_7,x_8|\tau)\stackrel{S}{\longrightarrow}\tau^2Y(x_2, x_1, x_8, x_5, x_4, x_7, x_6, x_3|\tau)\,.
\end{eqnarray}
We can construct a septet $Y_{\mathbf{7}}(\tau)=(Y'_1, Y'_2, Y'_3, Y'_4, Y'_5, Y'_6, Y'_7)^{T}$ by the modular function $Y(x_1$, $x_2$, $x_3$, $x_4, x_5, x_6, x_7, x_8|\tau)$, and $Y_{\mathbf{7}}(\tau)$ transforms as $\mathbf{7}$ under $\Gamma_7\cong \Sigma(168)$,
\begin{eqnarray}
Y_{\mathbf{7}}(-1/\tau)=\tau^2\rho_{\mathbf{7}}(S)Y_{\mathbf{7}}(\tau),\qquad Y_{\mathbf{7}}(\tau+1)=\rho_{\mathbf{7}}(T)Y_{\mathbf{7}}(\tau)\,.
\end{eqnarray}
We can then straightforwardly find the solutions for $Y'_{i}$ $(i=1,2,\ldots,7)$ are given by
\begin{eqnarray}
\nonumber&&Y'_1(\tau)=\frac{c}{2\sqrt{2}}\,Y\left(7, -1, -1, -1, -1, -1,-1,-1|\tau\right)\,,\\
\nonumber&&Y'_2(\tau)=cY\left(0, 1, \rho^{-1}, \rho^{-2}, \rho^{-3}, \rho^3, \rho^2, \rho|\tau\right)\,,\\
\nonumber&&Y'_3(\tau)=cY\left(0, 1, \rho^{-2}, \rho^3, \rho, \rho^{-1}, \rho^{-3}, \rho^2|\tau\right)\,,\\
\nonumber&&Y'_4(\tau)=cY\left(0, 1, \rho^{-3}, \rho, \rho^{-2}, \rho^2, \rho^{-1}, \rho^{3}|\tau\right)\,,\\
\nonumber&&Y'_5(\tau)=cY\left(0, 1, \rho^3, \rho^{-1}, \rho^{2}, \rho^{-2}, \rho, \rho^{-3}|\tau\right)\,,\\
\nonumber&&Y'_6(\tau)=cY\left(0, 1, \rho^2, \rho^{-3}, \rho^{-1}, \rho, \rho^3, \rho^{-2}|\tau\right)\,,\\
\label{eq:Y7_2nd_app}&&Y'_7(\tau)=cY\left(0, 1, \rho, \rho^{2}, \rho^{3}, \rho^{-3}, \rho^{-2}, \rho^{-1}|\tau\right)\,,
\end{eqnarray}
up to the overall constant $c$. We shall choose $c=-\frac{i}{\sqrt{2}\,\pi}$ for convenience. The $q-$expansion of $Y'_i$ reads
\begin{eqnarray}
\nonumber&&Y'_1(\tau)=1+4 q+12 q^2+16 q^3+28 q^4+24 q^5+\ldots\,, \\
\nonumber&&Y'_2(\tau)=-\sqrt{2}q^{1/7}\left(1+15 q+24 q^2+36 q^3+30 q^4+91 q^5+\ldots\right)\,,\\
\nonumber&&Y'_3(\tau)=-\sqrt{2}q^{2/7}\left(3+13 q+31 q^2+24 q^3+72 q^4+38 q^5+\ldots\right)\,,\\
\nonumber&&Y'_4(\tau)=-2\sqrt{2}q^{3/7}\left(2+9q+9q^2+30q^3+16q^4+30q^5+\ldots\right)\,,\\
\nonumber&&Y'_5(\tau)=-\sqrt{2}q^{4/7}\left(7+12q+39q^2+31q^3+63q^4+56q^5+\ldots\right)\,,\\
\nonumber&&Y'_6(\tau)=-2\sqrt{2}q^{5/7}\left(3+14 q+10 q^2+21 q^3+24 q^4+45 q^5+\ldots\right)\,,\\
&&Y'_7(\tau)=-2\sqrt{2}q^{6/7}\left(6+7 q+21 q^2+20 q^3+27 q^4+21 q^5+\ldots\right)\,.
\end{eqnarray}
Further we find that the above modular forms can be expressed in terms of Miller-like basis vectors in Eq.~\eqref{eq:miller-like_basis} as follows,
\begin{eqnarray}
\nonumber&&Y'_1(\tau)=b_4(\tau)+4b_{11}(\tau)+12b_{18}(\tau)+16b_{25}(\tau)+28b_{26}(\tau)\,,\\
\nonumber&&Y'_2(\tau)=-\sqrt{2}\left[b_5(\tau)+15b_{12}(\tau)+24b_{19}(\tau)\right]\,,\\
\nonumber&&Y'_3(\tau)=-\sqrt{2}\left[3b_6(\tau)+13b_{13}(\tau)+31b_{20}(\tau)\right]\,,\\
\nonumber&&Y'_4(\tau)=-2\sqrt{2}\left[2b_7(\tau)+9b_{14}(\tau)+9b_{21}(\tau)\right]\,,\\
\nonumber&&Y'_5(\tau)=-\sqrt{2}\left[7b_8(\tau)+12b_{15}(\tau)+39b_{22}(\tau)\right]\,,\\
\nonumber&&Y'_6(\tau)=-2\sqrt{2}\left[3b_9(\tau)+14b_{16}(\tau)+10b_{23}(\tau)\right]\,,\\
&&Y'_7(\tau)=-2\sqrt{2}\left[6b_{10}(\tau)+7b_{17}(\tau)+21b_{24}(\tau)\right]\,.
\end{eqnarray}
We see that the modular form $Y_{\mathbf{7}}(\tau)$ is exactly the same as $Y^{(2)}_{\mathbf{7}}(\tau)$ in Eq.~\eqref{eq:Y2_7}.

\section{\label{sec:theta}Constructing weight 2 modular forms of $\Gamma(7)$ by theta function method }

\begin{figure}[t!]
\centering
\includegraphics[width=0.98\linewidth]{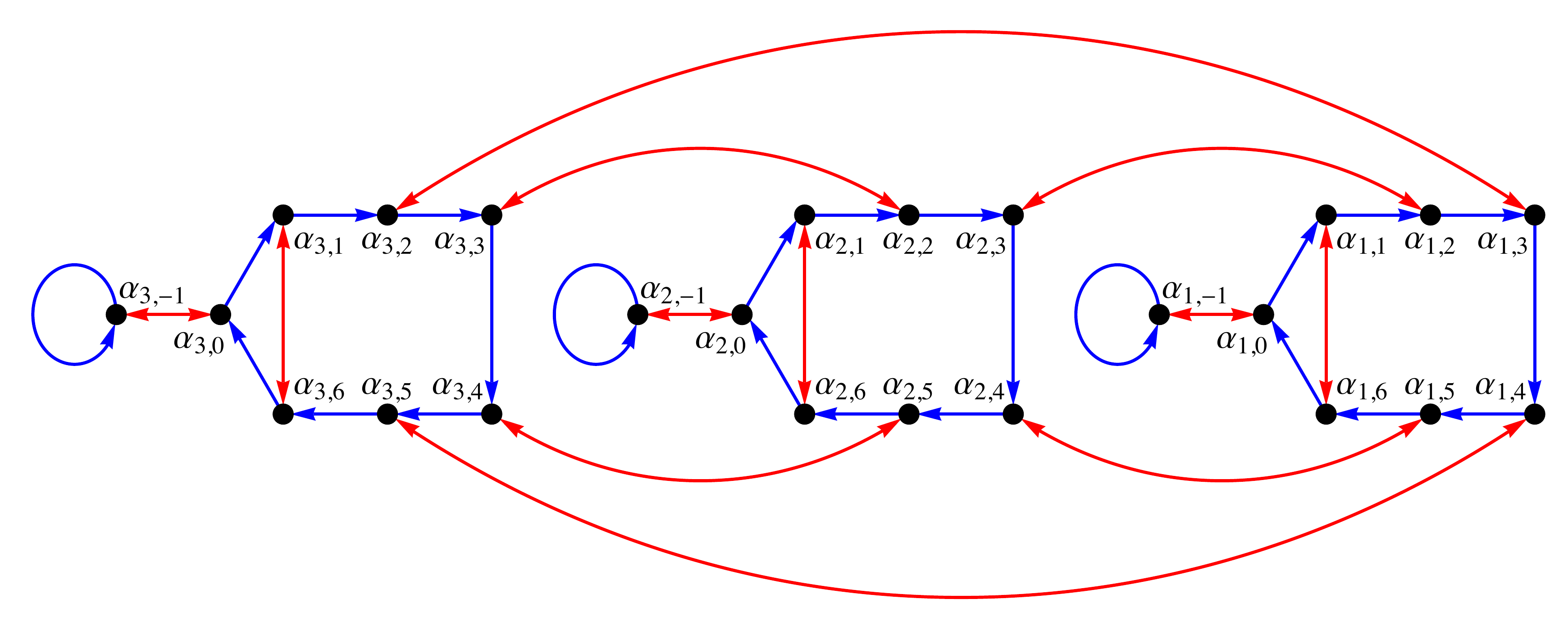}
\caption{\label{fig:graph} The graph illustrating the transformations
of the set of seed functions $\alpha_{i,j}(\tau)$,
defined in Eq.~\eqref{eq:seeds}, under the actions of $\Gamma_7\cong PSL(2,Z_{7})$ generators $S$ and $T$. The red and blue solid lines denote the transformations of $S$ and $T$ respectively.}
\end{figure}

The modular forms of level 5 can be constructed from the Jacobi theta functions~\cite{Novichkov:2018nkm}. In the following, we proceed to construct the weight 2 modular forms of level $N=7$ by using the Jacobi theta functions
$\theta_3(u, \tau)$ which is defined as
\begin{equation}\label{d1}
\theta_3(u,\tau )=\,\, \sum_{k\in\mathbb{Z}}q^{k^2/2}e^{2\pi iku}\,.
\end{equation}
It can also be expressed as the following infinite product,
\begin{equation}\label{eq:theta3}
\theta_3(u,\tau)=
\prod_{n=1}^\infty(1-q^{n})(1+q^{n-1/2}e^{2\pi i u})(1+q^{n-1/2}e^{-2\pi i u})\,.
\end{equation}
The theta function $\theta_3(u,\tau)$ has the following properies~\cite{Kharchev:2015tv}:
\begin{eqnarray}
\nonumber && \theta_3(-u,\tau)=\theta_3(u,\tau), \quad \theta_3(u+m,\tau+2n)=\theta_3(u,\tau), \quad
\theta_3(u+\frac{1}{2}+m,\tau+1+2n)=\theta_3(u,\tau)\,, \\
&&\theta_3(\tau-u,\tau)=e^{(2u-\tau)\pi i}\theta_{3}(u,\tau), \quad
\theta_3(u,\tau)=\frac{1}{\sqrt{-i\tau}}\,
e^{-\pi iu^2/\tau}\theta_3\left (u/\tau,-1/\tau\right)\,,
\end{eqnarray}
with $m,n \in \mathbb{Z}$. The lowest weight modular form with $k=2$ can be expressed as linear combinations of the logarithmic derivatives
of some ``seed'' functions $\alpha_{i,j}(\tau)$. We choose the closed set of the seed functions $\alpha_{i,j}(\tau)$ as follows,
\begin{eqnarray}
\nonumber\displaystyle \alpha_{1,-1}(\tau) = \theta_3 \left( \frac{\tau+1}{2}, 7\tau \right) ,  &
\displaystyle \alpha_{2,-1}(\tau) = e^{\frac{2\pi i\tau}{7}}\theta_3 \left( \frac{3 \tau+1}{2}, 7\tau \right) , &
\displaystyle \alpha_{3,-1}(\tau) = e^{\frac{6\pi i\tau}{7}}\theta_3 \left( \frac{5 \tau+1}{2}, 7\tau \right) , \\
\nonumber\displaystyle \alpha_{1,0}(\tau) = \theta_3 \left( \frac{\tau+1}{14}, \frac{\tau}{7} \right) ,~~~ &~~
\displaystyle \alpha_{2,0}(\tau) = \theta_3 \left( \frac{\tau+3}{14}, \frac{\tau}{7} \right) , &
\displaystyle \alpha_{3,0}(\tau) = \theta_3 \left( \frac{\tau+5}{14}, \frac{\tau}{7} \right) , \\
\nonumber\displaystyle \alpha_{1,1}(\tau) = \theta_3 \left( \frac{\tau+2}{14}, \frac{\tau+1}{7} \right) , &
\displaystyle \alpha_{2,1}(\tau) = \theta_3 \left( \frac{\tau+4}{14}, \frac{\tau+1}{7} \right)  , &
\displaystyle \alpha_{3,1}(\tau) = \theta_3 \left( \frac{\tau+6}{14}, \frac{\tau+1}{7} \right) , \\
\nonumber\displaystyle \alpha_{1,2}(\tau) = \theta_3 \left( \frac{\tau+3}{14}, \frac{\tau+2}{7} \right) , &
\displaystyle \alpha_{2,2}(\tau) = \theta_3 \left( \frac{\tau+5}{14}, \frac{\tau+2}{7} \right) , &
\nonumber\displaystyle \alpha_{3,2}(\tau) = \theta_3 \left( \frac{\tau+7}{14}, \frac{\tau+2}{7} \right) , \\
\nonumber\displaystyle \alpha_{1,3}(\tau) = \theta_3 \left( \frac{\tau+4}{14}, \frac{\tau+3}{7} \right) , &
\displaystyle \alpha_{2,3}(\tau) = \theta_3 \left( \frac{\tau+6}{14}, \frac{\tau+3}{7} \right) , &
\displaystyle \alpha_{3,3}(\tau) = \theta_3 \left( \frac{\tau+8}{14}, \frac{\tau+3}{7} \right) , \\
\nonumber\displaystyle \alpha_{1,4}(\tau) = \theta_3 \left( \frac{\tau+5}{14}, \frac{\tau+4}{7} \right) , &
\displaystyle \alpha_{2,4}(\tau) = \theta_3 \left( \frac{\tau+7}{14}, \frac{\tau+4}{7} \right) , &
\displaystyle \alpha_{3,4}(\tau) = \theta_3 \left( \frac{\tau+9}{14}, \frac{\tau+4}{7} \right) , \\
\nonumber\displaystyle \alpha_{1,5}(\tau) = \theta_3 \left( \frac{\tau+6}{14}, \frac{\tau+5}{7} \right) , &
\displaystyle \alpha_{2,5}(\tau) = \theta_3 \left( \frac{\tau+8}{14}, \frac{\tau+5}{7} \right) , &
\displaystyle \alpha_{3,5}(\tau) = \theta_3 \left( \frac{\tau+10}{14}, \frac{\tau+5}{7} \right) , \\
\label{eq:seeds}\displaystyle \alpha_{1,6}(\tau) = \theta_3 \left( \frac{\tau+7}{14}, \frac{\tau+6}{7} \right) , &
\displaystyle \alpha_{2,6}(\tau) = \theta_3 \left( \frac{\tau+9}{14}, \frac{\tau+6}{7} \right) , &
\displaystyle \alpha_{3,6}(\tau) = \theta_3 \left( \frac{\tau+11}{14}, \frac{\tau+6}{7} \right) \,.
\end{eqnarray}
Note the set of seed functions is not unique although the same results for modular forms are obtained. Under the action of the generators $S$ and $T$, we can check that each of these seed functions $\alpha_{i,j}(\tau)$ is mapped to another, up to some $\tau$-dependent multiplicative factor. The transformation properties of $\alpha_{i,j}(\tau)$ under $S$ and $T$ are shown in figure~\ref{fig:graph}. Hence we can start from any seed function $\alpha_{i,j}(\tau)$(e.g. $ \alpha_{1,-1}(\tau)$) to generate all the others. Moreover, we find that each seed function is mapped into itself under the actions of the modular transformations $S^2$, $(ST)^3$, $(ST^3)^4$ and $T^7$ up to some $\tau$ relevant factors. Taking logarithmic derivatives, we find
\begin{align}
\frac{d}{d\tau}\log\alpha_{i,j}(-1/\tau) &=
\frac{i\pi}{28}\left(1-\frac{1}{\tau^2}\right)+ \frac{1}{2\tau}
+\frac{d}{d\tau}\log\alpha^S_{i,j}(\tau)\,,
\\[2mm]
\frac{d}{d\tau}\log\alpha_{i,j}(\tau+1) &=
\frac{d}{d\tau}\log\alpha^T_{i,j}(\tau)\,,
\end{align}
where $\alpha^S_{i,j}$ and $\alpha^T_{i,j}$
are the images of $\alpha_{i,j}$
under the maps of $S$ and $T$ shown in figure~\ref{fig:graph},
respectively. As a consequence, the modular functions
\begin{align}
Y(x_{1,-1},\ldots,x_{1,6};x_{2,-1},\ldots,x_{2,6};x_{3,-1},\ldots,x_{3,6}|\tau) \equiv \sum_{i,j} x_{i,j}
\frac{d}{d\tau}\log\alpha_{i,j}(\tau)
\,,\quad \textrm{with } \sum_{i,j} x_{i,j} = 0\,,
\end{align}
span a 24-dimensional linear space of weight two modular forms of level $N=7$. Under $S$ and $T$, the modular function $Y$ transforms as follows,
\begin{eqnarray}
S: && Y(x_{1,-1}, ..., x_{1,6}; x_{2,-1}, ... x_{2,6}; x_{3,-1}, ... x_{3,6}| \tau) \stackrel{S}{\longmapsto}
Y(x_{1,-1}, ..., x_{1,6}; x_{2,-1}, ... x_{2,6}; x_{3,-1}, ... x_{3,6}| -1/\tau) \nonumber\\
&& \hspace{0.6cm} =
\tau^2 Y(x_{1,0}, x_{1,-1}, x_{1,6}, x_{2,3}, x_{3,2}, x_{3,5}, x_{2,4}, x_{1,1};
   x_{2,0}, x_{2,-1}, x_{2,6}, x_{3,3}, x_{1,2}, x_{1,5}, x_{3,4}, x_{2,1}; \nonumber\\
&& \hspace{8cm}
   x_{3,0}, x_{3,-1}, x_{3,6}, x_{1,3}, x_{2,2}, x_{2,5}, x_{1,4}, x_{3,1} | \tau) \,, \nonumber\\
T: && Y(x_{1,-1}, ..., x_{1,6}; x_{2,-1}, ... x_{2,6}; x_{3,-1}, ... x_{3,6}| \tau) \stackrel{T}{\longmapsto}
Y(x_{1,-1}, ..., x_{1,6}; x_{2,-1}, ... x_{2,6}; x_{3,-1}, ... x_{3,6}| \tau+1) \nonumber\\
&& \hspace{1cm} =
Y(x_{1,-1}, x_{1,6}, x_{1,0}, x_{1,1}, x_{1,2}, x_{1,3}, x_{1,4}, x_{1,5};
   x_{2,-1}, x_{2,6}, x_{2,0}, x_{2,1}, x_{2,2}, x_{2,3}, x_{2,4}, x_{2,5}; \nonumber\\
&& \hspace{8cm}
   x_{3,-1}, x_{3,6}, x_{3,0}, x_{3,1}, x_{3,2}, x_{3,3}, x_{3,4}, x_{3,5} | \tau) \,.
\end{eqnarray}
As shown in Eq.~\eqref{eq:irr_MFD}, we can always choose a basis such that the modular forms can be organized into different multiplets of $\Gamma_7$:
\begin{align}
\widetilde{Y}^{(2)}_\mathbf{7}(\tau)
&=\frac{-i}{2\sqrt{2}\pi }\left(\begin{array}{c}
\frac{1}{2\sqrt{2}}Y\left(7,-\mathbf{v}_0;7,-\mathbf{v}_0;7,-\mathbf{v}_0\middle|\tau\right)\\
Y(0,\mathbf{v}_1;\,0,\mathbf{v}_1;0,\mathbf{v}_1|\,\tau)\\
Y(0,\mathbf{v}_2;\,0,\mathbf{v}_2;0,\mathbf{v}_2|\,\tau)\\
Y(0,\mathbf{v}_3;\,0,\mathbf{v}_3;0,\mathbf{v}_3|\,\tau)\\
Y(0,\mathbf{v}_4;\,0,\mathbf{v}_4;0,\mathbf{v}_4|\,\tau)\\
Y(0,\mathbf{v}_5;\,0,\mathbf{v}_5;0,\mathbf{v}_5|\,\tau)\\
Y(0,\mathbf{v}_6;\,0,\mathbf{v}_6;0,\mathbf{v}_6|\,\tau)
\end{array}\right)\,,
\label{eq:mod_7_2} \\[2mm]
\widetilde{Y}^{(2)}_{\mathbf{8}a}(\tau)
&=\left(\begin{array}{c}
\frac{1}{\sqrt{2}}Y\left(2(c_1-c_3),\mathbf{0};2(c_2-c_1),\mathbf{v}_0;2(c_3-c_2),-\mathbf{v}_0\middle|\tau\right)\\
-\frac{1}{\sqrt{6}}Y\left(1+6c_2,-2\mathbf{v}_0;1+6c_3,\mathbf{v}_0;1+6c_1,\mathbf{v}_0\middle|\tau\right)\\
Y\left(0,\mathbf{v}_1;0,\mathbf{0};0,-\mathbf{v}_1\middle|\tau\right)\\
Y\left(0,\mathbf{0};0,-\mathbf{v}_2;0,\mathbf{v}_2\middle|\tau\right)\\
Y\left(0,-\mathbf{v}_3;0,\mathbf{v}_3;0,\mathbf{0}\middle|\tau\right)\\
Y\left(0,\mathbf{v}_4;0,-\mathbf{v}_4;0,\mathbf{0}\middle|\tau\right)\\
Y\left(0,\mathbf{0};0,\mathbf{v}_5;0,-\mathbf{v}_5\middle|\tau\right)\\
Y\left(0,-\mathbf{v}_6;0,\mathbf{0};0,\mathbf{v}_6\middle|\tau\right)
\end{array}\right)\,,
\label{eq:mod_81_2} \\[2mm]
\widetilde{Y}^{(2)}_{\mathbf{8}b}(\tau)
&=\left(\begin{array}{c}
\frac{1}{\sqrt{2}}Y\left(2(c_3-c_2),-\mathbf{v}_0;2(c_1-c_3),\mathbf{0};2(c_2-c_1),\mathbf{v}_0\middle|\tau\right)\\
-\frac{1}{\sqrt{6}}Y\left(1+6c_1,\mathbf{v}_0;1+6c_2,-2\mathbf{v}_0;1+6c_3,\mathbf{v}_0\middle|\tau\right)\\
Y\left(0,-\mathbf{v}_1;0,\mathbf{v}_1;0,\mathbf{0}\middle|\tau\right)\\
Y\left(0,\mathbf{v}_2;0,\mathbf{0};0,-\mathbf{v}_2\middle|\tau\right)\\
Y\left(0,\mathbf{0};0,-\mathbf{v}_3;0,\mathbf{v}_3\middle|\tau\right)\\
Y\left(0,\mathbf{0};0,\mathbf{v}_4;0,-\mathbf{v}_4\middle|\tau\right)\\
Y\left(0,-\mathbf{v}_5;0,\mathbf{0};0,\mathbf{v}_5\middle|\tau\right)\\
Y\left(0,\mathbf{v}_6;0,-\mathbf{v}_6;0,\mathbf{0}\middle|\tau\right)
\end{array}\right)\,,
\label{eq:mod_82_2}
\end{align}
where $\mathbf{v}_k\equiv\{1,\rho^{6k},\rho^{5k},\rho^{4k},\rho^{3k},\rho^{2k},\rho^{k}\}$ for $k=0,1,\cdots, 6$ and $\mathbf{0}\equiv\{0,0,0,0,0,0,0\}$.
From the expressions of $q-$expansion, we see that the modular form $\widetilde{Y}^{(2)}_\mathbf{7}(\tau)$ in the representation $\mathbf{7}$ coincides with $Y^{(2)}_{\mathbf{7}}(\tau)$ of Eq.~\eqref{eq:Y2_7}. Moreover, the two modular octets $Y^{(2)}_{\mathbf{8}a}(\tau)$ and $Y^{(2)}_{\mathbf{8}b}(\tau)$ in Eqs.~(\ref{eq:Y2_8a}, \ref{eq:mod})
are linear combinations of $\widetilde{Y}^{(2)}_{\mathbf{8}a}(\tau)$ and $\widetilde{Y}^{(2)}_{\mathbf{8}b}(\tau)$ as follow,
\begin{equation}
Y^{(2)}_{\mathbf{8}a}=\frac{y_1\widetilde{Y}^{(2)}_{\mathbf{8}a}(\tau)+y_2\widetilde{Y}^{(2)}_{\mathbf{8}b}(\tau)}{2\sqrt{2}\pi i},
~~~~~Y^{(2)}_{\mathbf{8}b}=\frac{z_1\widetilde{Y}^{(2)}_{\mathbf{8}a}(\tau)+z_2\widetilde{Y}^{(2)}_{\mathbf{8}b}(\tau)}{2\sqrt{2}\pi i}\,,
\end{equation}
with
\begin{eqnarray}
\nonumber &&y_1=-\frac{2}{7} (4 c_1+c_2-5 c_3), ~~~~ y_2=-\frac{2}{7}  (5 c_1-4 c_2-c_3),\\
&&z_1=\frac{1}{35} (2 c_1-3 c_2+c_3), ~~~~ z_2=\frac{1}{35} (-c_1-2 c_2+3 c_3)\,.
\end{eqnarray}
In short,  we can construct 23 modular forms of weight 2 and level 7, and they can be decomposed into one septuplet and two
octuplets of $\Gamma_7$. We can not build the modular multiplet $Y^{(2)}_{\mathbf{3}}(\tau)$ in the representation $\mathbf{3}$ of $\Gamma_7$ from the theta function $\theta_3(u, \tau)$.

\section{\label{appsec:w4_cons}Higher weight modular forms and constraints}

Through the tensor products of the modular forms $Y^{(2)}_\mathbf{3}$, $Y^{(2)}_\mathbf{7}$, $Y^{(2)}_{\mathbf{8}a}$ and $Y^{(2)}_{\mathbf{8}b}$,
one can find, at weight 4, the following modular multiplets:
\begin{equation}\label{eq:Yw4l_all}
Y^{(4)}_{\mathbf{1}a}    \,=\,\left(Y^{(2)}_{\mathbf{7}}Y^{(2)}_{\mathbf{7}}\right)_{\mathbf{1}},~
{Y^{(4)}_{\mathbf{1}b}} \,=\,\left(Y^{(2)}_{\mathbf{8}a}Y^{(2)}_{\mathbf{8}a}\right)_{\mathbf{1}},~
{Y^{(4)}_{\mathbf{1}c}} \,=\,\left(Y^{(2)}_{\mathbf{8}a}Y^{(2)}_{\mathbf{8}b}\right)_{\mathbf{1}},~
{Y^{(4)}_{\mathbf{1}d}} \,=\, \left(Y^{(2)}_{\mathbf{8}b}Y^{(2)}_{\mathbf{8}b}\right)_{\mathbf{1}} \,.
\end{equation}
\begin{eqnarray}
\nonumber&&{Y^{(4)}_{\mathbf{3}a}}  \,=\,\left(Y^{(2)}_{\mathbf{3}}Y^{(2)}_{\mathbf{8}a}\right)_{\mathbf{3}},~~~ {Y^{(4)}_{\mathbf{3}b}}  \,=\,\left(Y^{(2)}_{\mathbf{3}}Y^{(2)}_{\mathbf{8}b}\right)_{\mathbf{3}},~~~
{Y^{(4)}_{\mathbf{3}c}}\,=\,\left(Y^{(2)}_{\mathbf{7}}Y^{(2)}_{\mathbf{8}a}\right)_{\mathbf{3}}\,, \\
&&{Y^{(4)}_{\mathbf{3}d}}  \,=\,\left(Y^{(2)}_{\mathbf{7}}Y^{(2)}_{\mathbf{8}b}\right)_{\mathbf{3}}, ~~~
{Y^{(4)}_{\mathbf{3}e}}  \,=\,\left(Y^{(2)}_{\mathbf{8}a}Y^{(2)}_{\mathbf{8}b}\right)_{\mathbf{3_A}}\,. \end{eqnarray}
\begin{equation}
{Y^{(4)}_{\mathbf{\bar{3}}a}}  \,=\,\left(Y^{(2)}_{\mathbf{7}}Y^{(2)}_{\mathbf{8}a}\right)_{\mathbf{\bar{3}}}, \qquad
{Y^{(4)}_{\mathbf{\bar{3}}b}}  \,=\,\left(Y^{(2)}_{\mathbf{7}}Y^{(2)}_{\mathbf{8}b}\right)_{\mathbf{\bar{3}}}, \qquad
{Y^{(4)}_{\mathbf{\bar{3}}c}}  \,=\,
\left(Y^{(2)}_{\mathbf{8}a}Y^{(2)}_{\mathbf{8}b}\right)_{\mathbf{\bar{3}}}  \,.
\end{equation}
\begin{eqnarray}
\nonumber&&{Y^{(4)}_{\mathbf{6}a}}  \,=\,\left(Y^{(2)}_{\mathbf{3}}Y^{(2)}_{\mathbf{3}}\right)_{\mathbf{6}}, ~~~
{Y^{(4)}_{\mathbf{6}b}}\,=\,\left(Y^{(2)}_{\mathbf{3}}Y^{(2)}_{\mathbf{7}}\right)_{\mathbf{6}} ,~~~{Y^{(4)}_{\mathbf{6}c}}  \,=\,\left(Y^{(2)}_{\mathbf{3}}Y^{(2)}_{\mathbf{8}a}\right)_{\mathbf{6}} , \\
\nonumber&&{Y^{(4)}_{\mathbf{6}d}}  \,=\,\left(Y^{(2)}_{\mathbf{3}}Y^{(2)}_{\mathbf{8}b}\right)_{\mathbf{6}} ,~~~  {Y^{(4)}_{\mathbf{6}e}}  \,=\,\left(Y^{(2)}_{\mathbf{7}}Y^{(2)}_{\mathbf{7}}\right)_{\mathbf{6_{S,1}}}, ~~~{Y^{(4)}_{\mathbf{6}f}}  \,=\,\left(Y^{(2)}_{\mathbf{7}}Y^{(2)}_{\mathbf{8}a}\right)_{\mathbf{6_{1}}},\\
\nonumber&&{Y^{(4)}_{\mathbf{6}g}}\,=\,\left(Y^{(2)}_{\mathbf{7}}Y^{(2)}_{\mathbf{8}b}\right)_{\mathbf{6_{1}}}, ~~~{Y^{(4)}_{\mathbf{6}h}} \,=\,\left(Y^{(2)}_{\mathbf{8}a}Y^{(2)}_{\mathbf{8}a}\right)_{\mathbf{6_{S,1}}},~~~
{Y^{(4)}_{\mathbf{6}i}}  \,=\,\left(Y^{(2)}_{\mathbf{8}a}Y^{(2)}_{\mathbf{8}b}\right)_{\mathbf{6_{S,1}}}, \\
\nonumber&&{Y^{(4)}_{\mathbf{6}j}}  \,=\,\left(Y^{(2)}_{\mathbf{8}b}Y^{(2)}_{\mathbf{8}b}\right)_{\mathbf{6_{S,1}}}, ~~~{Y^{(4)}_{\mathbf{6}k}}  \,=\,\left(Y^{(2)}_{\mathbf{7}}Y^{(2)}_{\mathbf{7}}\right)_{\mathbf{6_{S,2}}}, ~~~
{Y^{(4)}_{\mathbf{6}l}}  \,=\,\left(Y^{(2)}_{\mathbf{7}}Y^{(2)}_{\mathbf{8}a}\right)_{\mathbf{6_{S,2}}}, \\
\nonumber&&{Y^{(4)}_{\mathbf{6}m}}  \,=\, \left(Y^{(2)}_{\mathbf{7}}Y^{(2)}_{\mathbf{8}b}\right)_{\mathbf{6_{S,2}}}, ~~~
{Y^{(4)}_{\mathbf{6}n}}  \,=\, \left(Y^{(2)}_{\mathbf{8}a}Y^{(2)}_{\mathbf{8}a}\right)_{\mathbf{6_{S,2}}}, ~~~
{Y^{(4)}_{\mathbf{6}o}}  \,=\,
\left(Y^{(2)}_{\mathbf{8}a}Y^{(2)}_{\mathbf{8}b}\right)_{\mathbf{6_{S,2}}}, \\
&&{Y^{(4)}_{\mathbf{6}p}}  \,=\,
\left(Y^{(2)}_{\mathbf{8}b}Y^{(2)}_{\mathbf{8}b}\right)_{\mathbf{6_{S,2}}}\,.
\end{eqnarray}

\begin{eqnarray}
\nonumber&&{Y^{(4)}_{\mathbf{7}a}}  \,=\, \left(Y^{(2)}_{\mathbf{3}}Y^{(2)}_{\mathbf{7}}\right)_{\mathbf{7}} ,~~~
{Y^{(4)}_{\mathbf{7}b}}  \,=\, \left(Y^{(2)}_{\mathbf{7}}Y^{(2)}_{\mathbf{7}}\right)_{\mathbf{7_S}} , ~~~
{Y^{(4)}_{\mathbf{7}c}}  \,=\, \left(Y^{(2)}_{\mathbf{3}}Y^{(2)}_{\mathbf{8}b}\right)_{\mathbf{7}}\,, \\
\nonumber&& {Y^{(4)}_{\mathbf{7}d}}  \,=\, \left(Y^{(2)}_{\mathbf{3}}Y^{(2)}_{\mathbf{8}a}\right)_{\mathbf{7}} ,~~~
{Y^{(4)}_{\mathbf{7}e}}  \,=\, \left(Y^{(2)}_{\mathbf{7}}Y^{(2)}_{\mathbf{8}a}\right)_{\mathbf{7_1}} , ~~~
{Y^{(4)}_{\mathbf{7}f}}  \,=\, \left(Y^{(2)}_{\mathbf{7}}Y^{(2)}_{\mathbf{8}b}\right)_{\mathbf{7_1}}\,, \\
\nonumber&& {Y^{(4)}_{\mathbf{7}g}}  \,=\, \left(Y^{(2)}_{\mathbf{8}a}Y^{(2)}_{\mathbf{8}a}\right)_{\mathbf{7_{S,1}}} , ~~~
{Y^{(4)}_{\mathbf{7}h}}  \,=\, \left(Y^{(2)}_{\mathbf{8}a}Y^{(2)}_{\mathbf{8}b}\right)_{\mathbf{7_{S,1}}} , ~~~
{Y^{(4)}_{\mathbf{7}i}}  \,=\, \left(Y^{(2)}_{\mathbf{8}b}Y^{(2)}_{\mathbf{8}b}\right)_{\mathbf{7_{S,1}}} \,, \\
\nonumber&& {Y^{(4)}_{\mathbf{7}j}}  \,=\, \left(Y^{(2)}_{\mathbf{7}}Y^{(2)}_{\mathbf{8}a}\right)_{\mathbf{7_2}} ,~~~ {Y^{(4)}_{\mathbf{7}k}}  \,=\, \left(Y^{(2)}_{\mathbf{7}}Y^{(2)}_{\mathbf{8}b}\right)_{\mathbf{7_2}} , ~~~
{Y^{(4)}_{\mathbf{7}l}}  \,=\, \left(Y^{(2)}_{\mathbf{8}a}Y^{(2)}_{\mathbf{8}b}\right)_{\mathbf{7_{S,2}}}\,, \\
&& {Y^{(4)}_{\mathbf{7}m}}  \,=\, \left(Y^{(2)}_{\mathbf{8}a}Y^{(2)}_{\mathbf{8}b}\right)_{\mathbf{7_{A}}} \,.
\end{eqnarray}
\begin{eqnarray}
\nonumber&&{Y^{(4)}_{\mathbf{8}a}}  \,=\,\left(Y^{(2)}_{\mathbf{3}}Y^{(2)}_{\mathbf{7}}\right)_{\mathbf{8}} , ~~~
{Y^{(4)}_{\mathbf{8}b}}  \,=\,\left(Y^{(2)}_{\mathbf{8}a}Y^{(2)}_{\mathbf{8}a}\right)_{\mathbf{8_{S,1}}} ,~~~
{Y^{(4)}_{\mathbf{8}c}}  \,=\,\left(Y^{(2)}_{\mathbf{8}b}Y^{(2)}_{\mathbf{8}b}\right)_{\mathbf{8_{S,2}}} \,, \\
\nonumber&&{Y^{(4)}_{\mathbf{8}d}}  \,=\,\left(Y^{(2)}_{\mathbf{7}}Y^{(2)}_{\mathbf{7}}\right)_{\mathbf{8_S}} , ~~~
{Y^{(4)}_{\mathbf{8}e}}  \,=\,\left(Y^{(2)}_{\mathbf{7}}Y^{(2)}_{\mathbf{8}a}\right)_{\mathbf{8_1}} , ~~~{Y^{(4)}_{\mathbf{8}f}}  \,=\,\left(Y^{(2)}_{\mathbf{7}}Y^{(2)}_{\mathbf{8}b}\right)_{\mathbf{8_1}},  \\
\nonumber&&Y^{(4)}_{\mathbf{8}g}  \,=\,\left(Y^{(2)}_{\mathbf{3}}Y^{(2)}_{\mathbf{8}a}\right)_{\mathbf{8}}  , ~~~
{Y^{(4)}_{\mathbf{8}h}}  \,=\,\left(Y^{(2)}_{\mathbf{8}a}Y^{(2)}_{\mathbf{8}b}\right)_{\mathbf{8_{S,1}}}, ~~~{Y^{(4)}_{\mathbf{8}i}}  \,=\,\left(Y^{(2)}_{\mathbf{8}b}Y^{(2)}_{\mathbf{8}b}\right)_{\mathbf{8_{S,1}}}\,, \\
\nonumber&&{Y^{(4)}_{\mathbf{8}j}}  \,=\,\left(Y^{(2)}_{\mathbf{7}}Y^{(2)}_{\mathbf{8}a}\right)_{\mathbf{8_2}} , ~~~
{Y^{(4)}_{\mathbf{8}k}}  \,=\,\left(Y^{(2)}_{\mathbf{7}}Y^{(2)}_{\mathbf{8}b}\right)_{\mathbf{8_2}} , ~~~
{Y^{(4)}_{\mathbf{8}l}}  \,=\,\left(Y^{(2)}_{\mathbf{8}a}Y^{(2)}_{\mathbf{8}a}\right)_{\mathbf{8_{S,2}}}, \\
\nonumber&&{Y^{(4)}_{\mathbf{8}m}}  \,=\,\left(Y^{(2)}_{\mathbf{8}a}Y^{(2)}_{\mathbf{8}b}\right)_{\mathbf{8_{S,2}}}, ~~~
{Y^{(4)}_{\mathbf{8}n}}  \,=\,\left(Y^{(2)}_{\mathbf{3}}Y^{(2)}_{\mathbf{8}b}\right)_{\mathbf{8}},~~~
{Y^{(4)}_{\mathbf{8}o}}  \,=\,\left(Y^{(2)}_{\mathbf{7}}Y^{(2)}_{\mathbf{8}a}\right)_{\mathbf{8_3}}, \\
&& {Y^{(4)}_{\mathbf{8}p}}  \,=\,\left(Y^{(2)}_{\mathbf{7}}Y^{(2)}_{\mathbf{8}b}\right)_{\mathbf{8_3}} , ~~~
{Y^{(4)}_{\mathbf{8}q}}  \,=\,\left(Y^{(2)}_{\mathbf{8}a}Y^{(2)}_{\mathbf{8}b}\right)_{\mathbf{8_A}} \,.
\end{eqnarray}
Notice that not all of the above modular multiplets are linearly independent. From the $q$-expansions of $Y_i(\tau)$ given
in Eq.~\eqref{eq:mod}, we find the following 297 constraints between the different weight 4 modular multiplets $Y^{(4)}_{\mathbf{r}}$,
\begin{equation}\label{eq:corn_dim1}
Y^{(4)}_{\mathbf{1}a}=Y^{(4)}_{\mathbf{1}b}=300Y^{(4)}_{\mathbf{1}d}, \qquad Y^{(4)}_{\mathbf{1}c}=0\,.
\end{equation}
\begin{equation}\label{eq:corn_dim3}
Y^{(4)}_{\mathbf{3}a}=-30Y^{(4)}_{\mathbf{3}b}=
-\frac{1}{\sqrt{6}}Y^{(4)}_{\mathbf{3}c}=\sqrt{6}Y^{(4)}_{\mathbf{3}d}
=-5\sqrt{6}Y^{(4)}_{\mathbf{3}e}\,.
\end{equation}
\begin{equation}\label{eq:corn_dim3p}
Y^{(4)}_{\mathbf{\bar{3}}a}=Y^{(4)}_{\mathbf{\bar{3}}b}
=Y^{(4)}_{\mathbf{\bar{3}}c}=(0,0,0)^T\,.
\end{equation}
\begin{eqnarray}
\nonumber&&
\displaystyle Y^{(4)}_{\mathbf{6}c}=\frac{1}{2} \left(3 \sqrt{2} Y^{(4)}_{\mathbf{6}a}+Y^{(4)}_{\mathbf{6}b}\right), ~~~~
\displaystyle Y^{(4)}_{\mathbf{6}d}=\frac{1}{20} \left(Y^{(4)}_{\mathbf{6}b}-5 \sqrt{2} Y^{(4)}_{\mathbf{6}a}\right)\,,~~~~
 Y^{(4)}_{\mathbf{6}e}=2 \left(3 Y^{(4)}_{\mathbf{6}a}+\sqrt{2} Y^{(4)}_{\mathbf{6}b}\right),\\
\nonumber&&
 \displaystyle Y^{(4)}_{\mathbf{6}f}=8 \sqrt{2} Y^{(4)}_{\mathbf{6}a}-4 Y^{(4)}_{\mathbf{6}b}\,,~~~~Y^{(4)}_{\mathbf{6}g}=\frac{2}{5} \left(Y^{(4)}_{\mathbf{6}b}-6 \sqrt{2} Y^{(4)}_{\mathbf{6}a}\right), ~~~~
\displaystyle Y^{(4)}_{\mathbf{6}h}=13 \sqrt{2} Y^{(4)}_{\mathbf{6}a}-8 Y^{(4)}_{\mathbf{6}b}\,, \\
\nonumber&&Y^{(4)}_{\mathbf{6}i}=\frac{1}{10} \left(2 Y^{(4)}_{\mathbf{6}b}-3 \sqrt{2} Y^{(4)}_{\mathbf{6}a}\right),~~~~
\displaystyle Y^{(4)}_{\mathbf{6}j}=\frac{Y^{(4)}_{\mathbf{6}a}}{10 \sqrt{2}}\,,~~~~Y^{(4)}_{\mathbf{6}k}=-2 \sqrt{2} Y^{(4)}_{\mathbf{6}a}, \\ \nonumber&&
\displaystyle Y^{(4)}_{\mathbf{6}l}=19 \sqrt{2} Y^{(4)}_{\mathbf{6}a}-3 Y^{(4)}_{\mathbf{6}b}\,,~~~~
Y^{(4)}_{\mathbf{6}m}=\frac{1}{10} \left(Y^{(4)}_{\mathbf{6}b}-\sqrt{2} Y^{(4)}_{\mathbf{6}a}\right), ~~~~
\displaystyle Y^{(4)}_{\mathbf{6}n}=-2 \left(7 Y^{(4)}_{\mathbf{6}a}+\sqrt{2} Y^{(4)}_{\mathbf{6}b}\right)\,, \\
\label{eq:corn_dim6}&& Y^{(4)}_{\mathbf{6}o}=\frac{1}{20} \left(14 Y^{(4)}_{\mathbf{6}a}+3 \sqrt{2} Y^{(4)}_{\mathbf{6}b}\right), ~~~~~~~
\displaystyle Y^{(4)}_{\mathbf{6}p}=-\frac{Y^{(4)}_{\mathbf{6}b}}{50 \sqrt{2}}\,.
\end{eqnarray}

\begin{eqnarray}
\nonumber&&\displaystyle Y^{(4)}_{\mathbf{7}c}=-\frac{1}{10 \sqrt{2}}Y^{(4)}_{\mathbf{7}a}\,, ~~~~~~
\displaystyle Y^{(4)}_{\mathbf{7}e}=-11 Y^{(4)}_{\mathbf{7}a}, ~~~~~~ \displaystyle Y^{(4)}_{\mathbf{7}d}=-\frac{5 }{\sqrt{2}}Y^{(4)}_{\mathbf{7}a}, \\
\nonumber&&\displaystyle Y^{(4)}_{\mathbf{7}f}=\frac{1}{2} Y^{(4)}_{\mathbf{7}a}, ~~~~~~
\displaystyle Y^{(4)}_{\mathbf{7}g}=5 Y^{(4)}_{\mathbf{7}a}+\frac{1}{2} Y^{(4)}_{\mathbf{7}b}, ~~~~~~
\displaystyle Y^{(4)}_{\mathbf{7}h}=-\frac{3}{10} Y^{(4)}_{\mathbf{7}a}\,, \\
\nonumber&&\displaystyle Y^{(4)}_{\mathbf{7}i}=\frac{1}{600} (14 Y^{(4)}_{\mathbf{7}a}+Y^{(4)}_{\mathbf{7}b}),~~~~~~
\displaystyle Y^{(4)}_{\mathbf{7}j}=-2 Y^{(4)}_{\mathbf{7}a}\,, ~~~~~~
\displaystyle Y^{(4)}_{\mathbf{7}k}=-\frac{1}{5} Y^{(4)}_{\mathbf{7}a}\,,  \\
\label{eq:corn_dim7}&&\displaystyle Y^{(4)}_{\mathbf{7}l}=\frac{1}{30 \sqrt{3}}(11 Y^{(4)}_{\mathbf{7}a}-2 Y^{(4)}_{\mathbf{7}b}), ~~~~~~
\displaystyle Y^{(4)}_{\mathbf{7}m}=\frac{7}{10} Y^{(4)}_{\mathbf{7}a}\,.
\end{eqnarray}

\begin{eqnarray}
\nonumber&&\displaystyle Y^{(4)}_{\mathbf{8}d}=2 \sqrt{2} Y^{(4)}_{\mathbf{8}a}, ~~~~~~
\displaystyle Y^{(4)}_{\mathbf{8}e}=\sqrt{\frac{3}{2}} Y^{(4)}_{\mathbf{8}a}+Y^{(4)}_{\mathbf{8}b}+450 Y^{(4)}_{\mathbf{8}c}\,, \\
\nonumber&&\displaystyle Y^{(4)}_{\mathbf{8}f}=\frac{1}{360} \left(-91 \sqrt{6} Y^{(4)}_{\mathbf{8}a}-16 Y^{(4)}_{\mathbf{8}b}+18600 Y^{(4)}_{\mathbf{8}c}\right),~~~~~~ \displaystyle Y^{(4)}_{\mathbf{8}g}=-\frac{1}{4} Y^{(4)}_{\mathbf{8}a}\,, \\
\nonumber&&\displaystyle Y^{(4)}_{\mathbf{8}h}=\frac{1}{180} \left(35 \sqrt{6} Y^{(4)}_{\mathbf{8}a}+2 Y^{(4)}_{\mathbf{8}b}-8400 Y^{(4)}_{\mathbf{8}c}\right), ~~~~~~
\displaystyle Y^{(4)}_{\mathbf{8}i}=\frac{1}{300} \left(-\sqrt{6} Y^{(4)}_{\mathbf{8}a}-Y^{(4)}_{\mathbf{8}b}\right), \\
\nonumber&&\displaystyle Y^{(4)}_{\mathbf{8}j}=-\frac{15}{4}  \left(\sqrt{2} Y^{(4)}_{\mathbf{8}a}-80 \sqrt{3} Y^{(4)}_{\mathbf{8}c}\right), ~~~~~~
\displaystyle Y^{(4)}_{\mathbf{8}k}=\frac{1}{360} \left(-51 \sqrt{2} Y^{(4)}_{\mathbf{8}a}-8 \sqrt{3} (Y^{(4)}_{\mathbf{8}b}-150 Y^{(4)}_{\mathbf{8}c})\right)\,, \\
\nonumber&&\displaystyle Y^{(4)}_{\mathbf{8}l}=\sqrt{6} Y^{(4)}_{\mathbf{8}a}-300 Y^{(4)}_{\mathbf{8}c},~~~~~~
\displaystyle Y^{(4)}_{\mathbf{8}m}=\frac{1}{45} \left(-4 \sqrt{6} Y^{(4)}_{\mathbf{8}a}-Y^{(4)}_{\mathbf{8}b}+150 Y^{(4)}_{\mathbf{8}c}\right), \\
\nonumber&&\displaystyle Y^{(4)}_{\mathbf{8}n}=\frac{1}{40} Y^{(4)}_{\mathbf{8}a}\,,~~~~~~
 \displaystyle Y^{(4)}_{\mathbf{8}o}=\frac{9 }{2 \sqrt{2}}Y^{(4)}_{\mathbf{8}a}\,, \\
\label{eq:corn_dim8}&&\displaystyle Y^{(4)}_{\mathbf{8}p}=-\frac{1}{20 \sqrt{2}}Y^{(4)}_{\mathbf{8}a},~~~~~~
\displaystyle Y^{(4)}_{\mathbf{8}q}=\frac{1}{10 \sqrt{2}}Y^{(4)}_{\mathbf{8}a}\,.
\end{eqnarray}
These constraints in Eqs.~\eqref{eq:corn_dim1}\,--\,\eqref{eq:corn_dim8}
imply that the linear space of modular forms of weight $k=4$ and level 7 has dimension $54$, as explicitly listed in Eqs.~\eqref{eq:Yw4l}\,--\,\eqref{eq:w4_54forms_p3}.

There are 82 linearly independent modular forms arising at weight 6 and level $N=7$ which may be necessary in model construction, we give them in the following,
\begin{equation}\label{eq:Yw6l}
Y^{(6)}_{\mathbf{1}}  =\left(Y^{(2)}_{\mathbf{7}}Y^{(4)}_{\mathbf{7}b}\right)_{\mathbf{1}}  \,=\,Y_{10} Y^{(4)}_{25}+Y_{4} Y^{(4)}_{24}+Y^{(4)}_{26} Y_{9}+Y^{(4)}_{27} Y_{8}+Y^{(4)}_{28} Y_{7}+Y^{(4)}_{29} Y_{6}+Y^{(4)}_{30} Y_{5} \,.
\end{equation}
%
\begin{eqnarray}
&&{Y^{(6)}_{\mathbf{3}a}}  =\left(Y^{(2)}_{\mathbf{3}}Y^{(4)}_{\mathbf{1}a}\right)_{\mathbf{3}} \,=\,
 Y^{(4)}_{1}\left(
\begin{array}{c}
Y_{1}\\
Y_{2} \\
Y_{3}
\end{array}
\right) \,,\\[2mm]
\label{eq:Yw63}&&{Y^{(6)}_{\mathbf{3}b}} =\left(Y^{(2)}_{\mathbf{3}}Y^{(4)}_{\mathbf{8}a}\right)_{\mathbf{3}}\,=\,
\left(
\begin{array}{c}
\sqrt{3} Y_{1} Y^{(4)}_{31}+Y_{1} Y^{(4)}_{32}-\sqrt{6} Y_{2} Y^{(4)}_{38}-\sqrt{6} Y_{3} Y^{(4)}_{36}\\
\sqrt{6} Y_{1} Y^{(4)}_{33}-\sqrt{3} Y_{2} Y^{(4)}_{31}+Y_{2} Y^{(4)}_{32}-\sqrt{6} Y_{3} Y^{(4)}_{37}\\
\sqrt{6} Y_{1} Y^{(4)}_{35}+\sqrt{6} Y_{2} Y^{(4)}_{34}-2 Y_{3} Y^{(4)}_{32}
\end{array}
\right)  \,.
\end{eqnarray}
\begin{equation}\label{eq:Yw63b}
{Y^{(6)}_{\mathbf{\bar{3}}}}
=\left(Y^{(2)}_{\mathbf{3}}Y^{(4)}_{\mathbf{3}a}\right)_{\mathbf{\bar{3}}}  \,=\,
\left(
\begin{array}{c}
Y_{2} Y^{(4)}_{4}-Y_{3} Y^{(4)}_{3}\\
Y_{3} Y^{(4)}_{2}-Y_{1} Y^{(4)}_{4}\\
Y_{1} Y^{(4)}_{3}-Y_{2} Y^{(4)}_{2}
\end{array}
\right)  \,.
\end{equation}
%
\begin{eqnarray}
&&{Y^{(6)}_{\mathbf{6}a}} =\left(Y^{(2)}_{\mathbf{3}}Y^{(4)}_{\mathbf{3}a}\right)_{\mathbf{6}}
\,=\,
\left(
\begin{array}{c}
\sqrt{2} Y_{3} Y^{(4)}_{4}\\
\sqrt{2} Y_{1} Y^{(4)}_{2}\\
Y_{1} Y^{(4)}_{3}+Y_{2} Y^{(4)}_{2}\\
\sqrt{2} Y_{2} Y^{(4)}_{3}\\
Y_{1} Y^{(4)}_{4}+Y_{3} Y^{(4)}_{2}\\
Y_{2} Y^{(4)}_{4}+Y_{3} Y^{(4)}_{3}
\end{array}
\right) \,,\\[2mm]
\label{eq:Yw66}&&{Y^{(6)}_{\mathbf{6}b}} =\left(Y^{(2)}_{\mathbf{3}}Y^{(4)}_{\mathbf{7}b}\right)_{\mathbf{6}}
\,=\,
\left(
\begin{array}{c}
2 Y_{1} Y^{(4)}_{24}-2 \sqrt{2} Y_{2} Y^{(4)}_{30}-\sqrt{2} Y_{3} Y^{(4)}_{28}\\
-\sqrt{2} Y_{1} Y^{(4)}_{25}+2 Y_{2} Y^{(4)}_{24}-2 \sqrt{2} Y_{3} Y^{(4)}_{29}\\
-Y_{1} Y^{(4)}_{26}+3 Y_{2} Y^{(4)}_{25}-2 Y_{3} Y^{(4)}_{30}\\
-2 \sqrt{2} Y_{1} Y^{(4)}_{27}-\sqrt{2} Y_{2} Y^{(4)}_{26}+2 Y_{3} Y^{(4)}_{24}\\
3 Y_{1} Y^{(4)}_{28}-2 Y_{2} Y^{(4)}_{27}-Y_{3} Y^{(4)}_{25}\\
-2 Y_{1} Y^{(4)}_{29}-Y_{2} Y^{(4)}_{28}+3 Y_{3} Y^{(4)}_{26}
\end{array}
\right)  \,.
\end{eqnarray}
%
\begin{eqnarray}
&&{Y^{(6)}_{\mathbf{7}a}}
=\left(Y^{(2)}_{\mathbf{3}}Y^{(4)}_{\mathbf{6}a}\right)_{\mathbf{7}}
\,=\,
\left(
\begin{array}{c}
2 Y_{1} Y^{(4)}_{10}+2 Y_{2} Y^{(4)}_{9}+2 Y_{3} Y^{(4)}_{7}\\
-2 \sqrt{2} Y_{2} Y^{(4)}_{10}-2 Y_{3} Y^{(4)}_{8}\\
-2 Y_{1} Y^{(4)}_{5}-2 \sqrt{2} Y_{3} Y^{(4)}_{9}\\
3 Y_{1} Y^{(4)}_{6}-Y_{2} Y^{(4)}_{5}-\sqrt{2} Y_{3} Y^{(4)}_{10}\\
-2 \sqrt{2} Y_{1} Y^{(4)}_{7}-2 Y_{2} Y^{(4)}_{6}\\
-Y_{1} Y^{(4)}_{8}-\sqrt{2} Y_{2} Y^{(4)}_{7}+3 Y_{3} Y^{(4)}_{5}\\
-\sqrt{2} Y_{1} Y^{(4)}_{9}+3 Y_{2} Y^{(4)}_{8}-Y_{3} Y^{(4)}_{6}
\end{array}
\right)  \,,\\[2mm]
&&{Y^{(6)}_{\mathbf{7}b}}
=\left(Y^{(2)}_{\mathbf{3}}Y^{(4)}_{\mathbf{6}b}\right)_{\mathbf{7}}
\,=\,
\left(
\begin{array}{c}
2 Y_{1} Y^{(4)}_{16}+2 Y_{2} Y^{(4)}_{15}+2 Y_{3} Y^{(4)}_{13}\\
-2 \sqrt{2} Y_{2} Y^{(4)}_{16}-2 Y_{3} Y^{(4)}_{14}\\
-2 Y_{1} Y^{(4)}_{11}-2 \sqrt{2} Y_{3} Y^{(4)}_{15}\\
3 Y_{1} Y^{(4)}_{12}-Y_{2} Y^{(4)}_{11}-\sqrt{2} Y_{3} Y^{(4)}_{16}\\
-2 \sqrt{2} Y_{1} Y^{(4)}_{13}-2 Y_{2} Y^{(4)}_{12}\\
-Y_{1} Y^{(4)}_{14}-\sqrt{2} Y_{2} Y^{(4)}_{13}+3 Y_{3} Y^{(4)}_{11}\\
-\sqrt{2} Y_{1} Y^{(4)}_{15}+3 Y_{2} Y^{(4)}_{14}-Y_{3} Y^{(4)}_{12}
\end{array}
\right)  \,,\\[2mm]
&&{Y^{(6)}_{\mathbf{7}c}} =\left(Y^{(2)}_{\mathbf{3}}Y^{(4)}_{\mathbf{7}b}\right)_{\mathbf{7}}
\,=\,
\left(
\begin{array}{c}
\sqrt{2} Y_{1} Y^{(4)}_{30}+\sqrt{2} Y_{2} Y^{(4)}_{29}+\sqrt{2} Y_{3} Y^{(4)}_{27}\\
-\sqrt{2} Y_{1} Y^{(4)}_{24}-2 Y_{3} Y^{(4)}_{28}\\
-\sqrt{2} Y_{2} Y^{(4)}_{24}-2 Y_{1} Y^{(4)}_{25}\\
-Y_{1} Y^{(4)}_{26}+Y_{2} Y^{(4)}_{25}+2 Y_{3} Y^{(4)}_{30}\\
-\sqrt{2} Y_{3} Y^{(4)}_{24}-2 Y_{2} Y^{(4)}_{26}\\
Y_{1} Y^{(4)}_{28}+2 Y_{2} Y^{(4)}_{27}-Y_{3} Y^{(4)}_{25}\\
2 Y_{1} Y^{(4)}_{29}-Y_{2} Y^{(4)}_{28}+Y_{3} Y^{(4)}_{26}
\end{array}
\right) \,,\\[2mm]
\label{eq:Yw67}&&{Y^{(6)}_{\mathbf{7}d}}
=\left(Y^{(2)}_{\mathbf{7}}Y^{(4)}_{\mathbf{1}a}\right)_{\mathbf{7}}
\,=\,
Y^{(4)}_{1}\left(
\begin{array}{c}
Y_{4} \\
 Y_{5}\\
 Y_{6}\\
 Y_{7}\\
 Y_{8}\\
 Y_{9}\\
Y_{10}
\end{array}
\right)=Y^{(4)}_{1}Y^{(2)}_{\mathbf{7}}(\tau)  \,.
\end{eqnarray}
%
\begin{eqnarray}
&&{Y^{(6)}_{\mathbf{8}a}}
=\left(Y^{(2)}_{\mathbf{3}}Y^{(4)}_{\mathbf{6}b}\right)_{\mathbf{7}}
\,=\,
\left(
\begin{array}{c}
Y_{1} Y^{(4)}_{16}+Y_{2} Y^{(4)}_{15}-2 Y_{3} Y^{(4)}_{13}\\
\sqrt{3} Y_{2} Y^{(4)}_{15}-\sqrt{3} Y_{1} Y^{(4)}_{16}\\
\sqrt{2} Y_{2} Y^{(4)}_{16}-2 Y_{3} Y^{(4)}_{14}\\
\sqrt{2} Y_{3} Y^{(4)}_{15}-2 Y_{1} Y^{(4)}_{11}\\
2 Y_{2} Y^{(4)}_{11}-\sqrt{2} Y_{3} Y^{(4)}_{16}\\
2 Y_{2} Y^{(4)}_{12}-\sqrt{2} Y_{1} Y^{(4)}_{13}\\
\sqrt{2} Y_{2} Y^{(4)}_{13}-2 Y_{1} Y^{(4)}_{14}\\
\sqrt{2} Y_{1} Y^{(4)}_{15}-2 Y_{3} Y^{(4)}_{12}
\end{array}
\right)  \,,\\[2mm]
&&{Y^{(6)}_{\mathbf{8}b}}
=\left(Y^{(2)}_{\mathbf{3}}Y^{(4)}_{\mathbf{7}b}\right)_{\mathbf{7}}
\,=\,
\left(
\begin{array}{c}
5 Y_{1} Y^{(4)}_{30}-Y_{2} Y^{(4)}_{29}-4 Y_{3} Y^{(4)}_{27}\\
-\sqrt{3} Y_{1} Y^{(4)}_{30}+3 \sqrt{3} Y_{2} Y^{(4)}_{29}-2 \sqrt{3} Y_{3} Y^{(4)}_{27}\\
-4 Y_{1} Y^{(4)}_{24}-3 \sqrt{2} Y_{2} Y^{(4)}_{30}+2 \sqrt{2} Y_{3} Y^{(4)}_{28}\\
2 \sqrt{2} Y_{1} Y^{(4)}_{25}-4 Y_{2} Y^{(4)}_{24}-3 \sqrt{2} Y_{3} Y^{(4)}_{29}\\
4 \sqrt{2} Y_{1} Y^{(4)}_{26}+2 \sqrt{2} Y_{2} Y^{(4)}_{25}+\sqrt{2} Y_{3} Y^{(4)}_{30}\\
3 \sqrt{2} Y_{1} Y^{(4)}_{27}-2 \sqrt{2} Y_{2} Y^{(4)}_{26}+4 Y_{3} Y^{(4)}_{24}\\
-2 \sqrt{2} Y_{1} Y^{(4)}_{28}-\sqrt{2} Y_{2} Y^{(4)}_{27}-4 \sqrt{2} Y_{3} Y^{(4)}_{25}\\
-\sqrt{2} Y_{1} Y^{(4)}_{29}-4 \sqrt{2} Y_{2} Y^{(4)}_{28}-2 \sqrt{2} Y_{3} Y^{(4)}_{26}
\end{array}
\right)  \,,\\[2mm]
&&{Y^{(6)}_{\mathbf{8}c}}=\left(Y^{(2)}_{\mathbf{8}a}Y^{(4)}_{\mathbf{1}a}\right)_{\mathbf{8}}
\,=\,
Y^{(4)}_{1}\left(
\begin{array}{c}
Y_{11} \\
Y_{12} \\
Y_{13} \\
Y_{14}\\
Y_{15} \\
Y_{16} \\
Y_{17} \\
Y_{18}
\end{array}
\right)=Y^{(4)}_{1}Y^{(2)}_{\mathbf{8}a}(\tau)  \,,\\[2mm]
\label{eq:Yw68}&&{Y^{(6)}_{\mathbf{8}d}} =\left(Y^{(2)}_{\mathbf{8}b}Y^{(4)}_{\mathbf{1}a}\right)_{\mathbf{8}}
\,=\,
Y^{(4)}_{1}\left(
\begin{array}{c}
Y_{19} \\
Y_{20} \\
Y_{21} \\
Y_{22} \\
Y_{23} \\
Y_{24} \\
Y_{25} \\
Y_{26}
\end{array}
\right)=Y^{(4)}_{1}Y^{(2)}_{\mathbf{8}b}(\tau)  \,.
\end{eqnarray}
For modular forms of weight 8 ($k=4$), we find
\begin{equation}\label{eq:Yw8l}
Y^{(8)}_{\mathbf{1}} =\left(Y^{(4)}_{\mathbf{1}a}Y^{(4)}_{\mathbf{1}a}\right)_{\mathbf{1}}   \,=\, (Y^{(4)}_{1})^2 \,.
\end{equation}
%
\begin{eqnarray}
&&{Y^{(8)}_{\mathbf{3}a}}
=\left(Y^{(4)}_{\mathbf{1}a}Y^{(4)}_{\mathbf{3}a}\right)_{\mathbf{3}}
\,=\,
 Y^{(4)}_{1}\left(
\begin{array}{c}
Y^{(4)}_{2}\\
Y^{(4)}_{3} \\
Y^{(4)}_{4}
\end{array}
\right)= Y^{(4)}_{1}{Y^{(4)}_{\mathbf{3}a}}
  \,,\\[2mm]
\label{eq:Yw83}&&{Y^{(8)}_{\mathbf{3}b}} =\left(Y^{(4)}_{\mathbf{3}a}Y^{(4)}_{\mathbf{8}a}\right)_{\mathbf{3}}
\,=\,
\left(
\begin{array}{c}
\sqrt{3} Y^{(4)}_{2} Y^{(4)}_{31}+Y^{(4)}_{2} Y^{(4)}_{32}-\sqrt{6} Y^{(4)}_{3} Y^{(4)}_{38}-\sqrt{6} Y^{(4)}_{36} Y^{(4)}_{4}\\
\sqrt{6} Y^{(4)}_{2} Y^{(4)}_{33}-\sqrt{3} Y^{(4)}_{3} Y^{(4)}_{31}+Y^{(4)}_{3} Y^{(4)}_{32}-\sqrt{6} Y^{(4)}_{37} Y^{(4)}_{4}\\
\sqrt{6} Y^{(4)}_{2} Y^{(4)}_{35}+\sqrt{6} Y^{(4)}_{3} Y^{(4)}_{34}-2 Y^{(4)}_{32} Y^{(4)}_{4}
\end{array}
\right) \,.
\end{eqnarray}
\begin{equation}\label{eq:Yw83b}
{Y^{(8)}_{\mathbf{\bar{3}}}}
=\left(Y^{(4)}_{\mathbf{3}a}Y^{(4)}_{\mathbf{6}a}\right)_{\mathbf{\bar{3}}}
 \,=\,
\left(
\begin{array}{c}
\sqrt{2} Y^{(4)}_{2} Y^{(4)}_{9}+Y^{(4)}_{3} Y^{(4)}_{8}+Y^{(4)}_{4} Y^{(4)}_{6}\\
Y^{(4)}_{2} Y^{(4)}_{8}+\sqrt{2} Y^{(4)}_{3} Y^{(4)}_{7}+Y^{(4)}_{4} Y^{(4)}_{5}\\
\sqrt{2} Y^{(4)}_{10} Y^{(4)}_{4}+Y^{(4)}_{2} Y^{(4)}_{6}+Y^{(4)}_{3} Y^{(4)}_{5}
\end{array}
\right)
  \,,\\[2mm]
\end{equation}
%
\begin{eqnarray}
&&{Y^{(8)}_{\mathbf{6}a}}
=\left(Y^{(4)}_{\mathbf{1}a}Y^{(4)}_{\mathbf{6}a}\right)_{\mathbf{6}}
\,=\,
Y^{(4)}_{1}\left(
\begin{array}{c}
Y^{(4)}_{5}\\
Y^{(4)}_{6}\\
Y^{(4)}_{7}\\
Y^{(4)}_{8}\\
Y^{(4)}_{9}\\
Y^{(4)}_{10}
\end{array}
\right)= Y^{(4)}_{1}{Y^{(4)}_{\mathbf{6}a}}
 \,,\\[2mm]
&&{Y^{(8)}_{\mathbf{6}b}} =\left(Y^{(4)}_{\mathbf{1}a}Y^{(4)}_{\mathbf{6}b}\right)_{\mathbf{6}}
\,=\,
Y^{(4)}_{1}\left(
\begin{array}{c}
Y^{(4)}_{11}\\
Y^{(4)}_{12}\\
Y^{(4)}_{13}\\
Y^{(4)}_{14}\\
Y^{(4)}_{15}\\
Y^{(4)}_{16}
\end{array}
\right)= Y^{(4)}_{1}{Y^{(4)}_{\mathbf{6}b}}
  \,, \\[2mm]
&&{Y^{(8)}_{\mathbf{6}c}} =\left(Y^{(4)}_{\mathbf{3}a}Y^{(4)}_{\mathbf{3}a}\right)_{\mathbf{6}}
\,=\,
\left(
\begin{array}{c}
\sqrt{2} (Y^{(4)}_{4})^2\\
\sqrt{2} (Y^{(4)}_{2})^2\\
2 Y^{(4)}_{2} Y^{(4)}_{3}\\
\sqrt{2} (Y^{(4)}_{3})^2\\
2 Y^{(4)}_{2} Y^{(4)}_{4}\\
2 Y^{(4)}_{3} Y^{(4)}_{4}
\end{array}
\right)
 \,, \\[2mm]
\label{eq:Yw86}&&{Y^{(8)}_{\mathbf{6}d}}
=\left(Y^{(4)}_{\mathbf{3}a}Y^{(4)}_{\mathbf{7}a}\right)_{\mathbf{6}}
\,=\,
\left(
\begin{array}{c}
2 Y^{(4)}_{17} Y^{(4)}_{2}-\sqrt{2} Y^{(4)}_{21} Y^{(4)}_{4}-2 \sqrt{2} Y^{(4)}_{23} Y^{(4)}_{3}\\
2 Y^{(4)}_{17} Y^{(4)}_{3}-\sqrt{2} Y^{(4)}_{18} Y^{(4)}_{2}-2 \sqrt{2} Y^{(4)}_{22} Y^{(4)}_{4}\\
3 Y^{(4)}_{18} Y^{(4)}_{3}-Y^{(4)}_{19} Y^{(4)}_{2}-2 Y^{(4)}_{23} Y^{(4)}_{4}\\
2 Y^{(4)}_{17} Y^{(4)}_{4}-\sqrt{2} Y^{(4)}_{19} Y^{(4)}_{3}-2 \sqrt{2} Y^{(4)}_{2} Y^{(4)}_{20}\\
-Y^{(4)}_{18} Y^{(4)}_{4}+3 Y^{(4)}_{2} Y^{(4)}_{21}-2 Y^{(4)}_{20} Y^{(4)}_{3}\\
3 Y^{(4)}_{19} Y^{(4)}_{4}-2 Y^{(4)}_{2} Y^{(4)}_{22}-Y^{(4)}_{21} Y^{(4)}_{3}
\end{array}
\right) \,.
\end{eqnarray}
%
\begin{eqnarray}
&&{Y^{(8)}_{\mathbf{7}a}}
=\left(Y^{(4)}_{\mathbf{1}a}Y^{(4)}_{\mathbf{7}a}\right)_{\mathbf{7}}
\,=\,
Y^{(4)}_{1}\left(
\begin{array}{c}
Y^{(4)}_{17}\\
Y^{(4)}_{18}\\
Y^{(4)}_{19}\\
Y^{(4)}_{20}\\
Y^{(4)}_{21}\\
Y^{(4)}_{22}\\
Y^{(4)}_{23}
\end{array}
\right)= Y^{(4)}_{1}{Y^{(4)}_{\mathbf{7}a}}
  \,,\\[2mm]
&&{Y^{(8)}_{\mathbf{7}b}}
=\left(Y^{(4)}_{\mathbf{1}a}Y^{(4)}_{\mathbf{7}b}\right)_{\mathbf{7}}
\,=\,
Y^{(4)}_{1}\left(
\begin{array}{c}
Y^{(4)}_{24}\\
Y^{(4)}_{25}\\
Y^{(4)}_{26}\\
Y^{(4)}_{27}\\
Y^{(4)}_{28}\\
Y^{(4)}_{29}\\
Y^{(4)}_{30}
\end{array}
\right)= Y^{(4)}_{1}Y^{(4)}_{\mathbf{7}b}
  \,,\\[2mm]
&&{Y^{(8)}_{\mathbf{7}c}}
=\left(Y^{(4)}_{\mathbf{3}a}Y^{(4)}_{\mathbf{6}a}\right)_{\mathbf{7}}
\,=\,
\left(
\begin{array}{c}
2 Y^{(4)}_{10} Y^{(4)}_{2}+2 Y^{(4)}_{3} Y^{(4)}_{9}+2 Y^{(4)}_{4} Y^{(4)}_{7}\\
-2 \sqrt{2} Y^{(4)}_{10} Y^{(4)}_{3}-2 Y^{(4)}_{4} Y^{(4)}_{8}\\
-2 Y^{(4)}_{2} Y^{(4)}_{5}-2 \sqrt{2} Y^{(4)}_{4} Y^{(4)}_{9}\\
-\sqrt{2} Y^{(4)}_{10} Y^{(4)}_{4}+3 Y^{(4)}_{2} Y^{(4)}_{6}-Y^{(4)}_{3} Y^{(4)}_{5}\\
-2 \sqrt{2} Y^{(4)}_{2} Y^{(4)}_{7}-2 Y^{(4)}_{3} Y^{(4)}_{6}\\
-Y^{(4)}_{2} Y^{(4)}_{8}-\sqrt{2} Y^{(4)}_{3} Y^{(4)}_{7}+3 Y^{(4)}_{4} Y^{(4)}_{5}\\
-\sqrt{2} Y^{(4)}_{2} Y^{(4)}_{9}+3 Y^{(4)}_{3} Y^{(4)}_{8}-Y^{(4)}_{4} Y^{(4)}_{6}
\end{array}
\right)
  \,,\\[2mm]
\label{eq:Yw87}&&{Y^{(8)}_{\mathbf{7}d}}
=\left(Y^{(4)}_{\mathbf{3}a}Y^{(4)}_{\mathbf{6}b}\right)_{\mathbf{7}}
\,=\,
\left(
\begin{array}{c}
2 Y^{(4)}_{13} Y^{(4)}_{4}+2 Y^{(4)}_{15} Y^{(4)}_{3}+2 Y^{(4)}_{16} Y^{(4)}_{2}\\
-2 Y^{(4)}_{14} Y^{(4)}_{4}-2 \sqrt{2} Y^{(4)}_{16} Y^{(4)}_{3}\\
-2 Y^{(4)}_{11} Y^{(4)}_{2}-2 \sqrt{2} Y^{(4)}_{15} Y^{(4)}_{4}\\
-Y^{(4)}_{11} Y^{(4)}_{3}+3 Y^{(4)}_{12} Y^{(4)}_{2}-\sqrt{2} Y^{(4)}_{16} Y^{(4)}_{4}\\
-2 Y^{(4)}_{12} Y^{(4)}_{3}-2 \sqrt{2} Y^{(4)}_{13} Y^{(4)}_{2}\\
3 Y^{(4)}_{11} Y^{(4)}_{4}-\sqrt{2} Y^{(4)}_{13} Y^{(4)}_{3}-Y^{(4)}_{14} Y^{(4)}_{2}\\
-Y^{(4)}_{12} Y^{(4)}_{4}+3 Y^{(4)}_{14} Y^{(4)}_{3}-\sqrt{2} Y^{(4)}_{15} Y^{(4)}_{2}
\end{array}
\right)\,.
\end{eqnarray}
%
\begin{eqnarray}
&&{Y^{(8)}_{\mathbf{8}a}}
=\left(Y^{(4)}_{\mathbf{1}a}Y^{(4)}_{\mathbf{8}a}\right)_{\mathbf{8}}
\,=\,
Y^{(4)}_{1}\left(
\begin{array}{c}
Y^{(4)}_{31}\\
Y^{(4)}_{32}\\
Y^{(4)}_{33}\\
Y^{(4)}_{34}\\
Y^{(4)}_{35}\\
Y^{(4)}_{36}\\
Y^{(4)}_{37}\\
Y^{(4)}_{38}
\end{array}
\right)= Y^{(4)}_{1}{Y^{(4)}_{\mathbf{8}a}}
  \,,\\[2mm]
&&{Y^{(8)}_{\mathbf{8}b}}
=\left(Y^{(4)}_{\mathbf{1}a}Y^{(4)}_{\mathbf{8}b}\right)_{\mathbf{8}}
\,=\,
Y^{(4)}_{1}\left(
\begin{array}{c}
Y^{(4)}_{39}\\
Y^{(4)}_{40}\\
Y^{(4)}_{41}\\
Y^{(4)}_{42}\\
Y^{(4)}_{43}\\
Y^{(4)}_{44}\\
Y^{(4)}_{45}\\
Y^{(4)}_{46}
\end{array}
\right)= Y^{(4)}_{1}{Y^{(4)}_{\mathbf{8}b}}
  \,,\\[2mm]
&&  {Y^{(8)}_{\mathbf{8}c}}=\left(Y^{(4)}_{\mathbf{1}a}Y^{(4)}_{\mathbf{8}c}\right)_{\mathbf{8}}
\,=\,
Y^{(4)}_{1}\left(
\begin{array}{c}
Y^{(4)}_{47}\\
Y^{(4)}_{48}\\
Y^{(4)}_{49}\\
Y^{(4)}_{50}\\
Y^{(4)}_{51}\\
Y^{(4)}_{52}\\
Y^{(4)}_{53}\\
Y^{(4)}_{54}
\end{array}
\right)= Y^{(4)}_{1}{Y^{(4)}_{\mathbf{8}c}}
  \,,\\[2mm]
&&{Y^{(8)}_{\mathbf{8}d}}
=\left(Y^{(4)}_{\mathbf{3}a}Y^{(4)}_{\mathbf{6}a}\right)_{\mathbf{8}}
\,=\,
\left(
\begin{array}{c}
Y^{(4)}_{10} Y^{(4)}_{2}+Y^{(4)}_{3} Y^{(4)}_{9}-2 Y^{(4)}_{4} Y^{(4)}_{7}\\
\sqrt{3} Y^{(4)}_{3} Y^{(4)}_{9}-\sqrt{3} Y^{(4)}_{10} Y^{(4)}_{2}\\
\sqrt{2} Y^{(4)}_{10} Y^{(4)}_{3}-2 Y^{(4)}_{4} Y^{(4)}_{8}\\
\sqrt{2} Y^{(4)}_{4} Y^{(4)}_{9}-2 Y^{(4)}_{2} Y^{(4)}_{5}\\
2 Y^{(4)}_{3} Y^{(4)}_{5}-\sqrt{2} Y^{(4)}_{10} Y^{(4)}_{4}\\
2 Y^{(4)}_{3} Y^{(4)}_{6}-\sqrt{2} Y^{(4)}_{2} Y^{(4)}_{7}\\
\sqrt{2} Y^{(4)}_{3} Y^{(4)}_{7}-2 Y^{(4)}_{2} Y^{(4)}_{8}\\
\sqrt{2} Y^{(4)}_{2} Y^{(4)}_{9}-2 Y^{(4)}_{4} Y^{(4)}_{6}
\end{array}
\right)
  \,, \\[2mm]
&&{Y^{(8)}_{\mathbf{8}e}}
=\left(Y^{(4)}_{\mathbf{3}a}Y^{(4)}_{\mathbf{6}b}\right)_{\mathbf{8}}
\,=\,
\left(
\begin{array}{c}
-2 Y^{(4)}_{13} Y^{(4)}_{4}+Y^{(4)}_{15} Y^{(4)}_{3}+Y^{(4)}_{16} Y^{(4)}_{2}\\
\sqrt{3} Y^{(4)}_{15} Y^{(4)}_{3}-\sqrt{3} Y^{(4)}_{16} Y^{(4)}_{2}\\
\sqrt{2} Y^{(4)}_{16} Y^{(4)}_{3}-2 Y^{(4)}_{14} Y^{(4)}_{4}\\
\sqrt{2} Y^{(4)}_{15} Y^{(4)}_{4}-2 Y^{(4)}_{11} Y^{(4)}_{2}\\
2 Y^{(4)}_{11} Y^{(4)}_{3}-\sqrt{2} Y^{(4)}_{16} Y^{(4)}_{4}\\
2 Y^{(4)}_{12} Y^{(4)}_{3}-\sqrt{2} Y^{(4)}_{13} Y^{(4)}_{2}\\
\sqrt{2} Y^{(4)}_{13} Y^{(4)}_{3}-2 Y^{(4)}_{14} Y^{(4)}_{2}\\
\sqrt{2} Y^{(4)}_{15} Y^{(4)}_{2}-2 Y^{(4)}_{12} Y^{(4)}_{4}
\end{array}
\right)
  \,, \\[2mm]
\label{eq:Yw88}&&{Y^{(8)}_{\mathbf{8}f}}
=\left(Y^{(4)}_{\mathbf{3}a}Y^{(4)}_{\mathbf{7}a}\right)_{\mathbf{8}}
\,=\,
\left(
\begin{array}{c}
5 Y^{(4)}_{2} Y^{(4)}_{23}-4 Y^{(4)}_{20} Y^{(4)}_{4}-Y^{(4)}_{22} Y^{(4)}_{3}\\
-\sqrt{3} Y^{(4)}_{2} Y^{(4)}_{23}-2 \sqrt{3} Y^{(4)}_{20} Y^{(4)}_{4}+3 \sqrt{3} Y^{(4)}_{22} Y^{(4)}_{3}\\
-4 Y^{(4)}_{17} Y^{(4)}_{2}+2 \sqrt{2} Y^{(4)}_{21} Y^{(4)}_{4}-3 \sqrt{2} Y^{(4)}_{23} Y^{(4)}_{3}\\
-4 Y^{(4)}_{17} Y^{(4)}_{3}+2 \sqrt{2} Y^{(4)}_{18} Y^{(4)}_{2}-3 \sqrt{2} Y^{(4)}_{22} Y^{(4)}_{4}\\
2 \sqrt{2} Y^{(4)}_{18} Y^{(4)}_{3}+4 \sqrt{2} Y^{(4)}_{19} Y^{(4)}_{2}+\sqrt{2} Y^{(4)}_{23} Y^{(4)}_{4}\\
4 Y^{(4)}_{17} Y^{(4)}_{4}-2 \sqrt{2} Y^{(4)}_{19} Y^{(4)}_{3}+3 \sqrt{2} Y^{(4)}_{2} Y^{(4)}_{20}\\
-4 \sqrt{2} Y^{(4)}_{18} Y^{(4)}_{4}-2 \sqrt{2} Y^{(4)}_{2} Y^{(4)}_{21}-\sqrt{2} Y^{(4)}_{20} Y^{(4)}_{3}\\
-2 \sqrt{2} Y^{(4)}_{19} Y^{(4)}_{4}+\left(-\sqrt{2}\right) Y^{(4)}_{2} Y^{(4)}_{22}-4 \sqrt{2} Y^{(4)}_{21} Y^{(4)}_{3}
\end{array}
\right)  \,.
\end{eqnarray}
Because of space limitation, we only present modular forms transforming as $\mathbf{3}$ and $\mathbf{\bar{3}}$ under $\Gamma_7$ in the following. There are three linearly independent triplet modular forms $Y^{(10)}_{\mathbf{3}a}$, $Y^{(10)}_{\mathbf{3}b}$ and $Y^{(10)}_{\mathbf{3}c}$ of weight 10 :
\begin{equation}
\begin{aligned}
{Y^{(10)}_{\mathbf{3}a}}  &\,=\,\left(Y^{(4)}_{\mathbf{1}a}Y^{(6)}_{\mathbf{3}a}\right)_{\mathbf{3}} =
(Y^{(4)}_{1})^2\left(
\begin{array}{c}
Y_{1} \\
Y_{2} \\
Y_{3}
\end{array}
\right)
  \,,\\[2mm]
{Y^{(10)}_{\mathbf{3}b}}  &\,=\,\left(Y^{(4)}_{\mathbf{1}a}Y^{(6)}_{\mathbf{3}b}\right)_{\mathbf{3}} =
Y^{(4)}_{1}\left(
\begin{array}{c}
 Y_{1} \left(\sqrt{3} Y^{(4)}_{31}+Y^{(4)}_{32}\right)-\sqrt{6} (Y_{2} Y^{(4)}_{38}+Y_{3} Y^{(4)}_{36})\\
 \sqrt{6} (Y_{1} Y^{(4)}_{33}-Y_{3} Y^{(4)}_{37})+Y_{2} \left(Y^{(4)}_{32}-\sqrt{3} Y^{(4)}_{31}\right)\\
 \sqrt{6} (Y_{1} Y^{(4)}_{35}+Y_{2} Y^{(4)}_{34})-2 Y_{3} Y^{(4)}_{32}
\end{array}
\right)
 \,, \\[2mm]
{Y^{(10)}_{\mathbf{3}c}}  &\,=\,\left(Y^{(4)}_{\mathbf{3}a}Y^{(6)}_{\mathbf{1}}\right)_{\mathbf{3}} =
(Y_{10} Y^{(4)}_{25}+Y_{4} Y^{(4)}_{24}+Y^{(4)}_{26} Y_{9}+Y^{(4)}_{27} Y_{8}+Y^{(4)}_{28} Y_{7}+Y^{(4)}_{29} Y_{6}+Y^{(4)}_{30} Y_{5})\left(
\begin{array}{c}
Y^{(4)}_{2} \\
Y^{(4)}_{3} \\
Y^{(4)}_{4}
\end{array}
\right) \,.
\end{aligned}
\end{equation}
We have two linearly independent triplet modular forms $Y^{(10)}_{\mathbf{\bar{3}}a}$ and $Y^{(10)}_{\mathbf{\bar{3}}b}$ of weight 10, which can be chosen as
{\footnotesize\begin{equation}
\begin{aligned}
{Y^{(10)}_{\mathbf{\bar{3}}a}}  &\,=\,\left(Y^{(4)}_{\mathbf{1}a}Y^{(6)}_{\mathbf{\bar{3}}}\right)_{\mathbf{\bar{3}}} =
Y^{(4)}_{1}\left(
\begin{array}{c}
Y_{2} Y^{(4)}_{4}-Y_{3} Y^{(4)}_{3}\\
Y_{3} Y^{(4)}_{2}-Y_{1} Y^{(4)}_{4}\\
Y_{1} Y^{(4)}_{3}-Y_{2} Y^{(4)}_{2}
\end{array}
\right)
  \,,\\[2mm]
{Y^{(10)}_{\mathbf{\bar{3}}b}}  &\,=\,\left(Y^{(4)}_{\mathbf{3}a}Y^{(6)}_{\mathbf{3}b}\right)_{\mathbf{\bar{3}}} \\
&=\left(
\begin{array}{c}
Y^{(4)}_{3} \left(\sqrt{6} Y_{1} Y^{(4)}_{35}+\sqrt{6} Y_{2} Y^{(4)}_{34}-2 Y_{3} Y^{(4)}_{32}\right)+Y^{(4)}_{4} \left(-\sqrt{6} Y_{1} Y^{(4)}_{33}+\sqrt{3} Y_{2} Y^{(4)}_{31}-Y_{2} Y^{(4)}_{32}+\sqrt{6} Y_{3} Y^{(4)}_{37}\right)\\
-\sqrt{6} Y^{(4)}_{2} (Y_{1} Y^{(4)}_{35}+Y_{2} Y^{(4)}_{34})+Y_{1} Y^{(4)}_{4} \left(\sqrt{3} Y^{(4)}_{31}+Y^{(4)}_{32}\right)-\sqrt{6} Y^{(4)}_{4} (Y_{2} Y^{(4)}_{38}+Y_{3} Y^{(4)}_{36})+2 Y_{3} Y^{(4)}_{2} Y^{(4)}_{32}\\
Y^{(4)}_{2} \left(\sqrt{6} Y_{1} Y^{(4)}_{33}-\sqrt{3} Y_{2} Y^{(4)}_{31}+Y_{2} Y^{(4)}_{32}-\sqrt{6} Y_{3} Y^{(4)}_{37}\right)+Y^{(4)}_{3} \left(-\sqrt{3} Y_{1} Y^{(4)}_{31}-Y_{1} Y^{(4)}_{32}+\sqrt{6} Y_{2} Y^{(4)}_{38}+\sqrt{6} Y_{3} Y^{(4)}_{36}\right)
\end{array}
\right)  \,.
\end{aligned}
\end{equation} }

\end{appendix}

\providecommand{\href}[2]{#2}\begingroup\raggedright\endgroup

\end{document}